\documentclass[11pt]{amsart}
\usepackage{amsmath,amssymb,amsthm,mathrsfs,graphicx}
\usepackage[utf8]{inputenc}
\usepackage[T1]{fontenc}
\usepackage{enumitem}
\usepackage[margin=1.1in]{geometry}
\usepackage{xurl}
\usepackage[font=small,skip=4pt]{caption}
\allowdisplaybreaks
\usepackage[colorlinks=true,linkcolor=blue,citecolor=blue,urlcolor=blue]{hyperref}

\numberwithin{equation}{section}
\theoremstyle{plain}
\newtheorem{theorem}{Theorem}[section]
\newtheorem{proposition}[theorem]{Proposition}
\newtheorem{lemma}[theorem]{Lemma}
\newtheorem{corollary}[theorem]{Corollary}
\newtheorem{fact}[theorem]{Fact}
\newtheorem*{theoremA}{Theorem A}
\newtheorem*{theoremB}{Theorem B}
\newtheorem*{corollaryC}{Corollary C}
\theoremstyle{definition}
\newtheorem{definition}[theorem]{Definition}
\newtheorem{assumption}[theorem]{Assumption}
\newtheorem{setting}[theorem]{Setting}
\theoremstyle{remark}
\newtheorem{remark}[theorem]{Remark}

\makeatletter\newcommand{\SerAnum}[1]{\@ifundefined{SerA@#1}{[A:#1]}{\@nameuse{SerA@#1}}}\makeatother
\makeatletter
\@namedef{SerA@sec:model}{Section~3}
\@namedef{SerA@sec:periodization}{Section~5}
\@namedef{SerA@thm:periodization}{Theorem~5.1}
\@namedef{SerA@eq:W}{condition~(W$\pm$)}
\@namedef{SerA@cor:hardy}{Corollary~6.7}
\@namedef{SerA@prop:configBO}{Proposition~6.9}
\@namedef{SerA@sec:finite-model}{Section~6.1}
\@namedef{SerA@lem:Qform}{Lemma~8.1}
\makeatother

\newcommand{\R}{\mathbb{R}}
\newcommand{\C}{\mathbb{C}}

\newcommand{\Z}{\mathbb{Z}}
\newcommand{\F}{\mathcal{F}}
\newcommand{\h}{\mathfrak h}
\newcommand{\abar}{\bar\alpha}
\newcommand{\dG}{\mathrm{d}\Gamma}
\newcommand{\Hf}{H_{\mathrm f}}
\newcommand{\Pf}{P_{\mathrm f}}
\newcommand{\hls}{h_{\mathrm{ls}}}

\newcommand{\iPF}{\iota_{\mathrm{PF}}}

\newcommand{\ee}{\mathrm e}

\newcommand{\Xh}{\hat X}

\newcommand{\Rea}{\mathrm{Re}\,}

\newcommand{\Qf}{\mathcal Q}
\newcommand{\vac}{{\rm vac}}
\newcommand{\Rk}{R(k)}
\newcommand{\ip}[2]{\langle #1,#2\rangle}
\newcommand{\Xw}{X^{(2)}_{w}}
\newcommand{\Rm}{\mathsf{R}}
\newcommand{\Dm}{\mathsf{D}}
\newcommand{\Wext}{W_{\mathrm{ext}}}
\newcommand{\Win}{W_{\mathrm{in}}}
\newcommand{\Ebox}{E_{\mathrm{box}}}
\newcommand{\Hbox}{H_{\mathrm{box}}}
\newcommand{\Hee}{H_{\mathrm{ee}}}
\newcommand{\Hie}{H_{\mathrm{ie}}}
\newcommand{\DEl}{D_{\mathrm{El}}}
\newcommand{\Lexc}{\mathcal{L}_{\mathrm{exc}}}
\newcommand{\GAg}{\Gamma^{\mathrm{Ag}}}
\newcommand{\Sch}{\check S}
\newcommand{\dd}{\,\mathrm{d}}
\newcommand{\diam}{\operatorname{diam}}
\newcommand{\tr}{\operatorname{tr}}
\newcommand{\supp}{\operatorname{supp}}
\newcommand{\dist}{\operatorname{dist}}
\newcommand{\vol}{\operatorname{vol}}
\newcommand{\transl}{\mathcal T}
\newcommand{\sat}{\varsigma}

\newcommand{\mAg}{\mu_{\mathrm A}}
\newcommand{\Mzero}{\mathcal M_0}
\newcommand{\rrel}{r_{\mathrm{rel}}}
\newcommand{\Kie}{{\omega^2_{\mathrm{ie}}}}
\newcommand{\Kosc}{\mathsf K}
\newcommand{\Tgrad}{T_\nabla}
\newcommand{\Tgradmax}{\overline T_\nabla}
\newcommand{\Eloss}{E_{\mathrm{loss}}}
\newcommand{\LipXi}{\mathrm{Lip}_\Xi}

\begin{document}

\title[The effective mass of the Pauli--Fierz electron]{The effective mass of the Pauli--Fierz electron:\\ a cutoff-independent lower bound and the breakdown\\ of mass renormalization}

\author{Shinichiro Futakuchi}
\address{Mugen Giken LLC, Tokyo, Japan}
\email{futakuchi0117@gmail.com}
\urladdr{https://orcid.org/0009-0000-5027-0517}


\begin{abstract}
We study the effective mass of a spinless non-relativistic electron coupled to the quantized transverse electromagnetic field (the Pauli--Fierz model without external potential), with bare mass $m$, fine structure constant $\alpha=e^2/4\pi$ and a sharp ultraviolet cutoff $\Lambda$. We define $1/(2m_{\rm eff}):=\limsup_{P\to0}(E(P)-E(0))/|P|^2$ for the fiber ground state energy $E(P)$; this agrees with the curvature mass whenever $E$ is twice differentiable at zero. By scaling, $h=m_{\rm eff}/m$ depends only on $\alpha$ and $\lambda=\Lambda/m$. For all $\alpha>0$ and $\lambda\ge1$ we prove $h\ge1+0.3441\,\alpha/(1+0.2985\sqrt\alpha+0.1252\,\alpha)$, a bound independent of the cutoff; with the band extended to the cutoff, its first order coefficient in $\alpha$ is the perturbative one of Hiroshima and Spohn. At fixed $\alpha>0$ we further prove $h\ge\lambda^{8/7-\epsilon}$ for every $\epsilon>0$ and all sufficiently large $\lambda$. Hence $\iota_{\rm PF}(\alpha):=\inf_{\lambda>0}h/\lambda>0$: at fixed coupling, a bare mass reproducing a prescribed physical mass $m_{\rm phys}$ can exist only if $\Lambda\le m_{\rm phys}/\iota_{\rm PF}(\alpha)$. This proves the obstruction to mass renormalization at fixed physical mass anticipated by Lieb and Loss (2002). The first bound uses a coherent translation of the field applied to the ground state as a trial state. The second uses the sharp self-energy asymptotics of a companion paper and a position estimator constructed from the local centre of a shallow level set of the photon cloud.
\end{abstract}

\maketitle

\setcounter{tocdepth}{1}
\tableofcontents

\section{Introduction}\label{sec:intro}

\subsection{The question}
The Pauli--Fierz model \cite{PF38} describes a non-relativistic charged particle coupled, through minimal coupling, to the quantized transverse electromagnetic field with an ultraviolet cutoff $\Lambda$; see Spohn's monograph \cite{Sp04} for its place in the theory of radiating charges. Its Hamiltonian is self-adjoint for all values of the coupling \cite{Hir02,HH08}, and in the translation invariant case it decomposes into fiber Hamiltonians $H(P)$ labelled by the total momentum $P$ \cite{Hi07,LMS07}. With a sharp cutoff and without spin, the fiber at $P=0$ has a ground state for all values of the coupling \cite{HSi23}, unique up to a phase \cite{Hi07}. Because the electron is dressed by the field, its free motion is no longer governed by its bare mass $m$. The effective mass is defined by the curvature of $P\mapsto E(P)=\inf\sigma(H(P))$ at $P=0$,
\[
E(P)-E(0)=\frac{P^2}{2m_{\mathrm{eff}}}+o(P^2),
\]
when $E$ is twice differentiable at $P=0$; in general we define $m_{\mathrm{eff}}$ through the $\limsup$ of second difference quotients (Definition~\ref{def:masses}), so that differentiability of $E$ is neither assumed nor proved.

We ask how $m_{\rm eff}$ depends on the cutoff. By scaling (Lemma~\ref{lem:scaling}), the ratio $h=m_{\rm eff}/m$ depends only on the fine structure constant $\alpha=e^2/4\pi$ and on $\lambda=\Lambda/m$; we write $h(\lambda)$ for fixed $\alpha$. Hiroshima and Spohn \cite{HSp05} formulated the question of mass renormalization in these terms. From second order perturbation theory with an infrared cutoff they derived a formula for $m/m_{\rm eff}$ (their (1.13)), and observed that the nature of mass renormalization is decided by the growth of $h(\lambda)$: if $h$ stays bounded the renormalization is finite, if $h$ grows linearly in $\lambda$ the theory cannot be renormalized in this way, and the intermediate case $h\simeq b_0\lambda^\gamma$ with $0<\gamma<1$ is the one of interest (their (1.14)--(1.20)). They also noted that no technique was available to determine the behaviour of $h$ for large $\lambda$. Their first order coefficient is
\begin{equation}\label{eq:HS-first-order}
h=1+\frac{8\alpha}{3\pi}\log\Bigl(1+\frac\lambda2\Bigr)+O(\alpha^2)
\end{equation}
(\cite[(1.22)]{HSp05}, with the infrared cutoff removed under an assumption stated there). Their main result is the second order term: in the same setting, $h(\lambda)=1+\frac{8\alpha}{3\pi}\log(1+\frac\lambda2)+c_0\alpha^2\sqrt\lambda+O(\alpha^3)$ with $c_0>0$, for $|\alpha|$ small depending on $\Lambda$ \cite[(1.26)]{HSp05}; this does not confirm the guess $h\simeq\lambda^{8\alpha/3\pi}$ suggested by the first order and, as they note, could suggest $\gamma=\frac12$.

Lieb and Loss \cite[Sect.~1]{LL02}, where upper bounds on binding energies are proved, compared two definitions of the renormalized mass. If the mass is fixed through the binding energy, a bare mass can be found for every $\Lambda$ and $m_{\mathrm{phys}}$. For the usual definition through the curvature of $E(P)$, they expected that the equation $m\,h(\Lambda/m)=m_{\rm phys}$ has a solution $m>0$ only if $\Lambda$ and $1/m_{\mathrm{phys}}$ are sufficiently small; they supported this by perturbation theory, classical electrodynamics, and van Kampen's exact solution of Kramers' dipole approximation model, in which $m_{\mathrm{phys}}=C'm+\alpha\Lambda$ up to conventions, and added that they could not prove it. The following elementary lemma makes the connection between the two formulations precise.

\begin{lemma}\label{lem:LL-equivalence}
Fix $\alpha$, let $h:(0,\infty)\to[1,\infty]$ and put $F_\Lambda(m):=m\,h(\Lambda/m)$.
\begin{enumerate}[label=(\roman*)]
\item If $\liminf_{\lambda\to\infty}h(\lambda)/\lambda>0$, then $g_*:=\inf_{\lambda>0}h(\lambda)/\lambda>0$, and for all pairs $(\Lambda,m_{\rm phys})$ with $m_{\rm phys}/\Lambda<g_*$ the equation $F_\Lambda(m)=m_{\rm phys}$ has no solution $m>0$.
\item If $h$ is continuous and $\liminf_{\lambda\to\infty}h(\lambda)/\lambda=0$, then the equation has a solution for all $(\Lambda,m_{\rm phys})$.
\end{enumerate}
In particular, for continuous $h$, the absence of solutions for large $\Lambda/m_{\rm phys}$ is equivalent to $\liminf_{\lambda\to\infty}h(\lambda)/\lambda>0$.
\end{lemma}

\begin{proof}
With $\lambda=\Lambda/m$ we have $F_\Lambda(m)=\Lambda\,h(\lambda)/\lambda$, and $\lambda$ runs through $(0,\infty)$ as $m$ does; hence the range of $F_\Lambda$ is $\Lambda\cdot\{h(\lambda)/\lambda:\lambda>0\}$. Since $h\ge1$, $h(\lambda)/\lambda\ge1/R$ for $\lambda\le R$, so $g_*>0$ if and only if $\liminf_{\lambda\to\infty}h/\lambda>0$. Part (i) follows because the range is contained in $[\Lambda g_*,\infty)$. For (ii), $h/\lambda$ is continuous, tends to $\infty$ as $\lambda\to0$, and has a sequence tending to $0$; by the intermediate value theorem its range is $(0,\infty)$.
\end{proof}

Thus the statement of Lieb and Loss is equivalent (given continuity of $h$) to $\iPF(\alpha):=\inf_{\lambda>0}h(\lambda)/\lambda>0$, that is, to a lower bound on the effective mass that grows linearly in the cutoff.

\subsection{Results}
Our first result is a lower bound that does not depend on the cutoff.

\begin{theoremA}[Theorem~\ref{thm:partI} and Corollaries~\ref{cor:partI-explicit}, \ref{cor:partI-first-order}]
For all $m,\Lambda>0$ and all real $e$ with $\lambda=\Lambda/m\ge1$,
\[
h\ \ge\ 1+\frac{0.3441\,\alpha}{1+0.2985\sqrt\alpha+0.1252\,\alpha}.
\]
Moreover, for every $\lambda>0$, $\liminf_{\alpha\to0}(h-1)/\alpha\ge\frac{8}{3\pi}\log(1+\frac\lambda2)$, the first order coefficient in \eqref{eq:HS-first-order}.
\end{theoremA}

The inequality $h\ge1$ follows from using the ground state as a trial state; Theorem~A improves it by an amount that is independent of the cutoff, for every $\alpha>0$. Our second result is that $h$ grows faster than linearly.

\begin{theoremB}[Theorem~\ref{thm:main}]
Fix $\alpha>0$. Then $h(\lambda)/\lambda\to\infty$ as $\lambda\to\infty$; more precisely, for every $\epsilon>0$ there is $\lambda_\epsilon$ such that $h(\lambda)\ge\lambda^{8/7-\epsilon}$ for all $\lambda\ge\lambda_\epsilon$. In particular $\iPF(\alpha)>0$.
\end{theoremB}

\begin{corollaryC}[Corollary~\ref{cor:LL}]
Fix $\alpha>0$ and a physical mass $m_{\rm phys}$. A bare mass $m>0$ with $m\,h(\Lambda/m)=m_{\rm phys}$ exists only if $\Lambda\le\Lambda_c:=m_{\rm phys}/\iPF(\alpha)$. Thus, at fixed coupling, mass renormalization with fixed physical mass is impossible above a finite cutoff. The condition $\Lambda\le\Lambda_c$ is necessary; we do not prove that it is sufficient (Remark~\ref{rem:nonclaims}(vi)).
\end{corollaryC}

Theorem~B yields every exponent below $8/7$; the present estimates do not reach the endpoint, and $8/7$ is not claimed to be the true growth exponent of $h$. An upper bound on the growth of $h(\lambda)$ as $\lambda\to\infty$ and a quantitative value of $\lambda_\epsilon$ remain open (a bound uniform in the cutoff is excluded by Theorem~B itself). Combining the two theorems, $h(\lambda)\ge1+\delta(\alpha)$ for all $\lambda\ge1$, with the explicit $\delta(\alpha)$ of Theorem~A, and $h(\lambda)\ge\lambda^{8/7-\epsilon}$ for $\lambda\ge\lambda_\epsilon$. Theorem~B is a mathematical asymptotic statement and gives no quantitative conclusion at physical values of the cutoff (of the order of the electron mass).

Physically, Theorem~B differs qualitatively from the dipole approximation, in which the mass grows linearly, $m_{\rm phys}=C'm+\alpha\Lambda$ (see above). In terms of the bare mass it reads $m_{\rm eff}=m\,h(\Lambda/m)\ge\Lambda^{8/7-\epsilon}m^{-1/7+\epsilon}$ for $\Lambda/m\ge\lambda_\epsilon$; for $\epsilon<1/7$ and fixed cutoff, the effective mass therefore tends to infinity as the bare mass tends to zero, so that a small physical mass cannot be produced by making the bare mass small. In the classification of Hiroshima and Spohn recalled above, if $h\simeq b_0\lambda^\gamma$ then Theorem~B forces $\gamma\ge8/7>1$, outside the cases $0\le\gamma\le1$ considered there; in particular the exponent $\gamma=\frac12$ suggested by their second order computation does not describe the behaviour at fixed $\alpha$, which is consistent since their expansion is not uniform in $\lambda$ (compare the window $\alpha\lambda^2\ll\log\lambda$ of the first order bound in Remark~\ref{rem:partI-values}(iii)). Hiroshima and Spohn also recall that in the Nelson model $h$ has a finite limit as $\lambda\to\infty$, so that mass renormalization is finite \cite[after (1.15)]{HSp05}; by Theorem~B the Pauli--Fierz model is not of this type.

\subsection{Methods}
Both parts bound the curvature of $E(P)$ at $P=0$ from above by a trial state obtained from the ground state $\psi_0$ of $H(0)$ by a unitary ``boost'', and the term linear in the boost parameter drops out: in Part~I because $\psi_0$ is a ground state, in Part~II because $E(p\hat z)\ge E(0)$ for all $p$. The two boosts are of a different nature, and the two parts are logically independent: Part~I uses only the model of Section~\ref{sec:model}, while Part~II uses the self-energy asymptotics of \cite{SerA} and, from Part~I, only the elementary bound $h\ge1$ (Definition~\ref{def:masses}). Part~I gives a bound at every finite cutoff with an explicit constant; Part~II gives the asymptotic growth.

\emph{Part~I.} The trial state is $W_\theta\psi_0$, where $W_\theta=\exp(\theta(a^*(f)-a(f)))$ translates the vector potential by a c-number field $\theta f$, and $f$ is a one-photon function in the band $|k|\le m$, essentially the first order perturbative correction to the electric field. Conjugation by $W_\theta$ turns the energy of $W_\theta\psi_0$ in the fiber $H(p\epsilon)$ into an exact quadratic polynomial in $(p,\theta)$ (Lemma~\ref{lem:boost}). The coupling of $p$ and $\theta$ comes from the c-number part of the commutator of the minimally coupled velocity $V=\Pf+eA(0)$ with creation operators, combined with transversality; for the linear coupling of the polaron, $[\Pf,a^*(f)]=a^*(kf)$ has no c-number part, and this mechanism is absent. The only remaining unknown is the annihilation amplitude $\lVert a(k_\mu f)\psi_0\rVert$ of the ground state (a single annihilation operator applied to $\psi_0$), which a naive bound would make of order $\sqrt\alpha\Lambda$. We bound it uniformly in $\Lambda$ by the pull-through formula, a comparison of resolvents of shifted fibers that follows from the diamagnetic inequality $E(P)\ge E(0)$ used for a whole one-parameter family of momenta, and the inequality part $\lVert(H(0)-E_0)^{-1/2}V_z\psi_0\rVert^2\le m/2$ of the sum rule, which appears in \cite[(1.13)]{HSp05} as an identity derived from perturbation theory.

\emph{Part~II.} The trial state is $\ee^{ip\Xh}\psi$, where $\Xh$ is a function of the field configuration, an \emph{estimator} of the position of the electron read off from the photon cloud. This gives
\[
\frac ch\ \le\ c\,\delta(\Xh)+\langle\Gamma_{\Xh}\rangle,
\]
where $c\asymp\lambda$ is the kinetic coefficient in our units, $\delta(\Xh)$ measures the failure of $\Xh$ to move rigidly under translations of the field, and $\Gamma_{\Xh}$ is the squared field gradient of $\Xh$ (its carr\'e du champ). Thus $h\ge c/C$ follows once we find an estimator with $c\,\delta\lesssim1$ and $\langle\Gamma_{\Xh}\rangle\lesssim1$; physically, $1/\langle\Gamma_{\Xh}\rangle$ is the inertia of the cloud conjugate to $\Xh$. In the strong ultraviolet regime the electron is localized on a length $\ell\asymp\lambda^{1/7}$, and the surrounding field forms a ``bubble'' whose shape is governed by a mean-field functional \cite{SerA}. The deep core of the bubble can be located precisely, but an estimator based on the core is sensitive to the steep wall of the core and yields only $\langle\Gamma\rangle\lesssim\ell$ up to logarithms. We therefore use the local centre $\Xw$ of a much shallower level set, at field level $g_w\asymp\ell^{-\kappa}$ with $0<\kappa<1$, in a window around the core (Definition~\ref{def:Xw}). Its fluctuation is bounded deterministically by the square of the window radius $\rho$ times the sensitivity $1/g_w\asymp\ell^\kappa$ of the level, divided by the volume $\gtrsim\ell^3$ of the shallow depressions in the window (Lemma~\ref{lem:Mw}); with $\rho\asymp\ell\log\lambda$ this gives $\langle\Gamma\rangle\lesssim\ell^{\kappa-1}(\log\lambda)^2\to0$. Since every $0<\kappa<1$ is admissible, this yields $h\gtrsim\lambda\,\ell^{1-\kappa}(\log\lambda)^{-2}$ and hence the exponent $8/7-\epsilon$ of Theorem~B.

The remaining estimate controls shallow depressions in the transition shell $[\rho/2,\rho]$ of the window. Such a depression couples the estimator to the fluctuations of the centre of the window, and the loss $\ell$ returns. Proposition~\ref{prop:shell} shows that a shallow depression in this shell occurs with probability at most $\ell^{-\kappa_p}$, for every $\kappa_p$; outside this event the derivative of the estimator through the window centre vanishes (Lemma~\ref{lem:Mw}). Its proof is an Agmon-type estimate based on an energy margin of configurations with such a depression. To extract the margin locally, one splits off a finite-dimensional ``external'' part of the field near the depression. This external field is also coupled to the electron; we choose it among divergence-free fields with vanishing normal component on the boundary of the external region, which have no gradient part, so that the decoupling costs only super-polynomially small errors once the depression is farther than the distance $\ell^{1/2}$ from the electron (Theorem~\ref{thm:gauge} and Corollary~\ref{cor:far}). The candidate locations of depressions are chosen on an absolute lattice determined by a localization cell of the electron alone, so that the sharpness of the selection does not multiply the sensitivity of the position (Lemma~\ref{lem:softmax}). The remaining localization of the depression mass of almost minimizing configurations is proved in Section~\ref{sec:constructive}.

\subsection{Relation to previous work}
For the ground state energy in the ultraviolet limit, Lieb and Loss \cite{LL99} proved bounds of order $\alpha^{1/2}\Lambda^{3/2}$ and $\alpha^{2/7}\Lambda^{12/7}$ and obtained the upper bound from product states; Bach and Hach \cite{BH22} studied the asymptotics of this upper bound, and \cite{SerA} identifies the sharp leading term $\alpha^{2/7}\Lambda^{12/7}$. Bach, Ballesteros and Mlinarzik \cite{BBM26}, Herdzik \cite{He25} and Bach and Herdzik \cite{BHe26} recently studied the ultraviolet behaviour through Bogoliubov--Hartree--Fock functionals. These works concern the ground state energy, not the dependence of the effective mass on $\Lambda$.

Rigorous results on the effective mass in non-relativistic QED concern small coupling at fixed cutoff. Hainzl and Seiringer \cite{HS02} (with spin $\frac12$) computed the leading corrections to the effective mass and to binding energies, and Hiroshima and Spohn \cite{HSp05} the perturbative coefficients of mass renormalization (for spin $\frac12$ see Hiroshima and Ito \cite{HIt07}). Bach, Chen, Fr\"ohlich and Sigal \cite{BCFS07}, Chen \cite{Ch08} and Fr\"ohlich and Pizzo \cite{FP10} established the regularity of $E(P)$ near $P=0$ and the existence and properties of the renormalized mass by renormalization group and infrared analysis; at fixed cutoff and small coupling, bounds of the form $1\le m_{\mathrm{ren}}/m\le1+c\alpha$ are proved in \cite[Thm.~2.1]{BCFS07} (spinless, with a smooth ultraviolet cutoff function, uniformly in an infrared regularization parameter $\sigma\ge0$) and \cite{Ch08} (spin $\frac12$). The existence of ground states for confined particles goes back to Bach, Fr\"ohlich and Sigal \cite{BFS98} and Griesemer, Lieb and Loss \cite{GLL01}. Hiroshima \cite{Hi25} studied the effective mass in a weak coupling limit, again at fixed cutoff; that preprint has since been withdrawn by its author. None of these results is uniform in the cutoff, so that the novelty of Theorem~A lies in its uniformity in $\Lambda$, its validity for all $\alpha$ and its explicit constant.

\emph{Comparison with the Fr\"ohlich polaron.} For the polaron at strong coupling, the divergence of the effective mass was established in stages: Spohn \cite{Sp87} gave a functional integral formulation of the Landau--Pekar conjecture $m_{\mathrm{eff}}\sim\alpha^4$ \cite{LP48}; Lieb and Seiringer \cite{LS20} proved divergence without a rate; Betz and Polzer \cite{BP23} proved $m_{\mathrm{eff}}\ge c\alpha^{2/5}$; Sellke \cite{Se24} proved an almost quartic lower bound, and Bazaes, Mukherjee, Sellke and Varadhan \cite{BMSV26} the bound $m_{\mathrm{eff}}\ge c\alpha^4$ of the predicted order; and Brooks and Seiringer \cite{BS24} (upper bound) and Brooks \cite{Br26} (matching lower bound) established the Landau--Pekar formula for the curvature mass. Dybalski and Spohn \cite{DS20} related the inverse effective mass to the diffusion constant of the path measure, and Lieb and Seiringer \cite{LS14} proved the equivalence of the curvature mass and a mass defined through binding energies for polaron type models; for the Pauli--Fierz model this relation is open, and all our statements concern the curvature mass. In the Pauli--Fierz model the role of the coupling constant is taken by the cutoff: Theorem~B plays the role of a divergence with a rate, with $\lambda\to\infty$ in place of $\alpha\to\infty$. At fixed $\lambda$, the bound of Part~I with a band shrinking like $\alpha^{-1/2}$ gives $h\ge(0.5865+o(1))\sqrt\alpha$ as $\alpha\to\infty$ (Remark~\ref{rem:partI-values}(ii)); the true rate is open. The general architecture of Part~II goes back to \cite{LS20}: a quantitative energy bound is transferred to a concentration statement for the state, and then to a curvature bound through a trial state, which in \cite{LS20} is built from a function of the field momentum and here from a function of the field configuration. Two differences from the polaron problem should be noted. First, in the Fr\"ohlich model the ultraviolet cutoff can be removed at fixed coupling, whereas in the Pauli--Fierz model at fixed bare mass the effective mass itself diverges as $\Lambda\to\infty$, so that a bound uniform in $\Lambda$, as in Theorem~A, is meaningful only at each finite $\Lambda$. Second, the shallow-level estimator and the localization of Sections~\ref{sec:surgery}--\ref{sec:shell} have no counterpart in the polaron problem, where the photon cloud has no steep core.

To our knowledge, neither Theorem~A nor the statement $\iPF(\alpha)>0$ has been proved before; we do not claim that our literature search is exhaustive.

\subsection{Caveats}
First, the proof of Theorem~B does not track $\lambda_\epsilon$ or $\lambda_*$, and we say nothing about physically accessible values of $\lambda$, where the apparent growth of $h$ may be pre-asymptotic. Second, Part~II has little room to spare: the decisive fluctuation bound decays only like $\ell^{\kappa-1}$ up to logarithms, and the proof needs $0<\kappa<1$ (Lemma~\ref{lem:Mw} and Appendix~\ref{app:terms}). Third, Part~II uses the sharp self-energy asymptotics of \cite{SerA} in a configuration-wise form; everything else is proved here, the more technical field lemmas and the properties of the core estimator in Appendices~\ref{app:cited} and~\ref{app:core}. Fourth, we do not determine the value of $\iPF$ (Section~\ref{sec:discussion} collects what is known).
\subsection{Organization}
Section~\ref{sec:model} introduces the model, the effective mass, the scaling relations and the main theorems. Part~I (Sections~\ref{sec:partI-result}--\ref{sec:partI-proof}) proves Theorem~A: Section~\ref{sec:partI-result} states the result, Section~\ref{sec:framework} proves abstract statements for fiber families with a diamagnetic inequality, and Section~\ref{sec:partI-proof} treats the coherent trial state. Part~II (Sections~\ref{sec:prelim}--\ref{sec:discussion}) proves Theorem~B: Section~\ref{sec:prelim} collects what we use from \cite{SerA} and proves the general operator tools, Section~\ref{sec:estimators} introduces the estimators, Section~\ref{sec:surgery} decouples the external field, Section~\ref{sec:shell} proves the rarity of shallow depressions in the window shell, Section~\ref{sec:proof} proves the fluctuation bound and Theorem~\ref{thm:main}, Section~\ref{sec:constructive} localizes the depression mass, and Section~\ref{sec:discussion} discusses where the difficulty went and open problems. Appendices~\ref{app:facts}--\ref{app:integrals} belong to Part~I (facts from the literature, the pull-through formula, the constants), and Appendices~\ref{app:terms}--\ref{app:core} to Part~II (bookkeeping of large factors and the technical lemmas).

\section{Model and main results}\label{sec:model}

\subsection{The Pauli--Fierz Hamiltonian and its fibers}
Let $\h:=L^2(\R^3\times\{1,2\})$ and let $\F:=\F_{\rm b}(\h)$ be the boson Fock space over $\h$, with annihilation and creation operators $a(k,r)$, $a^*(k,r)$. The polarization vectors $\varepsilon_r(k)$, $r=1,2$, depend only on $\hat k=k/|k|$, and $\{\hat k,\varepsilon_1(k),\varepsilon_2(k)\}$ is an orthonormal basis of $\R^3$; $\varepsilon_r$ is a measurable function of $\hat k$, and no statement below depends on its choice. The quantized vector potential in Coulomb gauge with sharp cutoff $|k|\le\Lambda$ is
\[
A_\Lambda(y)=\sum_{r=1,2}\int_{|k|\le\Lambda}\frac{\varepsilon_r(k)}{(2\pi)^{3/2}\sqrt{2|k|}}\Bigl(\ee^{ik\cdot y}a(k,r)+\ee^{-ik\cdot y}a^*(k,r)\Bigr)\,dk ,
\]
and $\Hf=\dG(\omega)$ with $\omega(k)=|k|$, $\Pf=\dG(k)$. We use units with $\hbar=c_{\rm light}=1$ and the Heaviside--Lorentz convention: the electron has charge $e\in\R$, and $\alpha:=e^2/4\pi$ is the fine structure constant, as in \cite[(1.21)]{HSp05}. In Part~II we abbreviate $\abar:=e^2=4\pi\alpha$.

For bare mass $m>0$ the spinless Pauli--Fierz Hamiltonian without external potential is $H=\frac1{2m}(p_x-eA_\Lambda(x))^2+\Hf$ on $L^2(\R^3)\otimes\F$. It commutes with the total momentum $p_x+\Pf$, and the Lee--Low--Pines transformation $\ee^{i\Pf\cdot x}$ decomposes it into the fibers
\begin{equation}\label{eq:fiber}
H_{m,\Lambda}(P)=\frac1{2m}\bigl(P-V\bigr)^2+\Hf,\qquad V:=\Pf+eA_\Lambda(0),\qquad D(H_{m,\Lambda}(P))=D(\Hf)\cap D(\Pf^2),
\end{equation}
$P\in\R^3$. We write $E_{m,\Lambda}(P):=\inf\sigma(H_{m,\Lambda}(P))$. The sign of $e$ plays no role: the unitary reflection $\Gamma(-1)$ of the field maps $A_\Lambda(y)$ to $-A_\Lambda(y)$ and commutes with $\Hf$ and $\Pf$, so it maps the fiber with charge $e$ onto the one with charge $-e$.

If the field is normalized as above and the coupling is written as $\sqrt{\alpha'}A$, then $\alpha'=\abar=4\pi\alpha$; if a factor $\sqrt{4\pi}$ is included in the normalization of the field, as in \cite{LL02}, then $\alpha'=\alpha$. Constants quoted from the literature must be compared together with the normalization of the field.

The following facts are known; the precise statements used and their sources are collected in Appendix~\ref{app:facts}.

\begin{theorem}[\cite{Hi07,LMS07,HSi23}]\label{thm:facts}
For all real $e$ and all $m,\Lambda>0$:
\begin{enumerate}[label=(\roman*)]
\item $H_{m,\Lambda}(P)$ is self-adjoint on $D(\Hf)\cap D(\Pf^2)$, and $H\cong\int^\oplus_{\R^3}H_{m,\Lambda}(P)\,dP$ \cite[Thm.~2.2]{Hi07}; Hiroshima attributes the self-adjointness to \cite{LMS07}.
\item $E_{m,\Lambda}(P)\ge E_{m,\Lambda}(0)$ for all $P$ \cite[Cor.~3.8(1)]{Hi07}, and $E_{m,\Lambda}(0)=\inf\sigma(H)$ \cite[Thm.~4.1(iv) and Prop.~7.1]{LMS07}, applied with $N=0$ (a single charge, in the notation of \cite{LMS07}) and $V=0$, for which (V.1) and (V.2) hold trivially and (E.I.) is \cite[Prop.~7.1]{LMS07}.
\item $H_{m,\Lambda}(0)$ has a ground state \cite[Thm.~2]{HSi23}, which is unique up to a phase \cite[Cor.~3.6]{Hi07}.
\end{enumerate}
\end{theorem}

The sources state these facts for bare mass $1$; the general case follows from Lemma~\ref{lem:scaling} (Appendix~\ref{app:facts}).

\subsection{Effective mass and scaling}

\begin{definition}\label{def:masses}
The \emph{effective mass ratio} is
\[
\hls(m,\Lambda):=\Bigl(2m\limsup_{p\to0}\frac{E_{m,\Lambda}(p\hat z)-E_{m,\Lambda}(0)}{p^2}\Bigr)^{-1}\in[1,\infty]
\]
(with $\hls=\infty$ if the $\limsup$ vanishes); we set $m_{\rm eff}:=m\,\hls$. If $E_{m,\Lambda}$ is twice differentiable at $P=0$, then $\hls$ is the ratio of the curvature mass to the bare mass; in general a lower bound on $\hls$ implies the same lower bound for the versions defined with the $\liminf$ or, when it exists, with the second derivative. That $\hls\ge1$ follows from $\ip{\psi_0}{H_{m,\Lambda}(p\hat z)\psi_0}=E_{m,\Lambda}(0)+p^2/2m$, where $\psi_0$ is the ground state and $\ip{\psi_0}{V\psi_0}=0$ (Lemma~\ref{lem:velocity-zero}).
\end{definition}

We use the $\limsup$ of the second difference quotient, so that differentiability of $E_{m,\Lambda}$ is neither assumed nor proved; by rotation invariance, $E_{m,\Lambda}(\mathcal OP)=E_{m,\Lambda}(P)$ for $\mathcal O\in SO(3)$ (Lemma~\ref{lem:rotation}), the direction $\hat z$ plays no role.

\begin{lemma}\label{lem:scaling}
For $a>0$ let $U_a:=\Gamma(u_a)$ with $(u_af)(k,r):=a^{-3/2}f(k/a,r)$. Then
\[
U_a\,H_{m,\Lambda}(P)\,U_a^*=a^{-1}H_{am,\,a\Lambda}(aP)
\]
with equality of domains. Consequently (i) $E_{m,\Lambda}(P)=a^{-1}E_{am,a\Lambda}(aP)$, and existence and uniqueness of ground states hold for both sides simultaneously; (ii) $\hls(m,\Lambda)=\hls(am,a\Lambda)$; (iii) $\alpha$ is unchanged.
\end{lemma}

\begin{proof}
Since $u_a\omega u_a^*=a^{-1}\omega$ and $u_aku_a^*=a^{-1}k$ as multiplication operators, $U_a\Hf U_a^*=a^{-1}\Hf$ and $U_a\Pf U_a^*=a^{-1}\Pf$; moreover $U_aa^\sharp(f)U_a^*=a^\sharp(u_af)$, and $U_a$ preserves $D(\Hf)\cap D(\Pf^2)$. Because $\varepsilon_r(k/a)=\varepsilon_r(k)$ and $\mathbf 1_{\{|k/a|\le\Lambda\}}=\mathbf 1_{\{|k|\le a\Lambda\}}$, the coupling function $f_\Lambda$ of $A_\Lambda(0)$ satisfies $u_af_\Lambda=a^{-3/2}a^{1/2}f_{a\Lambda}=a^{-1}f_{a\Lambda}$, so $U_aA_\Lambda(0)U_a^*=a^{-1}A_{a\Lambda}(0)$. Substituting,
\[
U_aH_{m,\Lambda}(P)U_a^*=\frac1{2ma^2}\bigl(aP-\Pf-eA_{a\Lambda}(0)\bigr)^2+a^{-1}\Hf=a^{-1}H_{am,a\Lambda}(aP).
\]
Part (i) is immediate. For (ii), with $E'=E_{am,a\Lambda}$,
\[
\frac{E_{m,\Lambda}(p\hat z)-E_{m,\Lambda}(0)}{p^2}=a\,\frac{E'(ap\hat z)-E'(0)}{(ap)^2},
\]
so the $\limsup$ is multiplied by $a$ and $\hls(m,\Lambda)=\hls(am,a\Lambda)$.
\end{proof}

\emph{Units.} Two normalizations are used. In \emph{physical units} (used in Section~\ref{sec:intro} and Part~I) the bare mass is $m$, the cutoff is $\Lambda$, and
\[
h(\lambda):=\hls(1,\lambda)=\hls(m,\lambda m),\qquad\lambda=\Lambda/m,\qquad m_{\rm eff}(m,\Lambda)=m\,h(\Lambda/m).
\]
In Part~II we write the fibers with the \emph{kinetic coefficient} $c=1/m$ and the opposite sign of the charge,
\[
H_{c,\Lambda}(P):=\frac c2\bigl(P-\Pf+eA_\Lambda(0)\bigr)^2+\Hf ,
\]
which is $\Gamma(-1)H_{1/c,\Lambda}(P)\Gamma(-1)^*$; accordingly $\hls(c,\Lambda)$ denotes $\hls(1/c,\Lambda)$ there, and the scaling of Lemma~\ref{lem:scaling} reads $U_aH_{c,\Lambda}(P)U_a^*=a^{-1}H_{c/a,a\Lambda}(aP)$. The \emph{computational units} of Part~II fix the band to $|k|\le\pi$ and set
\[
c=\abar\ell^7,\qquad\ell:=\Bigl(\frac\lambda{\pi\abar}\Bigr)^{1/7},\qquad H(P):=H_{\abar\ell^7,\pi}(P),
\]
so that $h(\lambda)=\hls(\abar\ell^7,\pi)$ by Lemma~\ref{lem:scaling} with $a=\pi/\lambda$, and $c=\lambda/\pi$. The map $\lambda\mapsto\ell(\lambda)$ is a continuous increasing bijection of $(0,\infty)$; since $\alpha$ is fixed, $\ell\asymp\lambda^{1/7}$ and $\log\ell\asymp\log\lambda$. The length $\ell$ is the scale on which the electron localizes for large $\lambda$: there the kinetic energy $c/\ell^2$ balances the diamagnetic energy, the field frequencies inside the photon cloud, of order $\abar\ell^2$, are much larger than the band edge $\pi$, and the mean-field energy is of order $\abar\ell^5$ \cite{SerA}. In Part~II we write $\psi$ for the normalized ground state of $H(0)$, $E:=\inf\sigma(H(0))$, and $\langle\cdot\rangle$ for expectations in $\psi$.

\subsection{Main results}

The lower bound uniform in the cutoff is Theorem~\ref{thm:partI} in Section~\ref{sec:partI-result}; for $\lambda\ge1$ it gives $h(\lambda)\ge1+0.3441\,\alpha/(1+0.2985\sqrt\alpha+0.1252\,\alpha)$ (Corollary~\ref{cor:partI-explicit}). The growth in the cutoff is the following.

\begin{definition}\label{def:iota}
$\iPF(\alpha):=\inf_{\lambda>0}h(\lambda)/\lambda\in[0,\infty]$. It is finite for small $\alpha$ (Section~\ref{sec:discussion}); since no upper bound on $h$ is known in general, finiteness for all $\alpha$ is not known.
\end{definition}

\begin{theorem}\label{thm:main}
Fix $\alpha>0$. Then $h(\lambda)/\lambda\to\infty$ as $\lambda\to\infty$. More precisely, for every $\epsilon>0$ there is $\lambda_\epsilon\ge1$ such that
\[
h(\lambda)\ge\lambda^{8/7-\epsilon}\qquad(\lambda\ge\lambda_\epsilon).
\]
Consequently $\iPF(\alpha)>0$: there is $\lambda_*\ge1$ with $h(\lambda)\ge\lambda$ for $\lambda\ge\lambda_*$, and $h(\lambda)\ge\lambda_*^{-1}\lambda$ for all $\lambda>0$.
\end{theorem}

The exponent $8/7=1+1/7$ is the limit, as $\kappa\to0$, of $1+(1-\kappa)/7$, where $1-\kappa$ is the margin in Lemma~\ref{lem:Mw}, set by the ratio of the squared window radius ($\asymp\ell^2$ up to logarithms) to the volume ($\gtrsim\ell^3$) of the shallow depressions in the window and by the sensitivity $1/g_w\asymp\ell^\kappa$ of the level, and $1/7$ comes from $\ell\asymp\lambda^{1/7}$; we do not claim that it is the true growth exponent of $h$. The proof does not track $\lambda_\epsilon$, which may grow rapidly as $\epsilon\to0$, nor $\lambda_*$.

\begin{corollary}[Cutoff bound for mass renormalization at fixed coupling]\label{cor:LL}
Fix $\alpha>0$. For every $(\Lambda,m_{\mathrm{phys}})$ with $m_{\mathrm{phys}}/\Lambda<\iPF(\alpha)$, the equation $m\,h(\Lambda/m)=m_{\mathrm{phys}}$ has no solution $m>0$. Thus the family of renormalized theories with fixed $m_{\mathrm{phys}}$ exists only for $\Lambda\le\Lambda_c:=m_{\mathrm{phys}}/\iPF(\alpha)$ (with $\Lambda_c=0$ if $\iPF(\alpha)=\infty$).
\end{corollary}

\begin{proof}
With $\lambda=\Lambda/m$ we have $m\,h(\Lambda/m)=\Lambda\,h(\lambda)/\lambda\ge\Lambda\,\iPF(\alpha)>m_{\mathrm{phys}}$ by Theorem~\ref{thm:main}. This is Lemma~\ref{lem:LL-equivalence}(i).
\end{proof}

We do not use the converse part of Lemma~\ref{lem:LL-equivalence} (existence of solutions for small $\Lambda/m_{\mathrm{phys}}$ under continuity of $h$); Corollary~\ref{cor:LL} is the half of the statement of Lieb and Loss asserting non-existence.

\begin{remark}[What we do not claim]\label{rem:nonclaims}
(i) Upper bounds on $h$, or the true growth exponent of $h$; Theorem~\ref{thm:main} is a lower bound, and the present estimates do not reach the exponent $8/7$ itself. (ii) The value of $\iPF$ or of $\Lambda_c$ (see Section~\ref{sec:discussion}). (iii) The size of $\lambda_\epsilon$ or $\lambda_*$, or anything at physically accessible $\lambda$; at the values of $\lambda$ that can be reached in practice the apparent growth of $h$ may be slower than linear; this would be a pre-asymptotic effect and is not in conflict with Theorem~\ref{thm:main}. (iv) The spin-$\frac12$ model, external potentials, smooth ultraviolet cutoffs in the model, lattice models. (v) The relation between the curvature mass and the binding-energy mass of \cite{LL02}. (vi) Existence, uniqueness or continuity in $\Lambda$ of a bare mass solving $m\,h(\Lambda/m)=m_{\rm phys}$ for $\Lambda\le\Lambda_c$, or optimality of $\Lambda_c$; if $h$ were known to be continuous, $h/\lambda$ would attain its infimum (as $h\ge1$ and by Theorem~\ref{thm:main}), and solutions would exist exactly for $\Lambda\le\Lambda_c$. Nor do we treat renormalization schemes in which $\alpha$ also depends on $\Lambda$.
\end{remark}

\begin{remark}\label{rem:units}
In computational units we prove, for each $0<\kappa<1$, $c/h(\lambda)\le C_\kappa\ell^{\kappa-1}(\log\lambda)^2$ for $\lambda\ge\lambda_1(\kappa)$, with $C_\kappa$ independent of $\lambda$ (Section~\ref{sec:proof}); since $c=\lambda/\pi$ and $\ell\asymp\lambda^{1/7}$, this is Theorem~\ref{thm:main}.
\end{remark}

\part*{Part I. A lower bound uniform in the cutoff}
\section{Statement and idea of the proof}\label{sec:partI-result}

\subsection{Setting}
In Part~I we work with general $m,\Lambda>0$, write $\lambda=\Lambda/m$, and use the fibers $H(P)=H_{m,\Lambda}(P)$ of \eqref{eq:fiber}. The symbols introduced in Part~I ($K$, $S$, $Q$, $f$, $\sigma$, $\rho_K$, $\dots$) are local to Sections~\ref{sec:partI-result}--\ref{sec:partI-proof} and Appendices~\ref{app:facts}--\ref{app:integrals}; Part~II has its own table of symbols (Table~\ref{tab:symbols}). It is convenient to describe the one-photon space without polarization vectors, as the space of transverse vector fields $\h_\perp:=\{f\in L^2(\R^3;\C^3):k\cdot f(k)=0\text{ a.e.}\}$; the identification $\h_\perp\cong L^2(\R^3\times\{1,2\})$, $f(k,r)=\varepsilon_r(k)\cdot f(k)$, depends on the choice of polarization vectors, but none of the definitions below does. Rotations $\mathcal O\in SO(3)$ act on $\h_\perp$ by $(u_{\mathcal O}f)(k):=\mathcal Of(\mathcal O^{-1}k)$, and $U_{\mathcal O}:=\Gamma(u_{\mathcal O})$, so that $U_{\mathcal O}A(\zeta)U_{\mathcal O}^*=A(u_{\mathcal O}\zeta)$. We write $P_\perp(\hat k)=1-\hat k\hat k^{\sf T}$, $A(\zeta):=a(\zeta)+a^*(\zeta)$ for $\zeta\in\h_\perp$,
\[
\varkappa_k:=(2\pi)^{-3/2}(2|k|)^{-1/2},\qquad u_\mu(k):=\varkappa_k\mathbf 1_{\{|k|\le\Lambda\}}P_\perp(\hat k)e_\mu\in\h_\perp ,
\]
where $e_\mu$ is the standard basis, so that $A_{\Lambda,\mu}(0)=A(u_\mu)$, $V_\mu=P_{{\rm f},\mu}+eA(u_\mu)$, and
\[
E(P)=\inf\sigma(H(P)),\qquad E_0=E(0),\qquad D_0:=H(0)-E_0\ge0 .
\]
We write $\psi_0$ for the normalized ground state of $H(0)$ (Theorem~\ref{thm:facts}(iii)) and $\langle\cdot\rangle$ for expectations in $\psi_0$. For a real unit vector $\epsilon$ let $\phi_\epsilon:=\epsilon\cdot V\psi_0$. The effective mass ratio is $h=\hls(m,\Lambda)$ of Definition~\ref{def:masses},
\[
\frac1h=\limsup_{p\to0}\frac{2m\,(E(p\hat z)-E_0)}{p^2}.
\]
We use the $\limsup$ because our lower bounds on $h$ are upper bounds on the curvature; a lower bound on $h$ in this form implies the same bound for the versions with $\liminf$ or with the second derivative, when it exists. ``Uniform in the cutoff'' does not refer to a limit $\Lambda\to\infty$: the bounds below hold for every finite $\Lambda$, with right-hand sides that do not contain $\Lambda$.

\subsection{Statement}
For $0<K\le\Lambda$ put $\sigma:=K/2m$ and
\begin{gather*}
x_K:=\frac{4\alpha}{3\pi}\log(1+\sigma),\qquad t_K:=\frac{3\pi\alpha}{16\log(1+\sigma)}\,\mathsf L(\sigma)^2,\qquad\rho_K:=2\sqrt{q_Kt_K}+2t_K,\\
q_K:=\frac{2}{\log(1+\sigma)}\Bigl(\log(1+\sigma)-\frac\sigma{1+\sigma}\Bigr),\qquad\mathsf L(\sigma):=\int_0^\sigma\sqrt{\frac t{1+t}}\,dt=\sqrt{\sigma(1+\sigma)}-\mathrm{arsinh}\sqrt\sigma .
\end{gather*}
Here $x_K$ is the leading term, produced by the c-number part of the commutator of the velocity with the field, $q_K$ is a free photon recoil term, and $t_K$ bounds the annihilation amplitude of the ground state (Section~\ref{sec:partI-proof} and Appendix~\ref{app:integrals}). The quantities $x_K$ and $t_K$ are proportional to $\alpha$, $q_K$ does not depend on $\alpha$, and none of them contains $\Lambda$ except through $K$.

\begin{theorem}\label{thm:partI}
For all $m>0$, $\Lambda>0$, $e\in\R$ and $0<K\le\Lambda$,
\[
E(P)-E_0\ \le\ \frac{P^2}{2m}\Bigl(1+\frac{2x_K}{1+\rho_K}\Bigr)^{-1}\quad\text{for all }P\in\R^3,\qquad\text{hence}\qquad h\ \ge\ 1+\frac{2x_K}{1+\rho_K}.
\]
In particular, with $K=\min(\Lambda,m)$ the right-hand side does not contain $\Lambda$ when $\lambda\ge1$, and for every $\lambda>0$ it is strictly larger than $1$ when $e\ne0$.
\end{theorem}

The first inequality is a global quadratic upper bound on the energy--momentum relation; only its consequence for the curvature at $P=0$ is used below.

\begin{corollary}\label{cor:partI-explicit}
If $\lambda\ge1$, then for all real $e$
\[
h\ \ge\ 1+\frac{0.3441\,\alpha}{1+0.2985\sqrt\alpha+0.1252\,\alpha}.
\]
\end{corollary}

\begin{proof}
Take $K=m$, i.e.\ $\sigma=\frac12$. Then $2x_m=0.3441694\ldots\,\alpha$ and $\rho_m=0.2984322\ldots\sqrt\alpha+0.1251579\ldots\,\alpha$ (Appendix~\ref{app:constants}, interval arithmetic). The displayed constants are rounded in the safe direction.
\end{proof}

For $\alpha=1/137.036$ this gives $h-1\ge0.002446$, and as $\alpha\to\infty$ the right-hand side tends to $1+0.3441/0.1252=3.748\ldots$ (and the bound of Theorem~\ref{thm:partI} with $K=m$ to $1+2x_m/(2t_m)=3.749\ldots$).

\begin{corollary}\label{cor:partI-first-order}
For every $\lambda>0$,
\[
\liminf_{\alpha\to0}\frac{h-1}\alpha\ \ge\ \frac8{3\pi}\log\Bigl(1+\frac\lambda2\Bigr).
\]
The right-hand side is the first order coefficient $a_1(\lambda,0)$ of \cite[(1.22) and Lemma~3.8]{HSp05} (there $m=1$, so that their $\Lambda$ is our $\lambda$).
\end{corollary}

\begin{proof}
Take $K=\Lambda$ in Theorem~\ref{thm:partI}. Then $2x_\Lambda=\frac{8\alpha}{3\pi}\log(1+\frac\lambda2)$ is proportional to $\alpha$, $q_\Lambda$ does not depend on $\alpha$, and $t_\Lambda=O_\lambda(\alpha)$, so $\rho_\Lambda=O_\lambda(\sqrt\alpha)$ and $(h-1)/\alpha\ge(2x_\Lambda/\alpha)/(1+O_\lambda(\sqrt\alpha))$.
\end{proof}

We do not claim the matching upper bound $\limsup_{\alpha\to0}(h-1)/\alpha\le a_1(\lambda,0)$: Part~I gives no upper bound on $h$, and the perturbative expansion of \cite{HSp05} is rigorous only with an infrared cutoff. The statement ``sharp to first order'' is meant only in the sense that the first order coefficient of our nonperturbative lower bound coincides with the perturbative one. The error term $O_\lambda(\sqrt\alpha)$ is not uniform in $\lambda$.

\begin{remark}\label{rem:partI-values}
(i) For $\lambda\to0$ and $K=\Lambda$ one has $2x_\Lambda\sim\frac{4\alpha}{3\pi}\lambda$, in agreement with the one-loop value $h-1\simeq\frac{4\alpha}{3\pi}\lambda$.

(ii) As $\alpha\to\infty$ the bound of Corollary~\ref{cor:partI-explicit} saturates at about $3.75$; this is an artefact of the fixed band $K=m$. If the band shrinks with $\alpha$, Theorem~\ref{thm:partI} gives $h\to\infty$: for every $\lambda>0$,
\[
\liminf_{\alpha\to\infty}\frac h{\sqrt\alpha}\ \ge\ \frac4{3\pi}\sqrt{\frac6\pi}=0.5865\ldots .
\]
Indeed, take $\sigma=\sqrt{6/(\pi\alpha)}$, i.e.\ $K=2m\sigma$, which satisfies $K\le\Lambda$ and $\sigma\le1$ once $\alpha\ge\max(24/(\pi\lambda^2),6/\pi)$. For $0<\sigma\le1$ one has $\log(1+\sigma)\ge\sigma-\sigma^2/2$, $\log(1+\sigma)\le\sigma$ (hence $q_K\le2\sigma/(1+\sigma)\le2\sigma$) and $\mathsf L(\sigma)\le\int_0^\sigma\sqrt t\,dt=\frac23\sigma^{3/2}$, so that $t_K\le\frac{\pi\alpha\sigma^2}{12(1-\sigma/2)}=\frac1{2-\sigma}$ and $\rho_K\le2\sqrt{2\sigma/(2-\sigma)}+\frac2{2-\sigma}$. Hence
\[
h\ \ge\ 1+\frac{8\alpha}{3\pi}\,\frac{\sigma(1-\sigma/2)}{1+\rho_K},
\]
and since $\sigma\to0$, $\rho_K\le1+o(1)$ and $\alpha\sigma=\sqrt{6\alpha/\pi}$, the claim follows. (Within the family of Theorem~\ref{thm:partI} this choice of $\sigma$ is asymptotically optimal: for $\sigma\to0$ with $u:=\pi\alpha\sigma^2/12$ fixed the bound is $1+\frac8{3\pi}\sqrt{12\alpha/\pi}\,\frac{\sqrt u}{1+2u}(1+o(1))$, which is largest at $u=\frac12$, while for $\sigma$ bounded away from $0$ it stays bounded as $\alpha\to\infty$, since $2x_K/(1+\rho_K)\le x_K/t_K$.) The true rate of growth of $h$ as $\alpha\to\infty$ at fixed $\lambda$ is open; for the Fr\"ohlich polaron the divergence of the effective mass was proved in \cite{LS20}, and the Landau--Pekar rate $\alpha^4$ is now established \cite{BMSV26,BS24,Br26} (see the comparison in Section~\ref{sec:intro}).

(iii) For $K=m$ the bound does not increase with $\lambda$. For $K=\Lambda$ it is sharp to first order, but $t_\Lambda$ grows like $\alpha\lambda^2/\log\lambda$ and the bound deteriorates outside the window $\alpha\lambda^2\ll\log\lambda$. The growth of $h$ as $\lambda\to\infty$ is the subject of Part~II.
\end{remark}

\subsection{The idea}
Let $f$ be a transverse one-photon function supported in the band $k_1\le|k|\le K$, essentially the first order perturbative correction $\Omega_k^{-1}\,\epsilon\cdot u$ (Section~\ref{sec:partI-proof}), and let $W_\theta=\exp(\theta(a^*(f)-a(f)))$ be the corresponding coherent translation of the field. We use $W_\theta\psi_0$ as a trial state for $H(p\epsilon)$. Since $W_\theta$ translates the vector potential by a c-number, conjugation gives exactly
\[
\ip{W_\theta\psi_0}{(H(p\epsilon)-E_0)W_\theta\psi_0}=\frac{p^2}{2m}-\frac{2\theta peS}m+\theta L+\theta^2Q
\]
with $S=\ip{\epsilon\cdot u}{f}$ and explicit $L$, $Q$ (Lemma~\ref{lem:boost}). Three points make the resulting bound uniform in $\Lambda$.

\emph{The linear term.} At $p=0$ the left-hand side is nonnegative for all $\theta$, because $\psi_0$ is a ground state; hence $L=0$. No commutator of $H(0)$ with field operators has to be computed, and no domain question about double commutators arises.

\emph{The vertex.} The numerator $2eS=2e\sum_\mu\epsilon_\mu\ip{u_\mu}{f}$ comes from the c-number part $2e\ip{u_\mu}{f}$ of the commutator of the velocity $V_\mu$ with $a^*(f)-a(f)$; the operator parts have mean zero in $\psi_0$ by rotation invariance. Electron recoil does not enter the numerator.

\emph{The annihilation amplitude.} The denominator $Q$ contains $\frac1{2m}\sum_\mu\lVert a(k_\mu f)\psi_0\rVert^2$, the squared norm of the annihilation amplitude $a(k_\mu f)\psi_0$ of the ground state (a single annihilation operator applied to $\psi_0$). A naive bound would bring in $\lVert V\psi_0\rVert^2\le2mE_0$, of order $\alpha\Lambda^2$. Instead, the pull-through formula expresses $a(k)\psi_0$ through the resolvent of the shifted fiber $H(-k)$, and the diamagnetic inequality, used for the whole one-parameter family $E(-tq)\ge E_0$, compares this resolvent with $D_0^{-1}$ (Corollary~\ref{cor:resolvent}); the remaining quantity $\lVert D_0^{-1/2}\phi_\epsilon\rVert^2$ is at most $m/2$ by an inequality version of the sum rule (Lemma~\ref{lem:sum-rule}). Finally, since the estimate of this term grows like $\alpha(K/m)^2$ for $K\gg m$ while the leading term $x_K$ grows only logarithmically, we keep $K/m$ fixed.

\emph{Scope.} Part~I does not give upper bounds on $m_{\rm eff}$ and does not treat spin $\frac12$, external potentials or smooth cutoff functions (the existence of $\psi_0$ that we use is proved for the sharp cutoff). Section~\ref{sec:framework} proves the abstract statements, which use only the fiber structure and the diamagnetic inequality, and Section~\ref{sec:partI-proof} proves Theorem~\ref{thm:partI}.

\section{Fiber families with a diamagnetic inequality}\label{sec:framework}

The statements of this section do not use the specific structure of the Pauli--Fierz model (Fock space, polarizations, cutoff). We only assume the following.

\begin{assumption}\label{ass:AF}
\begin{enumerate}[label=(\Alph*)]
\item\emph{(Fiber family.)} $\mathcal H$ is a Hilbert space, $m>0$, $d\ge1$. For each $P\in\R^d$, $H(P)$ is a self-adjoint operator bounded from below, whose form domain $\Qf:=\Qf(H(P))$ does not depend on $P$. There are symmetric operators $V_1,\dots,V_d$ with $\Qf\subseteq D(V_\mu)$ such that, for all $\psi,\psi'\in\Qf$,
\[
\ip{\psi}{H(P)\psi'}=\ip{\psi}{H(0)\psi'}-\frac1m\sum_{\mu=1}^dP_\mu\ip{\psi}{V_\mu\psi'}+\frac{P^2}{2m}\ip{\psi}{\psi'},
\]
where the first two terms are values of the closed forms.
\item\emph{(Diamagnetic inequality.)} $E(P):=\inf\sigma(H(P))$ satisfies $E(P)\ge E(0)=:E_0$ for all $P$.
\item\emph{(Ground state.)} $E_0$ is an eigenvalue of $H(0)$ with normalized eigenvector $\psi_0$. We write $D_0:=H(0)-E_0\ge0$.
\end{enumerate}
\end{assumption}

Simplicity of $E_0$ is not needed in this section. For a real unit vector $\epsilon$ let $\phi_\epsilon:=\epsilon\cdot V\psi_0$ (note $\psi_0\in D(H(0))\subseteq\Qf\subseteq D(V_\mu)$). The only consequence of (B) that we use is the following form inequality.

\begin{lemma}[Form inequality for the velocity]\label{lem:qV}
Assume (A) and (B). For $q\in\R^d$, $t>0$ and $\psi\in\Qf$,
\begin{equation}\label{eq:qV-lower}
\Bigl\langle\psi,\frac{q\cdot V}m\psi\Bigr\rangle\ \ge\ -\frac1t\ip{\psi}{D_0\psi}-\frac{t\,q^2}{2m}\lVert\psi\rVert^2 .
\end{equation}
Consequently $\ip{\psi}{\epsilon\cdot V\psi}^2\le2m\lVert\psi\rVert^2\ip{\psi}{D_0\psi}$ for every real unit vector $\epsilon$.
\end{lemma}

\begin{proof}
By the variational principle and (B), $\ip{\psi}{H(-tq)\psi}\ge E(-tq)\lVert\psi\rVert^2\ge E_0\lVert\psi\rVert^2$. Expanding the left-hand side by (A) and dividing by $t$ gives \eqref{eq:qV-lower}. For the second statement take $q=\pm\epsilon$ and multiply by $t$: the quadratic polynomial $\tau\mapsto\ip{\psi}{D_0\psi}+\frac\tau m\ip{\psi}{\epsilon\cdot V\psi}+\frac{\tau^2}{2m}\lVert\psi\rVert^2$ is nonnegative on $\R$ (at $\tau=\pm t$, and at $\tau=0$ since $D_0\ge0$), so its discriminant is nonpositive.
\end{proof}

\begin{lemma}[The velocity has mean zero]\label{lem:velocity-zero}
Assume (A)--(C). Then $\ip{\psi_0}{V_\mu\psi_0}=0$; in particular $\phi_\epsilon\perp\psi_0$.
\end{lemma}

\begin{proof}
Apply the second statement of Lemma~\ref{lem:qV} to $\psi=\psi_0$, for which $\ip{\psi_0}{D_0\psi_0}=0$.
\end{proof}

\begin{lemma}[Sum rule inequality]\label{lem:sum-rule}
Assume (A)--(C). For every real unit vector $\epsilon$ and every $\chi\in\Qf$,
\[
|\ip{\phi_\epsilon}{\chi}|^2\ \le\ \frac m2\,\ip{\chi}{D_0\chi}.
\]
\end{lemma}

\begin{proof}
Replacing $\chi$ by $\ee^{i\vartheta}\chi$ we may assume $\ip{\phi_\epsilon}{\chi}\in\R$. For $s\in\R$ apply the second statement of Lemma~\ref{lem:qV} to $\psi=\psi_0+s\chi\in\Qf$. By $D_0\psi_0=0$, Lemma~\ref{lem:velocity-zero} and the symmetry of $\epsilon\cdot V$,
\[
\ip{\psi}{\epsilon\cdot V\psi}=2s\ip{\phi_\epsilon}{\chi}+s^2\ip{\chi}{\epsilon\cdot V\chi},\qquad\ip{\psi}{D_0\psi}=s^2\ip{\chi}{D_0\chi},
\]
so $s^2\bigl(2\ip{\phi_\epsilon}{\chi}+s\ip{\chi}{\epsilon\cdot V\chi}\bigr)^2\le2m\,s^2\lVert\psi_0+s\chi\rVert^2\ip{\chi}{D_0\chi}$. Divide by $s^2$ and let $s\to0$.
\end{proof}

In the notation $G:=\sup_\chi|\ip{\phi_\epsilon}{\chi}|^2/\ip{\chi}{D_0\chi}=\lVert D_0^{-1/2}\phi_\epsilon\rVert^2$, Lemma~\ref{lem:sum-rule} states $G\le\frac m2$. This is the inequality part of the sum rule $m/m_{\rm eff}=1-2G/m$, which appears in \cite[(1.13)]{HSp05} as an identity derived from second order perturbation theory with an infrared cutoff; in physical terms $1-2G/m$ is the defect in the Thomas--Reiche--Kuhn sum rule \cite{Th25,Ku25,RT25}. Only the inequality $G\le\frac m2$ is used below.

\begin{lemma}[Operator inequality for shifted fibers]\label{lem:shift}
Assume (A) and (B). For $q\in\R^d\setminus\{0\}$ and $\tau>0$, in the sense of quadratic forms on $\Qf$,
\[
H(-q)-E_0+\tau\ \ge\ \mathsf t_\tau(q)\,D_0,\qquad\mathsf t_\tau(q):=\frac{2m\tau}{2m\tau+q^2}.
\]
\end{lemma}

\begin{proof}
By (A), $H(-q)-E_0+\tau=D_0+q\cdot V/m+q^2/2m+\tau$. Inserting \eqref{eq:qV-lower} gives, for all $t>0$, $H(-q)-E_0+\tau\ge(1-\frac1t)D_0+\tau-\frac{(t-1)q^2}{2m}$. The choice $t=1+2m\tau/q^2$ removes the constant term, and then $1-\frac1t=\mathsf t_\tau(q)$.
\end{proof}

With $\tau=|q|$, the photon dispersion, $\mathsf t=2m/(2m+|q|)$. The argument is of the same type as the lower bound on $E(P-k)-E(P)$ in \cite[Thm.~4.1(iii)]{LMS07}.

\begin{corollary}[Comparison of resolvents]\label{cor:resolvent}
Assume (A)--(C), let $q\in\R^d\setminus\{0\}$, $\tau>0$ and $A:=H(-q)-E_0+\tau$. Then $A\ge\tau$, so $A$ is invertible with $\lVert A^{-1}\rVert\le1/\tau$, and for every real unit vector $\epsilon$
\[
\ip{\phi_\epsilon}{A^{-1}\phi_\epsilon}\le\frac m2\Bigl(1+\frac{q^2}{2m\tau}\Bigr),\qquad\lVert A^{-1}\phi_\epsilon\rVert^2\le\frac1\tau\ip{\phi_\epsilon}{A^{-1}\phi_\epsilon}.
\]
\end{corollary}

\begin{proof}
$A\ge\tau$ follows from (B). Let $\chi:=A^{-1}\phi_\epsilon\in D(A)\subseteq\Qf$ and $a:=\ip{\phi_\epsilon}{A^{-1}\phi_\epsilon}=\ip{\phi_\epsilon}{\chi}=\ip{\chi}{A\chi}\ge0$. By Lemma~\ref{lem:sum-rule} and Lemma~\ref{lem:shift},
\[
a^2=|\ip{\phi_\epsilon}{\chi}|^2\le\frac m2\ip{\chi}{D_0\chi}\le\frac m{2\,\mathsf t_\tau(q)}\ip{\chi}{A\chi}=\frac m{2\,\mathsf t_\tau(q)}\,a,
\]
which is the first inequality. The second is $A^{-2}\le\tau^{-1}A^{-1}$.
\end{proof}

\begin{remark}[Scope of the framework]\label{rem:framework-scope}
(i) \emph{The Pauli--Fierz model.} The fibers \eqref{eq:fiber} satisfy (A) with $V=\Pf+eA_\Lambda(0)$: the domain $D(H(P))=D(H_{00})$, $H_{00}:=\Pf^2/2m+\Hf$, does not depend on $P$ or $e$ (Fact~\ref{fact:domain}), so the graph norms are equivalent by the closed graph theorem, and the L\"owner--Heinz inequality with exponent $\frac12$ gives $\Qf=\Qf(H_{00})=D(\Hf^{1/2})\cap D(|\Pf|)$. The bound $\lVert V_\mu\psi\rVert\le C\lVert(H_{00}+1)^{1/2}\psi\rVert$ (trivial for $\Pf$, and from \eqref{eq:rel-bound} for $A(0)$) gives $\Qf\subseteq D(V_\mu)$. On $D(H_{00})$ one has $\ip{\psi}{H(P)\psi}=\frac1{2m}\sum_\mu\lVert(P_\mu-V_\mu)\psi\rVert^2+\lVert\Hf^{1/2}\psi\rVert^2$; both sides are continuous in the form norm of $H_{00}$ and $D(H_{00})$ is a form core, so this identity holds on $\Qf$ for the closed form, and it implies the identity in (A). Assumption (B) is Fact~\ref{fact:diamagnetic} and (C) is Fact~\ref{fact:gs}.

(ii) \emph{Other models.} For Nelson and Fr\"ohlich type models with ultraviolet cutoff (linear coupling, $V=\Pf$), (A) holds as well, and Lemmas~\ref{lem:qV}--\ref{lem:shift} apply once (B) is known. For spin $\frac12$, (B) for all couplings is described as an open question in \cite[Sect.~2, after Thm.~2]{HSi23}. Nothing in this section depends on the dimension $d$.
\end{remark}

\section{The coherent trial state and the proof of Theorem~\ref{thm:partI}}\label{sec:partI-proof}

We use the notation of Section~\ref{sec:partI-result}. By Remark~\ref{rem:framework-scope}(i) the fibers satisfy Assumption~\ref{ass:AF}, and we write $q_P(\psi):=\frac1{2m}\sum_\mu\lVert(P_\mu-V_\mu)\psi\rVert^2+\lVert\Hf^{1/2}\psi\rVert^2$ for the closed form of $H(P)$ on $\Qf=D(\Hf^{1/2})\cap D(|\Pf|)$. For $k\ne0$ let
\[
\Rk:=(H(-k)-E_0+|k|)^{-1},
\]
the operator $A^{-1}$ of Corollary~\ref{cor:resolvent} with $q=k$ and $\tau=|k|$.

\subsection{The trial function}
Fix a real unit vector $\epsilon$, a band $0<k_1<K\le\Lambda$, and let
\[
f(k):=\varkappa_k\,g(|k|)\,P_\perp(\hat k)\epsilon\in\h_\perp,\qquad g:=\Omega_k^{-1}\mathbf 1_{[k_1,K]}(|k|),\qquad\Omega_k:=|k|+\frac{k^2}{2m}.
\]
Then $f$ is real, bounded, supported in the band $B:=\{k_1\le|k|\le K\}$, and $f(-k)=f(k)$; in particular
\begin{equation}\label{eq:f-even}
\ip{f}{k_\mu f}=\int k_\mu|f(k)|^2\,d^3k=0\qquad(\mu=1,2,3).
\end{equation}
The infrared cutoff $k_1>0$ is necessary, since $\lVert f\rVert^2=\frac1{6\pi^2}\int kg^2\,dk$ diverges logarithmically for $k_1=0$; we let $k_1\downarrow0$ at the end. With $\varpi=\angle(k,\epsilon)$ the angular integrals are
\[
\int\varkappa_k^2\bigl(1-(\epsilon\cdot\hat k)^2\bigr)\Phi(|k|)\,d^3k=\frac1{6\pi^2}\int_0^\infty k\,\Phi(k)\,dk,\qquad\int\varkappa_k^2\sin\varpi\,\Phi(|k|)\,d^3k=\frac1{16\pi}\int_0^\infty k\,\Phi(k)\,dk .
\]
Let $S:=\int\varkappa_k^2(1-(\epsilon\cdot\hat k)^2)g\,d^3k$ and $x:=e^2S/m$. Since $\int P_\perp(\hat k)\,d\Omega(\hat k)=\frac{8\pi}3\mathbf 1$ and $K\le\Lambda$, $\ip{u_\mu}{f}=\epsilon_\mu S$, and since $g\Omega_k=1$ on the band,
\begin{equation}\label{eq:S-values}
\ip{f}{\Omega_kf}=S=\frac m{3\pi^2}\log\frac{2m+K}{2m+k_1},\qquad x=\frac{4\alpha}{3\pi}\log\frac{2m+K}{2m+k_1}.
\end{equation}

\emph{Why this $f$ (heuristics, not used in the proofs).} At $e=0$ the ground state is the Fock vacuum, and to first order in $e$ the vector $D_0^{-1}\phi_\epsilon$ that saturates $G=\lVert D_0^{-1/2}\phi_\epsilon\rVert^2$ is $e\,a^*(\Omega_k^{-1}\epsilon\cdot u)\Omega_\vac$, since $D_0$ acts on one-photon states as multiplication by $\Omega_k$. For $K=\Lambda$ and $k_1\downarrow0$, our $f$ is $\Omega_k^{-1}\epsilon\cdot u$, and the first order coefficient of the resulting bound is the perturbative one (Corollary~\ref{cor:partI-first-order}). The restriction $K=m$ used for uniformity in $\Lambda$ gives up this exactness for the hard photons.

\subsection{Rotation invariance and the annihilation amplitude}

\begin{lemma}\label{lem:rotation}
$U_{\mathcal O}VU_{\mathcal O}^*=\mathcal O^{-1}V$ and $U_{\mathcal O}H(P)U_{\mathcal O}^*=H(\mathcal OP)$ for $\mathcal O\in SO(3)$, and $U_{\mathcal O}\psi_0=\chi(\mathcal O)\psi_0$ with $|\chi(\mathcal O)|=1$.
\end{lemma}

\begin{proof}
We have $u_{\mathcal O}k_\mu u_{\mathcal O}^*=(\mathcal O^{-1}k)_\mu$ as multiplication operators and $u_{\mathcal O}u_\mu=\sum_\nu\mathcal O_{\nu\mu}u_\nu$, by $P_\perp(\mathcal O^{-1}\hat k)=\mathcal O^{-1}P_\perp(\hat k)\mathcal O$. Hence $U_{\mathcal O}\Pf U_{\mathcal O}^*=\mathcal O^{-1}\Pf$, $U_{\mathcal O}A(0)U_{\mathcal O}^*=\mathcal O^{-1}A(0)$, $U_{\mathcal O}\Hf U_{\mathcal O}^*=\Hf$, and $U_{\mathcal O}H(P)U_{\mathcal O}^*=(P-\mathcal O^{-1}V)^2/2m+\Hf=H(\mathcal OP)$. Thus $U_{\mathcal O}\psi_0$ is a ground state of $H(0)$, and by uniqueness (Theorem~\ref{thm:facts}(iii)) $U_{\mathcal O}\psi_0=\chi(\mathcal O)\psi_0$ with $|\chi(\mathcal O)|=1$. The phase cancels in expectations: $\ip{\psi_0}{U_{\mathcal O}XU_{\mathcal O}^*\psi_0}=\ip{\psi_0}{X\psi_0}$.
\end{proof}

By the pointwise pull-through formula (Fact~\ref{fact:pullthrough}), for a.e.\ $k$ with $0<|k|\le\Lambda$,
\begin{equation}\label{eq:PT}
a_r(k)\psi_0=-\frac{e\,\varkappa_k}m\,\Rk\,\varepsilon_r(k)\cdot V\psi_0 ,
\end{equation}
and $a_r(k)\psi_0=0$ for a.e.\ $|k|>\Lambda$.

\begin{lemma}\label{lem:amplitude}
(i) $\ip{\psi_0}{A(k_\mu f)\psi_0}=0$ for $\mu=1,2,3$.
(ii) For every real unit vector $\hat e$ and $k\ne0$, $\lVert\Rk\phi_{\hat e}\rVert\le\beta(k):=\sqrt{\frac m{2|k|}\bigl(1+\frac{|k|}{2m}\bigr)}$.
(iii) $\sum_\mu\lVert a(k_\mu f)\psi_0\rVert^2\le\mathcal A^2$, where $\mathcal A:=\frac{|e|}{16\pi m}\int_0^Kk^2g\,\beta\,dk$, and $\mathcal A\le\frac{|e|m}{8\pi}\mathsf L(\sigma)$.
\end{lemma}

\begin{proof}
(i) Write $f_\epsilon$ for $f$ as a function of $\epsilon\in\R^3$ (it is real and linear in $\epsilon$), and $T_\mu(\epsilon):=\ip{\psi_0}{A(k_\mu f_\epsilon)\psi_0}$, which is defined since $k_\mu f_\epsilon$ vanishes for $|k|<k_1$ and $\psi_0\in D(\Hf^{1/2})$. Then $T(\epsilon)=M\epsilon$ with a real $3\times3$ matrix $M$. By $P_\perp(\mathcal O^{-1}\hat k)=\mathcal O^{-1}P_\perp(\hat k)\mathcal O$ we have $u_{\mathcal O}(k_\mu f_\epsilon)=\sum_\nu\mathcal O_{\nu\mu}k_\nu f_{\mathcal O\epsilon}$, and $\ip{\psi_0}{A(u_{\mathcal O}\zeta)\psi_0}=\ip{\psi_0}{A(\zeta)\psi_0}$ by Lemma~\ref{lem:rotation}. Hence $T(\epsilon)=\mathcal O^{\sf T}T(\mathcal O\epsilon)$, that is, $M$ commutes with every $\mathcal O\in SO(3)$. Commuting with the half-turns about the coordinate axes makes $M$ diagonal, and commuting with the quarter-turns about them makes its diagonal entries equal; hence $M=c\mathbf 1$. Transversality gives $\sum_\mu k_\mu f_{e_\mu}(k)=\varkappa_kg(|k|)P_\perp(\hat k)k=0$, so $3c=\sum_\mu T_\mu(e_\mu)=0$.

(ii) Corollary~\ref{cor:resolvent} with $q=k$, $\tau=|k|$.

(iii) Let $v(k):=\sum_rf(k,r)a_r(k)\psi_0$. By \eqref{eq:PT} and $\sum_r(\varepsilon_r\cdot f(k))\varepsilon_r=f(k)=\varkappa_kg\,|P_\perp(\hat k)\epsilon|\,\hat e(k)$ with $|P_\perp(\hat k)\epsilon|=\sin\varpi$ and the unit vector $\hat e(k):=P_\perp(\hat k)\epsilon/\sin\varpi$,
\[
v(k)=-\frac{e\varkappa_k^2g(|k|)\sin\varpi}m\,\Rk\,\phi_{\hat e(k)},\qquad\lVert v(k)\rVert\le\frac{|e|\varkappa_k^2g(|k|)\sin\varpi}m\,\beta(k)
\]
by (ii). Since $f$ is real, $a(k_\mu f)\psi_0=\int k_\mu v(k)\,d^3k$ as Bochner integrals (Fact~\ref{fact:pullthrough}(i)), and Minkowski's integral inequality with values in $\ell^2(\{1,2,3\};\F)$ gives $(\sum_\mu\lVert a(k_\mu f)\psi_0\rVert^2)^{1/2}\le\int|k|\lVert v(k)\rVert\,d^3k\le\mathcal A$ by the second angular integral. The value of the integral is computed in Appendix~\ref{app:integrals}.
\end{proof}

\subsection{The coherent translation}
For $\zeta\in\h_\perp$ let $W(\zeta)$ be the Weyl operator, the unitary group generated by the closure $B_\zeta$ of $a^*(\zeta)-a(\zeta)$, $W(\zeta)=\ee^{B_\zeta}$. We use the standard facts (see e.g.\ \cite{Ar18}) $W(\zeta)^*W(\eta)=\ee^{i\,{\rm Im}\ip{\zeta}{\eta}}W(\eta-\zeta)$, $\Gamma(u)W(\zeta)\Gamma(u)^*=W(u\zeta)$ for unitary $u$, and $\ee^{isA(\zeta)}=W(is\zeta)$. If $\zeta$ is supported in $B$ and $N_B:=\dG(\mathbf 1_B)$, then $\lVert B_\zeta\chi\rVert\le2\lVert\zeta\rVert\lVert(N_B+1)^{1/2}\chi\rVert$, and $N_B\le k_1^{-1}\Hf$.

\begin{lemma}[Coherent translation]\label{lem:weyl}
Let $W_\theta:=W(\theta f)$, $\theta\in\R$. Then $W_\theta$ maps $\Qf$ onto $\Qf$, and for $\psi\in\Qf$ and $\mu=1,2,3$,
\begin{gather*}
V_\mu W_\theta\psi=W_\theta(V_\mu+\theta C_\mu)\psi,\qquad C_\mu:=A(k_\mu f)+2e\ip{u_\mu}{f}=A(k_\mu f)+2e\epsilon_\mu S,\\
\lVert\Hf^{1/2}W_\theta\psi\rVert^2=\lVert\Hf^{1/2}\psi\rVert^2+\theta\ip{\psi}{A(\omega f)\psi}+\theta^2\ip{f}{\omega f}\lVert\psi\rVert^2 .
\end{gather*}
\end{lemma}

\begin{proof}
\emph{The vector potential.} By the Weyl relations, $W(\zeta)^*W(\eta)W(\zeta)=\ee^{2i\,{\rm Im}\ip{\zeta}{\eta}}W(\eta)$. With $\zeta=\theta f$ and $\eta=isu_\mu$ this gives $W_\theta^*\ee^{isA(u_\mu)}W_\theta=\ee^{is(A(u_\mu)+2\theta\ip{u_\mu}{f})}$, so $W_\theta^*A(u_\mu)W_\theta=A(u_\mu)+2\theta\ip{u_\mu}{f}$ as self-adjoint operators, with equal domains.

\emph{Field momentum and field energy.} Let $T$ be multiplication by $k_\mu$ or by $\omega=|k|$, so that $\Gamma(\ee^{isT})=\ee^{is\dG(T)}$. By the Weyl relations,
\[
U(s):=W_\theta^*\ee^{is\dG(T)}W_\theta=\ee^{i\theta^2\int\sin(sT)|f|^2}\,W(g_s)\,\ee^{is\dG(T)},\qquad g_s:=\theta(\ee^{isT}-1)f .
\]
$U$ is a unitary group with generator $iW_\theta^*\dG(T)W_\theta$. Let $\psi\in D(\dG(T))\cap D(\Hf^{1/2})$ and $\chi_s:=\ee^{is\dG(T)}\psi$. Since $g_s$ is supported in $B$, $(W(g_s)-1)\chi_s=\int_0^1W(rg_s)B_{g_s}\chi_s\,dr$, and since $g_s/s\to i\theta Tf$ in $\h_\perp$ (dominated convergence) while $\chi_s\to\psi$ in the graph norm of $(N_B+1)^{1/2}$, we get $s^{-1}(W(g_s)-1)\chi_s\to B_{i\theta Tf}\psi=i\theta A(Tf)\psi$. Hence $U(s)\psi$ is differentiable at $s=0$ with derivative $i(\dG(T)+\theta A(Tf)+\theta^2\ip{f}{Tf})\psi$, and by Stone's theorem $W_\theta\psi\in D(\dG(T))$ and
\[
W_\theta^*\dG(T)W_\theta\psi=\bigl(\dG(T)+\theta A(Tf)+\theta^2\ip{f}{Tf}\bigr)\psi .
\]
For $T=k_\mu$ the last term vanishes by \eqref{eq:f-even}, which gives $P_{{\rm f},\mu}W_\theta\psi=W_\theta(P_{{\rm f},\mu}+\theta A(k_\mu f))\psi$ for $\psi\in D(P_{{\rm f},\mu})\cap D(\Hf^{1/2})$. For $T=\omega$ we obtain the identity for $\Hf^{1/2}$ on $D(\Hf)$; its right-hand side is continuous in the graph norm of $\Hf^{1/2}$ (since $\sqrt\omega f\in\h_\perp$, $A(\omega f)$ is $\Hf^{1/2}$-bounded), and $D(\Hf)$ is a core for $\Hf^{1/2}$, so the closedness of $\Hf^{1/2}$ extends it to $D(\Hf^{1/2})$, and $W_\theta$ maps $D(\Hf^{1/2})$ into itself.

Since $D(|\Pf|)=\bigcap_\mu D(P_{{\rm f},\mu})$, $W_\theta$ maps $\Qf$ into $\Qf$, and onto since $W_\theta^{-1}=W_{-\theta}$. Adding the identities for $P_{{\rm f},\mu}$ and $eA(u_\mu)$ gives the one for $V_\mu$.
\end{proof}

\begin{lemma}[The boosted trial state]\label{lem:boost}
For all $p,\theta\in\R$,
\[
E(p\epsilon)-E_0\ \le\ \frac{p^2}{2m}-\frac{2\theta peS}m+\theta^2Q,\qquad Q:=\ip{f}{\omega f}+\frac1{2m}\sum_\mu\lVert C_\mu\psi_0\rVert^2 .
\]
\end{lemma}

\begin{proof}
Since $\psi_0\in D(H(0))\subseteq\Qf$, Lemma~\ref{lem:weyl} and the unitarity of $W_\theta$ give
\[
q_{p\epsilon}(W_\theta\psi_0)=\frac1{2m}\sum_\mu\lVert(p\epsilon_\mu-V_\mu-\theta C_\mu)\psi_0\rVert^2+\lVert\Hf^{1/2}\psi_0\rVert^2+\theta\ip{\psi_0}{A(\omega f)\psi_0}+\theta^2\ip{f}{\omega f}.
\]
Here $\psi_0\in D(C_\mu)$ and $C_\mu$ is symmetric. Expanding the square and using $q_0(\psi_0)=E_0$ and $\langle V\rangle=0$ (Lemma~\ref{lem:velocity-zero}),
\begin{equation}\label{eq:boost-poly}
q_{p\epsilon}(W_\theta\psi_0)-E_0=\frac{p^2}{2m}-\frac{\theta p}m\,\epsilon\cdot\langle C\rangle+\theta L+\theta^2Q,\qquad L:=\frac1m\sum_\mu\Rea\ip{V_\mu\psi_0}{C_\mu\psi_0}+\langle A(\omega f)\rangle .
\end{equation}
At $p=0$ the left-hand side is $q_0(W_\theta\psi_0)-E_0\ge0$ for all $\theta$, since $\lVert W_\theta\psi_0\rVert=1$ and $E_0=\inf\sigma(H(0))$. A polynomial $\theta L+\theta^2Q$ that is nonnegative on $\R$ has $L=0$. By Lemma~\ref{lem:amplitude}(i), $\langle A(k_\mu f)\rangle=0$, so $\epsilon\cdot\langle C\rangle=2eS$. The variational principle $E(p\epsilon)\le q_{p\epsilon}(W_\theta\psi_0)$ gives the claim.
\end{proof}

The vanishing of $L$ replaces the computation of the commutator $[H(0),a^*(f)-a(f)]$ and of the double commutator that would otherwise express $Q$; only $D_0\ge0$ and $D_0\psi_0=0$ are used.

\subsection{Proof of Theorem~\ref{thm:partI}}
\emph{Bound on $Q$.} By Lemma~\ref{lem:amplitude}(i) the cross terms vanish and $\sum_\mu\lVert C_\mu\psi_0\rVert^2=\sum_\mu\lVert A(k_\mu f)\psi_0\rVert^2+4e^2S^2$. From $\lVert a^*(\zeta)\psi_0\rVert^2=\lVert\zeta\rVert^2+\lVert a(\zeta)\psi_0\rVert^2$ we get $\lVert A(\zeta)\psi_0\rVert\le\lVert\zeta\rVert+2\lVert a(\zeta)\psi_0\rVert$, and Minkowski's inequality in $\ell^2$ together with Lemma~\ref{lem:amplitude}(iii) gives $\sum_\mu\lVert A(k_\mu f)\psi_0\rVert^2\le(\lVert kf\rVert+2\mathcal A)^2$, where $\lVert kf\rVert^2:=\sum_\mu\lVert k_\mu f\rVert^2$. Since $\ip{f}{\omega f}+\frac1{2m}\lVert kf\rVert^2=\ip{f}{\Omega_kf}=S$ by \eqref{eq:S-values},
\[
Q\ \le\ S+\frac2m\bigl(\lVert kf\rVert\mathcal A+\mathcal A^2\bigr)+\frac{2e^2S^2}m .
\]
By Appendix~\ref{app:integrals}, $\lVert kf\rVert^2\le\mathsf N_K:=\frac{2m^2}{3\pi^2}\bigl(\log(1+\sigma)-\frac\sigma{1+\sigma}\bigr)$ and $\mathcal A\le\mathsf a_K:=\frac{|e|m}{8\pi}\mathsf L(\sigma)$, uniformly in $k_1$. Hence $Q\le S(1+2x+\rho(k_1))$ with $\rho(k_1):=\frac2{mS}(\sqrt{\mathsf N_K}\,\mathsf a_K+\mathsf a_K^2)$.

\emph{Minimization.} We have $Q\ge\ip{f}{\omega f}>0$. Taking $\theta=peS/(mQ)$ in Lemma~\ref{lem:boost},
\[
E(p\epsilon)-E_0\le\frac{p^2}{2m}\Bigl(1-\frac{2e^2S^2}{mQ}\Bigr)\le\frac{p^2}{2m}\Bigl(1-\frac{2x}{1+2x+\rho(k_1)}\Bigr)=\frac{p^2}{2m}\Bigl(1+\frac{2x}{1+\rho(k_1)}\Bigr)^{-1}.
\]
The left-hand side does not depend on $k_1$. As $k_1\downarrow0$, $S\to\frac m{3\pi^2}\log(1+\sigma)$ by \eqref{eq:S-values}, so $x\to x_K$ and $\rho(k_1)\to2\sqrt{q_Kt_K}+2t_K=\rho_K$, since $\mathsf N_K/(mS)\to q_K$ and $\mathsf a_K^2/(mS)\to t_K$. This gives the first inequality of Theorem~\ref{thm:partI} for $P=p\epsilon$; since $\epsilon$ is arbitrary and $x_K,\rho_K$ do not depend on it, it holds for all $P$. With $P=p\hat z$, dividing by $p^2/2m$ and taking the $\limsup$ gives the bound on $h$. For $K=\min(\Lambda,m)$ and $\Lambda\ge m$, $K=m$ does not depend on $\Lambda$. \qed

\part*{Part II. Superlinear growth in the cutoff}
\section{Preliminaries}\label{sec:prelim}

We collect what is used from \cite{SerA}, show that the multipliers we need preserve the form domain, reduce the lower bound on $h$ to two properties of an estimator by a boost of the ground state (Lemma~\ref{lem:consumer}), and prove the general operator tools used later. The logical structure of the proof of Theorem~\ref{thm:main} is outlined in the guide below; the constants are tabulated at the beginning of Section~\ref{sec:proof}. Computational units (Section~\ref{sec:model}) are used throughout. We write $A\lesssim B$ if $A\le CB$ with $C$ depending only on $\alpha$, $\kappa$, the design constants and the window constants, and not on $\lambda$, and $A\asymp B$ if $A\lesssim B\lesssim A$ (Remark~\ref{rem:order}).

\emph{Guide to Part~II.} Part~II involves two locators, two levels, several windows and thresholds and three exceptional sets, but the proof of Theorem~\ref{thm:main} passes through only four steps: three inequalities and the choice of constants. The core locator $X_c$ (core level, threshold $g_s\asymp\ell^{-2}$) enters the final estimate only as the centre of the window of the level-$\kappa$ estimator $\Xw$ (threshold $g_w\asymp\ell^{-\kappa}$) and as the source of its exact covariance; $\Xw$ is the estimator inserted into the boost bound. Each step below hands one inequality to the final estimate; once it is granted, the rest of that step may be forgotten.
\begin{enumerate}[label=(\arabic*),leftmargin=2.2em]
\item \emph{Boost bound.} For every estimator $\hat X\in\mathscr C$, $c/h\le c\,\delta(\hat X)+\langle\Gamma_{\hat X}\rangle$ with $\delta(\hat X)\le\langle\|\Rm(\hat X)\|^2\rangle$ (Lemma~\ref{lem:consumer}). Since $c=\abar\ell^7$, the covariance defect must vanish outside sets of very small probability, while the fluctuation $\langle\Gamma\rangle$ carries no large factor. The multiplier calculus (Section~\ref{sec:tools-multiplier}, Lemma~\ref{lem:phase-mult}) supplies the domain statements behind this lemma and behind the Agmon and localization identities used in step~(3).
\item \emph{Deterministic bounds on the good set.} On the core good set $\mathcal G_\theta$, $\Rm(X_c)=0$ (Proposition~\ref{prop:Mprime}(b)) and hence $\Rm(\Xw)=0$ (Remark~\ref{rem:cov}). On $\mathcal G_\theta\cap(\Gamma'_\rho)^c$, where $\Gamma'_\rho$ is the shell event, one has $M\ge2M^w_*$ (Remark~\ref{rem:Mw}) and $\Gamma_{\Xw}\lesssim\ell^{\kappa-1}(\log\lambda)^2$ (Lemma~\ref{lem:Mw} with $M_*=M^w_*$, evaluated in the proof of Proposition~\ref{prop:expect}). For all configurations, $\Gamma_{\Xw}\le K_G\ell^D$ and $\|\Rm(\Xw)\|^2\le K_G\ell^D$ (Proposition~\ref{prop:global}). Of the window and the design, the bound on $\Gamma_{\Xw}$ uses only $\rho\asymp\ell\log\lambda$, $g_w\asymp\ell^{-\kappa}$, $M^w_*\asymp\ell^3/b^3$ and $K_{\mathrm S}=O(1)$; conditions (D2)--(D4) and the precise form of $\rho$ in (D5) serve the proofs of step~(3).
\item \emph{The three bad sets are rare to every power.} Outside $\mathcal G_\theta\cap(\Gamma'_\rho)^c$ one of the following occurs: the mass event $S_M$ (Proposition~\ref{prop:SM}, proved in Section~\ref{sec:constructive} from the input of \cite{SerA}), the rare event $\mathcal B_\theta$ of the core level (Proposition~\ref{prop:events}(b), proved in Appendix~\ref{app:core}), or the shell event $\Gamma'_\rho$ (Proposition~\ref{prop:shell}, Section~\ref{sec:shell}, using the external field of Section~\ref{sec:surgery}). For every $\kappa_p>0$, each has probability at most $\ell^{-\kappa_p}$ once the window constants are at least a threshold $K_0(\kappa_p)$ and $\lambda\ge\lambda_1$. Only this statement is used: with $\kappa_p>D+8$ the polynomial bounds of step~(2) make the contribution of the bad sets to both terms of the boost bound $O(\ell^{-1})$. The proofs of these three propositions contain the bulk of the work and most of the auxiliary constants; those of Propositions~\ref{prop:SM} and~\ref{prop:events} use neither Proposition~\ref{prop:shell} nor Proposition~\ref{prop:expect} (Section~\ref{sec:estimators}). The fluctuation bound of $X_c$ (Proposition~\ref{prop:Mprime}(a)) is used inside these proofs and, in its global form, for the polynomial bounds of Proposition~\ref{prop:global} in step~(2).
\item \emph{Choice of constants.} Steps~(2) and~(3) give $c\,\delta(\Xw)\lesssim\ell^{-1}$ and $\langle\Gamma_{\Xw}\rangle\lesssim\ell^{\kappa-1}(\log\lambda)^2$ for $\lambda\ge\lambda_1(\kappa)$ (Proposition~\ref{prop:expect}), hence $c/h\le C_\kappa\ell^{\kappa-1}(\log\lambda)^2$ and, with $\kappa=\min(7\epsilon/2,\frac12)$, Theorem~\ref{thm:main} (Section~\ref{sec:proof}). The constants are chosen in the order of Remark~\ref{rem:order}: all constants other than the window constants first, which fix $D$ and the polynomial degrees and hence the required power $\kappa_p^*$; then the window constants, which enter only the implicit constants and $\lambda_1$, not the powers of $\ell$.
\end{enumerate}

\emph{Symbols.} Each symbol in Table~\ref{tab:symbols} has one meaning throughout Part~II. Letters used only inside one construction or one proof (such as $L$, $\sigma$, $t$ in Section~\ref{sec:constructive}, the shell width $w$ of Lemma~\ref{lem:edge}, a parameter of that lemma distinct from the edge width $w$ of $\chi_W$, and the box side $n$ in Appendix~\ref{app:core}, and integration variables) are defined where they occur; a subscript $w$ (as in $g_w$, $d_w$, $\tau_w$, $\mu_w$, $M^w_*$) refers to the level $\kappa$ of Definition~\ref{def:Xw}. The correspondence with the notation of \cite{SerA} is the dictionary at the end of Section~\ref{ssec:inputA}.
\begin{center}\small
\captionof{table}{Symbols of Part~II.}\label{tab:symbols}
\begin{tabular}{p{0.2\textwidth}p{0.5\textwidth}p{0.2\textwidth}}
symbol & meaning & first used\\\hline
$m$, $\Lambda$, $\lambda=\Lambda/m$ & bare mass, ultraviolet cutoff, their ratio & Sect.~\ref{sec:model}\\
$e$, $\alpha=e^2/4\pi$, $\abar=e^2$ & charge, fine structure constant & Sect.~\ref{sec:model}\\
$h$ & effective mass ratio & Sect.~\ref{sec:model}\\
$c=\abar\ell^7$, $\ell$ & kinetic coefficient, localization length & Sect.~\ref{sec:model}\\
$\omega(k)=|k|$; $\Omega_\vac$ & dispersion; Fock vacuum & Sect.~\ref{sec:model}\\
$\Gamma(u)$, $\dG$ & second quantization (covariance matrices are $\Gamma$ with an index, carr\'e du champ $\Gamma_F$ with a function) & Sect.~\ref{sec:model}\\
$\mathrm e$ & Euler's number & Sect.~\ref{sec:prelim}\\
$F$, $F[1]$ & mean-field energy, minimum of $\mathcal E_1$ (with a subscript, $F_{\max}$, $F_{\mathrm{net}}$, and as the argument of Lemma~\ref{lem:agmon} and Corollary~\ref{cor:agmonquad}, $F$ denotes an Agmon level) & Sect.~\ref{sec:prelim}\\
$J$, $J$-real & complex conjugation in position space; vectors with $Jf=f$ & Sect.~\ref{sec:tools-multiplier}\\
$\transl_a$ & translation of the field by $a$ & Sect.~\ref{sec:prelim}\\
$\Rm$, $\Dm$ & covariance defect, translation derivative of an estimator & Sect.~\ref{sec:prelim}\\
$\psi$ & ground state of the fiber $H(0)$; in Proposition~\ref{prop:MF}, Section~\ref{sec:constructive} and Appendix~\ref{app:core}, an electronic wave function in the box $\mathcal C_e$ & Sect.~\ref{sec:model}\\
$\mAg$ & margin in the Agmon estimate & Lemma~\ref{lem:agmon}\\
$\rrel$ & relative localization length of electron and bubble & Lemma~\ref{lem:ims}\\
$\beta_{\mathrm{sel}}$ & sharpness of the selection by the field & Lemma~\ref{lem:softmax}\\
$\Kie$; $\omega_{\mathrm{ee}}$, $\Omega_{\mathrm{ee}}$ & seam matrix $P_{\mathrm{in}}\omega^2P_{\mathrm{ext}}$; frequency matrix and ground state of $\Hee$ & Lemma~\ref{lem:seam}\\
\end{tabular}
\end{center}
\begin{center}\small
\captionof*{table}{Table~\ref{tab:symbols} (continued)}
\begin{tabular}{p{0.2\textwidth}p{0.5\textwidth}p{0.2\textwidth}}
symbol & meaning & first used\\\hline
$s$, $b=C_bs$ & block and cube sides & Sect.~\ref{sec:estimators}\\
$\tau$, $\tau_w$; $\sat_\tau$ & saturation heights; saturation function $\sat_\tau(u)=\tau(1-\mathrm e^{-u/\tau})$ & Sect.~\ref{sec:estimators}\\
$d$, $d_w$ & depression indicators (core level, level $\kappa$) & Sect.~\ref{sec:estimators}\\
$r_a$ & smooth lifting of a denominator below $2a$ & Sect.~\ref{sec:estimators}\\
$X_c$, $W_{\mathrm{ms}}$, $W_*$; $\Mzero$ & core locator, its mass, its mass threshold; threshold of the local depression mass & Definition~\ref{def:core}\\
$\chi_W$, $r_W$, $w$ & window of the core locator, its radius and edge width & Definition~\ref{def:core}\\
$R_c$ & core radius: a physical length $\asymp\ell$ in Definition~\ref{def:core}, Proposition~\ref{prop:events} and Appendix~\ref{app:core}; in Section~\ref{sec:constructive} a number in units of $R_*$ (physical radius $R_cR_*$) & Definition~\ref{def:core}\\
$\Xw$, $M$; $\Gamma'_\rho$ & level-$\kappa$ estimator, its mass (the matrix $0\le M\le1$ of Lemma~\ref{lem:hardy}(iii), used only in Section~\ref{sec:surgery} and its proof in Appendix~\ref{app:cited}, is a different object); shell event & Definition~\ref{def:Xw}\\
$\chi_\rho$, $\rho$ & window of $\Xw$ and its radius & Definition~\ref{def:Xw}\\
$\mathcal B_\theta$, $\Gamma'_R$ & rare event of the core level; edge event of the core locator & Proposition~\ref{prop:events}\\
$K$, $K_d$ & window constants (lengths $K\ell\log\lambda$, $\rrel=K_d\ell\log\lambda$) & Proposition~\ref{prop:events}\\
$c_{\mathrm{ev}}$ & rate constant of the rare events & Proposition~\ref{prop:events}\\
$K_{\rm t}$ & tail constant in the seam threshold $b_\Sigma^2$ & Proposition~\ref{prop:shell}, Step 7\\
$\Eloss$ & loss in the configuration-wise mean-field bound & Proposition~\ref{prop:MF}\\
$\Kosc$ & frequency matrix of an oscillator & Lemma~\ref{lem:hardy}\\
$\varepsilon_{\mathrm H}$ & fraction of the external kinetic energy used in the Hardy margin & Lemma~\ref{lem:hardy}\\
$\mu_w$, $\mu$ & energy margins of the external field & Sect.~\ref{sec:surgery}\\
$T_{\mathcal E}$ & the operator $\omega^{-1/2}S\Pi1_{\mathcal E}$ & Lemma~\ref{lem:607N}\\
$\Tgrad$, $\Tgradmax$ & $\|\nabla\varphi\|^2$ and its bound ($\frac12\Tgradmax$ also bounds $\|\varphi\|_1$) & Lemma~\ref{lem:cubes}, Sect.~\ref{sec:constructive}\\
$\mathcal C_e$, $n_e$, $K_e$ & electron box (boxes are $\mathcal C$, the cutoff is $\Lambda$), its side $n_e=2s\lceil K_e\ell/(2s)\rceil$ & (D$'$), Appendix~\ref{app:core}\\
$\LipXi$ & Lipschitz constant of $\Xi_0$ & Lemma~\ref{lem:Wpm}\\
$\beta_{\mathrm{MF}}$, $c_{\mathrm{lev}}$ & mean-field constant of \cite{SerA}; constant in the core-level margin & Appendix~\ref{app:core}\\
\end{tabular}
\end{center}

\subsection{Input from \cite{SerA}}\label{ssec:inputA}
The paper \cite{SerA} proves $E=(1+o(1))(6\pi^2)^{-4/7}F[1]\,\abar^{2/7}\Lambda^{12/7}$ (bare mass $1$, $\abar=e^2=4\pi\alpha$). What we use is the configuration-wise block Born--Oppenheimer bound behind its lower bound, \cite[\SerAnum{prop:configBO}]{SerA}: before the infimum over the electronic wave function is taken, it bounds the fiber energy below by a multiplication operator in the field configuration, block by block. Together with the removal of the periodization \cite[\SerAnum{thm:periodization}(a)]{SerA} it enters only through Proposition~\ref{prop:MF} and Lemma~\ref{lem:Wpm}. We use it only to localize the mass of the depression set of almost minimizing configurations on the scale $\asymp\ell$ of the bubble (Section~\ref{sec:constructive}); no information on the shape of the bubble is needed, and no rate of convergence in \cite{SerA} is used: all that enters is $\varepsilon_1+\Eloss/F\to0$ as $\ell\to\infty$, together with the upper bound $E\le(1+o(1))F$ of \cite{SerA} (Section~\ref{sec:constructive}).

\emph{Dictionary of notation.} The paper \cite{SerA} uses different conventions, which we do not rewrite. There the letter $\lambda$ denotes a \emph{kinetic coefficient}: the working Hamiltonian is $\frac\lambda2(p+eA_\pi)^2+\Hf$ with band $|k|<\pi$, $\lambda=\Lambda/\pi$, length $\ell=(\lambda/e^2)^{1/7}$ and mean-field constant $F[1]$. Translated to the present paper: the kinetic coefficient $\lambda$ of \cite{SerA} is our $c=\abar\ell^7=\lambda/\pi$ (with our $\lambda=\Lambda/m$), the length $\ell$ and the constant $F[1]$ are the same, and $e^2=\abar$; the block constants $h_Q$ of \cite{SerA}, which we keep, are unrelated to the effective mass ratio $h$, and the mean-field constant $\beta$ of \cite{SerA} is written $\beta_{\mathrm{MF}}$. The sign of $e$ is immaterial, since $\Gamma(-1)$ maps $A$ to $-A$ and commutes with $\Hf$ and $\Pf$.

\subsection{Functions of the field configuration as multipliers}\label{sec:tools-multiplier}
All trial states and all Agmon and localization identities of Part~II multiply the fiber ground state by functions of the field configuration. This subsection determines which differentiability assumptions on such a function $\hat X$ ensure, on the Fock space with infinitely many modes, that (i) $\hat X$ preserves the form domain, (ii) the double commutator identity holds in the sense of forms, and (iii) the derivative of $\hat X$ under translations of the field equals $i[\Pf,\hat X]$. In Segal's representation the annihilation operators act as derivatives on Gaussian space, and the statements below are the Fock space form of the chain rule and of the closability of the Malliavin derivative and of the carr\'e du champ of the Ornstein--Uhlenbeck Dirichlet form; see \cite[Sect.~1.2]{Nu06}, \cite{BHi91}, and for Fock space and Segal fields \cite{Ar18}. We include the proofs in order to verify the domains for the Pauli--Fierz fibers.

\emph{Notation.} Let $\mathfrak h$ be the one-particle space, realized as the transverse $\mathbb C^3$-valued functions on the band, and $(JF)(k):=\overline{F(-k)}$. The real subspace $\mathfrak h_{\R}:=\{JF=F\}$ consists of the Fourier transforms of real transverse band-limited fields, and the coupling functions of $A(y)$ lie in $\mathfrak h_{\R}$. Let $\varphi(F)=2^{-1/2}(a(F)+a^*(F))$ and $W(F)=\mathrm e^{i\varphi(F)}$. The algebra $\mathfrak A:=\{W(F):F\in\mathfrak h_{\R}\}''$ is commutative (it is the algebra of multiplication operators in Segal's representation) and contains the spectral projections of $A_j(y)$. The \emph{total annihilation map} $\mathfrak a:\F\to\mathfrak h\otimes\F$ is $(\mathfrak a\Psi)^{(n)}(k;k_1,\dots,k_n):=\sqrt{n+1}\,\Psi^{(n+1)}(k,k_1,\dots,k_n)$, and $\mathfrak a_\omega:=(\omega^{1/2}\otimes1)\mathfrak a$, defined on finite-particle vectors; it is closable, and we use the same symbol for its closure. Then $Q(\Hf)=D(\mathfrak a_\omega)$ and $q_{\Hf}[\Psi]=\|\mathfrak a_\omega\Psi\|^2$: on finite-particle vectors $\|\mathfrak a_\omega\Psi\|^2=\sum_n(n+1)\langle\Psi^{(n+1)},(\omega\otimes1)\Psi^{(n+1)}\rangle=\langle\Psi,\mathrm d\Gamma(\omega)\Psi\rangle$ by symmetry, and one takes closures.

\emph{Cylinder operators.} For $g\in L^1\cap L^2(\R^3;\R^3)$ let $A(g):=\int g\cdot A(y)\dd y=\sqrt2\,\varphi(F_g)$, with $F_g\in\mathfrak h_{\R}\cap D(\omega^{-1/2})\cap D(\omega^{1/2})$ and $2\langle F_g,\omega F_{g'}\rangle=\langle\Pi g,\Pi g'\rangle_{L^2}$ by Plancherel, where $\Pi$ is the projection onto transverse band-limited fields. For real $F\in C_b^1(\R^n)$, $\hat X=F(A(g_1),\dots,A(g_n))$ (functional calculus of a commuting family) is called a \emph{cylinder operator}, with gradient
\[
\nabla\hat X:=\sum_{i=1}^n(\partial_iF)(A(g_1),\dots,A(g_n))\,\Pi g_i,
\]
a multiplication operator with values in $L^2(\R^3;\R^3)$.

\begin{definition}[The class $\mathscr C$]\label{def:class-C}
A bounded self-adjoint $\hat X\in\mathfrak A$ belongs to $\mathscr C$ if:
\begin{itemize}
\item[(C1)] \emph{(Gradient.)} There are cylinder operators $\hat X_N$ with $\sup_N\|\hat X_N\|<\infty$, $\hat X_N\to\hat X$ strongly, and gradients $\nabla\hat X_N$ uniformly bounded (as operators $\F\to L^2(\R^3;\R^3)\otimes\F$) and strongly convergent; the limit is denoted $\nabla\hat X$. It does not depend on the approximating sequence: the operator $G$ of Lemma~\ref{lem:multiplier}(i) is determined on $Q(\Hf)$ as $\mathfrak a_\omega\hat X-(1\otimes\hat X)\mathfrak a_\omega$, and $g\mapsto\omega^{1/2}F_g$ is injective on the range of $\Pi$.
\item[(C2)] \emph{(Translation.)} $\transl_a(\hat X):=\mathrm e^{ia\cdot\Pf}\hat X\mathrm e^{-ia\cdot\Pf}$ is differentiable in operator norm at $a=0$; $D_j:=\partial_{a_j}\transl_a(\hat X)|_{a=0}$.
\end{itemize}
A vector $\hat X=(\hat X_1,\hat X_2,\hat X_3)$ belongs to $\mathscr C$ if its components do. In Segal's representation $\transl_a(\hat X)$ is the multiplication by $\hat X(\transl_aA)$, the functional evaluated at the translated configuration (Setting~\ref{set:full}).
\end{definition}

\begin{lemma}[Multiplier lemma]\label{lem:multiplier}
Let $\hat X\in\mathscr C$.
\begin{itemize}
\item[(i)] $\hat X$ preserves $Q(\Hf)$, and there is a bounded operator $G:\F\to\mathfrak h\otimes\F$ with $\mathfrak a_\omega\hat X\Psi=(1\otimes\hat X)\mathfrak a_\omega\Psi+G\Psi$ and $\|G\Psi\|^2=\frac12\langle\Psi,\Gamma_{\hat X}\Psi\rangle$ for $\Psi\in Q(\Hf)$. The carr\'e du champ $\Gamma_{\hat X}:=(\nabla\hat X)^*(\nabla\hat X)\in\mathfrak A$ is bounded and nonnegative; for cylinder operators $\Gamma_{\hat X}=\|\nabla\hat X\|_{L^2}^2=\sum_{i,i'}\partial_iF\,\partial_{i'}F\,\langle\Pi g_i,\Pi g_{i'}\rangle$.
\item[(ii)] $\hat X$ preserves $D(P_{\mathrm f,j})$ and $i[P_{\mathrm f,j},\hat X]=D_j$; $D_j=D_j^*\in\mathfrak A$ is bounded.
\item[(iii)] For every fiber $H=H(P)$ of Section~\ref{sec:model}, with $V^P_j:=P_{\mathrm f,j}-eA_j(0)-P_j$, one has $Q(H)=Q(\Hf)\cap\bigcap_jD(\bar V^P_j)$ and $q_H=\frac c2\sum_j\|\bar V^P_j\cdot\|^2+q_{\Hf}$; in particular $Q(H(P))$ does not depend on $P$. $\hat X$ preserves $Q(H)$, and for $\Psi\in Q(H)$
\[
q_H[\hat X\Psi]=\mathrm{Re}\,q_H(\hat X^2\Psi,\Psi)+\frac12\bigl\langle\Psi,\bigl(c|D|^2+\Gamma_{\hat X}\bigr)\Psi\bigr\rangle .
\]
\item[(iv)] \emph{(Sufficient condition for (C1).)} Let $B\subset\R^3$ be a bounded domain and $\Theta:L^2(B;\R^3)\to\R$ bounded and Fr\'echet $C^1$ with $\sup\|\nabla\Theta\|_{L^2(B)}<\infty$. Then $\hat X:=\Theta(\mathbf 1_BA)$ (in Segal's representation, multiplication by $\Theta$ composed with the restriction of the configuration $A$ to $B$, which lies in $L^2(B)$ almost surely) satisfies (C1), with $\nabla\hat X=\Pi\nabla\Theta(\mathbf 1_BA)$ and $\Gamma_{\hat X}\le\sup\|\nabla\Theta\|^2$.
\item[(v)] \emph{(Sufficient condition for (C2).)} If $a\mapsto\hat X(\transl_aA)$ is $C^2$ near $a=0$ with first and second derivatives bounded uniformly in the configuration, then (C2) holds and $D$ is the multiplication by $\partial_a\hat X(\transl_aA)|_{a=0}$.
\item[(vi)] \emph{(Functional calculus.)} Let $\phi\in C^1(\R)$ be real. Then $\phi(\hat X)\in\mathscr C$, with $\nabla\phi(\hat X)=\phi'(\hat X)\nabla\hat X$, $\Gamma_{\phi(\hat X)}=\phi'(\hat X)^2\Gamma_{\hat X}$, translation derivative $\phi'(\hat X)D$, and operator $G_{\phi(\hat X)}=(1\otimes\phi'(\hat X))G$ in (i).
\end{itemize}
\end{lemma}

\begin{proof}
\emph{Step 1 (Weyl operators).} For $F\in\mathfrak h_{\R}\cap D(\omega^{1/2})$, on finite-particle vectors $\mathfrak a\,a^*(F)=F\otimes\cdot+(1\otimes a^*(F))\mathfrak a$ and $\mathfrak a\,a(F)=(1\otimes a(F))\mathfrak a$, so $\mathfrak a\varphi(F)=(1\otimes\varphi(F))\mathfrak a+2^{-1/2}F\otimes\cdot$. Finite-particle vectors are analytic for $\varphi(F)$; differentiating $t\mapsto(1\otimes W(tF))^*\mathfrak aW(tF)\Psi$ gives $\mathfrak aW(F)=(1\otimes W(F))(\mathfrak a+\frac i{\sqrt2}F\otimes\cdot)$, and multiplying by $\omega^{1/2}\otimes1$ gives the version for $\mathfrak a_\omega$. Closedness of $\mathfrak a_\omega$ extends it to all of $Q(\Hf)$.

\emph{Step 2 (cylinder operators).} If $F\in\mathcal S(\R^n)$, then $\hat X=(2\pi)^{-n/2}\int\hat F(\xi)W(\sqrt2\,\xi\cdot F_g)\dd\xi$ as a strong Bochner integral, the $\varphi(F_{g_i})$ commuting. Integrating Step~1 against $\hat F$ (note $\int|\hat F|(1+|\xi|)\dd\xi<\infty$) and using closedness, $\mathfrak a_\omega\hat X\Psi=(1\otimes\hat X)\mathfrak a_\omega\Psi+\sum_i(\omega^{1/2}F_{g_i})\otimes(\partial_iF)\Psi$, and the squared norm of the last term is $\langle\Psi,\sum\partial_iF\partial_{i'}F\langle\omega^{1/2}F_{g_i},\omega^{1/2}F_{g_{i'}}\rangle\Psi\rangle=\frac12\langle\Psi,\|\nabla\hat X\|^2\Psi\rangle$. Thus $G=2^{-1/2}(\iota\otimes1)\nabla\hat X$ with the isometry $\iota:\Pi g\mapsto\sqrt2\,\omega^{1/2}F_g$; in particular $G$ is linear in the gradient. For $F\in C^1_b$, the functions $F_\delta:=(F*j_\delta)\chi(\delta\,\cdot)$ converge to $F$ together with $\nabla F$ locally uniformly and boundedly; dominated convergence in Segal's representation gives strong convergence of $\hat X_\delta$ and of the gradients, and the closedness argument of Step~3 applies.

\emph{Step 3 (limits).} For the sequence of (C1) and $\Psi\in Q(\Hf)$, $\hat X_N\Psi\to\hat X\Psi$ and $\mathfrak a_\omega\hat X_N\Psi=(1\otimes\hat X_N)\mathfrak a_\omega\Psi+G_N\Psi$ converges, where $G_N$ is the operator of Step~2 with $\|G_N\Psi\|^2=\frac12\langle\Psi,\|\nabla\hat X_N\|^2\Psi\rangle$; closedness of $\mathfrak a_\omega$ gives (i), and $G_N\to G$ strongly. Since $G_N^*G_N=\frac12\|\nabla\hat X_N\|^2\in\mathfrak A$ converges weakly and $\mathfrak A$ is weakly closed, $\Gamma_{\hat X}\in\mathfrak A$. Moreover $G_N\hat X_N=(1\otimes\hat X_N)G_N$ (both are multiplications), and in the limit $G\hat X=(1\otimes\hat X)G$. $\hat X_N^2$ is a cylinder operator with gradient $2\hat X_N\nabla\hat X_N\to2\hat X\nabla\hat X$; hence $\hat X^2$ satisfies (C1) and $\mathfrak a_\omega\hat X^2\Psi=(1\otimes\hat X^2)\mathfrak a_\omega\Psi+2(1\otimes\hat X)G\Psi$. With $Y:=\mathfrak a_\omega\Psi$,
\begin{align*}
q_{\Hf}[\hat X\Psi]-\mathrm{Re}\,q_{\Hf}(\hat X^2\Psi,\Psi)&=\|(1\otimes\hat X)Y+G\Psi\|^2-\mathrm{Re}\langle(1\otimes\hat X^2)Y+2(1\otimes\hat X)G\Psi,Y\rangle\\
&=\|G\Psi\|^2=\tfrac12\langle\Psi,\Gamma_{\hat X}\Psi\rangle,
\end{align*}
the cross terms cancelling by self-adjointness of $\hat X$.

\emph{Step 4 (translations).} For $\Psi\in D(P_{\mathrm f,j})$,
\[
\tfrac1t\bigl(\mathrm e^{itP_{\mathrm f,j}}\hat X\Psi-\hat X\Psi\bigr)=\tfrac1t\bigl(\transl_{te_j}(\hat X)-\hat X\bigr)\mathrm e^{itP_{\mathrm f,j}}\Psi+\hat X\,\tfrac1t\bigl(\mathrm e^{itP_{\mathrm f,j}}\Psi-\Psi\bigr)\to D_j\Psi+i\hat XP_{\mathrm f,j}\Psi ;
\]
by Stone's theorem $\hat X\Psi\in D(P_{\mathrm f,j})$ and $iP_{\mathrm f,j}\hat X\Psi=D_j\Psi+i\hat XP_{\mathrm f,j}\Psi$, which is (ii). $\transl_a$ preserves $\mathfrak A$, since $\transl_aW(F)=W(\mathrm e^{ia\cdot k}F)$ and $\mathrm e^{ia\cdot k}F\in\mathfrak h_{\R}$; taking norm limits, $D_j\in\mathfrak A$.

\emph{Step 5 (form domain of $H$).} Let $t:=\frac c2\sum_j\|\bar V^P_j\cdot\|^2+q_{\Hf}$ on $D(t):=Q(\Hf)\cap\bigcap_jD(\bar V^P_j)$; as a sum of closed nonnegative forms, $t$ is closed. For $\Psi\in D(H)=D(\Hf)\cap D(\Pf^2)$ we have $V^P_j\Psi\in D(V^P_j)$, since $P_{\mathrm f,j}$ commutes with $\Hf$ and $A_j(0)$, $[P_{\mathrm f,j},A_j(0)]$ are $\Hf^{1/2}$-bounded (Fact~\ref{fact:rel-bound}), and $H(P)$ acts on this domain as $\frac c2\sum_j(V^P_j)^2+\Hf$ \cite[Sect.~2]{Hi07}; hence $t(\Psi',\Psi)=\langle\Psi',H\Psi\rangle$ for $\Psi'\in D(t)$. The self-adjoint operator associated with $t$ is therefore an extension of $H$, and since self-adjoint operators have no proper self-adjoint extensions the two coincide: $Q(H)=D(t)$ and $q_H=t$. As $\bar V^P_j=\bar V^0_j-P_j$, the domain $D(t)$ does not depend on $P$. $\hat X$ commutes with $A_j(0)$, which is affiliated with the same commutative algebra, and preserves $D(A_j(0))$; by (ii), $V^P_j\hat X=\hat XV^P_j-iD_j$ on $D(P_{\mathrm f,j})\cap D(A_j(0))$, and by closedness this extends to $D(\bar V^P_j)$. Together with (i), $\hat XD(t)\subset D(t)$. From $\bar V^P_j\hat X^2=\hat X^2\bar V^P_j-2i\hat XD_j$ and the self-adjointness and commutativity of $D_j$ and $\hat X$, $\|\bar V^P_j\hat X\Psi\|^2-\mathrm{Re}\langle\bar V^P_j\hat X^2\Psi,\bar V^P_j\Psi\rangle=\|D_j\Psi\|^2$ (the cross terms $2\,\mathrm{Re}(i\langle\hat X\bar V^P_j\Psi,D_j\Psi\rangle)$ agree on both sides). Adding Step~3 gives (iii).

\emph{Step 6 (sufficient condition for (C1)).} Let $\{e_k\}$ be an orthonormal basis of $L^2(B;\R^3)$ with $e_k\in L^1\cap L^2$, and $\Pi_N$ the projection onto the first $N$ elements. For every configuration with $\mathbf 1_BA\in L^2(B)$ (almost surely for the vacuum measure), $\Pi_N(\mathbf 1_BA)=\sum_{k\le N}A(e_k)e_k\to\mathbf 1_BA$ in $L^2(B)$. $\hat X_N:=\Theta(\Pi_N\mathbf 1_BA)$ is a cylinder operator ($F_N(q)=\Theta(\sum q_ke_k)\in C^1_b$) with gradient $\Pi\Pi_N\nabla\Theta(\Pi_N\mathbf 1_BA)$. Continuity of $\Theta$ and $\nabla\Theta$ gives almost sure convergence, and uniform boundedness and dominated convergence give strong convergence; this is (iv).

\emph{Step 7 (sufficient condition for (C2)).} By Taylor's formula at each configuration, $\|\transl_a(\hat X)-\hat X-a\cdot\partial_a\hat X(\transl_aA)|_{a=0}\|\le\frac12|a|^2\sup_{|a'|\le|a|}\sup_A|\partial_a^2\hat X(\transl_{a'}A)|$, since the operator norm of a multiplication operator is the essential supremum. This is (v).

\emph{Step 8 (functional calculus).} All operators involved lie in the commutative algebra $\mathfrak A$. Let $I$ be a bounded interval containing the spectra of $\hat X$ and of all $\hat X_N$. If $\hat X_N\to\hat X$ strongly with $\sup_N\|\hat X_N\|<\infty$, then $p(\hat X_N)\to p(\hat X)$ strongly for polynomials $p$, hence, by uniform approximation on $I$, $\psi(\hat X_N)\to\psi(\hat X)$ strongly for every continuous $\psi$. Now $\phi(\hat X_N)=(\phi\circ F_N)(A(g_1),\dots)$ is a cylinder operator with gradient $\phi'(\hat X_N)\nabla\hat X_N$, a product of uniformly bounded strongly convergent sequences, which converges strongly to $\phi'(\hat X)\nabla\hat X$; this is (C1) for $\phi(\hat X)$. By the linearity of $G$ in the gradient (Step~2), the operator of Step~2 for $\phi(\hat X_N)$ is $(1\otimes\phi'(\hat X_N))G_N$, which converges strongly to $(1\otimes\phi'(\hat X))G$. For (C2), $\transl_a$ is a $*$-automorphism of $\mathfrak A$, so $\transl_a(\phi(\hat X))=\phi(Y_a)$ with $Y_a:=\transl_a(\hat X)$; since $Y_a$ and $Y_0$ commute, the mean value theorem on the joint spectrum gives $\|\phi(Y_a)-\phi(Y_0)-\phi'(Y_0)(Y_a-Y_0)\|\le m_{\phi'}(\|Y_a-Y_0\|)\,\|Y_a-Y_0\|$, with $m_{\phi'}$ the modulus of continuity of $\phi'$ on $I$. As $Y_a$ is norm differentiable at $a=0$ with derivative $D$, so is $\phi(Y_a)$, with derivative $\phi'(\hat X)D$. This is (vi).
\end{proof}

In particular $\mathrm e^{f}\in\mathscr C$ for real $f\in\mathscr C$, which is used in the Agmon estimate (Lemma~\ref{lem:agmon}), and $\cos(p\hat X),\sin(p\hat X)\in\mathscr C$ for $\hat X\in\mathscr C$ and $p\in\R$, which is used for the phase $\mathrm e^{ip\hat X}$ in Lemma~\ref{lem:phase-mult}.

\subsection{Estimators and covariance defects}
An \emph{estimator} is a functional $\hat X$ of the field configuration with values in $\R^3$, read as a position. For an estimator in $\mathscr C$ we write $\Dm_{ij}:=i[P_{\mathrm f,j},\hat X_i]$ for its translation derivative (Lemma~\ref{lem:multiplier}(ii)) and call $\Rm:=I-\Dm$ the \emph{covariance defect}: $\Rm=0$ if a rigid translation of the field by $a$ moves $\hat X$ exactly by $a$. We set $\Gamma_{\hat X}:=\sum_i\Gamma_{\hat X_i}$.

All our estimators belong to $\mathscr C$. They are bounded functionals of the restriction of the configuration to a bounded ball, built from the local quantities $u'_{Q(z)}$ of Section~\ref{sec:estimators} by integrals against normalized kernels, smooth functions with bounded derivatives and denominators $r_a(\cdot)\ge a>0$. Since the saturation $\sat_\tau(u)=\tau(1-\mathrm e^{-u/\tau})$ has $|\partial_A\sat_\tau(u'_Q/|Q|)|\le(2\tau/(\mathrm e|Q|))^{1/2}$, they are Fr\'echet $C^1$ with bounded gradient, which gives (C1) by Lemma~\ref{lem:multiplier}(iv). By covariance of the local levels, the translation parameter $a$ of $\transl_a$ enters $\hat X(\transl_aA)$ only through the windows $\chi_W$, $\chi_\rho$ and through $X_c(\transl_aA)-a$ (Definitions~\ref{def:core} and~\ref{def:Xw}). The windows are of class $C^2$, so $a\mapsto\hat X(\transl_aA)$ is $C^2$ with first and second derivatives bounded uniformly in the configuration (by a polynomial in $\ell$), which gives (C2) by Lemma~\ref{lem:multiplier}(v).

\subsection{The boost bound}
\begin{lemma}[Phase multipliers]\label{lem:phase-mult}
Let $\hat X\in\mathscr C$ be real, with translation derivative $D=(D_1,D_2,D_3)$ and operator $G$ as in Lemma~\ref{lem:multiplier}, and $U_p:=\mathrm e^{ip\hat X}$, $p\in\R$. Then (i) $U_p$ preserves $Q(\Hf)$ and $\mathfrak a_\omega U_p\Psi=(1\otimes U_p)(\mathfrak a_\omega\Psi+ip\,G\Psi)$; (ii) for every $P$, $U_p$ preserves $D(\bar V^P_j)$ and $\bar V^P_jU_p\Psi=U_p(\bar V^P_j+pD_j)\Psi$; (iii) $U_p$ preserves $Q(H(P))$.
\end{lemma}
\begin{proof}
By Lemma~\ref{lem:multiplier}(vi), $\cos(p\hat X)$ and $\sin(p\hat X)$ lie in $\mathscr C$, with operators $-p(1\otimes\sin p\hat X)G$ and $p(1\otimes\cos p\hat X)G$ in (i) and translation derivatives $-p\sin(p\hat X)D$ and $p\cos(p\hat X)D$. Adding the identities of Lemma~\ref{lem:multiplier}(i) for $\cos(p\hat X)$ and $i\sin(p\hat X)$ gives (i). Lemma~\ref{lem:multiplier}(ii) gives $i[P_{\mathrm f,j},U_p]=ipU_pD_j$, i.e.\ $P_{\mathrm f,j}U_p=U_p(P_{\mathrm f,j}+pD_j)$ on $D(P_{\mathrm f,j})$. $U_p$ commutes with $A_j(0)$ and preserves $D(A_j(0))$, so $V^P_jU_p=U_p(V^P_j+pD_j)$ on $D(P_{\mathrm f,j})\cap D(A_j(0))$, and closedness of $\bar V^P_j$ gives (ii). (iii) follows from (i), (ii) and the description of $Q(H(P))$ in Lemma~\ref{lem:multiplier}(iii).
\end{proof}

\begin{lemma}[Boost bound]\label{lem:consumer}
Let $\hat X=(\hat X_1,\hat X_2,\hat X_3)\in\mathscr C$ be an estimator, let $D_j:=i[P_{\mathrm f,j},\hat X_z]$ be the translation derivative of its component $\hat X_z$, and set $\delta(\hat X):=\langle|D-e_z|^2\rangle$. Then
\begin{equation}\label{eq:consumer}
\frac ch\le c\,\delta(\hat X)+\langle\Gamma_{\hat X}\rangle.
\end{equation}
Moreover $\delta(\hat X)\le\langle\|\Rm\|^2\rangle$.
\end{lemma}
\begin{proof}
Let $U_p:=\mathrm e^{ip\hat X_z}$ and let $G$ be the operator of Lemma~\ref{lem:multiplier}(i) for $\hat X_z$. By Lemma~\ref{lem:phase-mult}, $U_p\psi\in Q(H(p\hat z))$ for every $p\in\R$, and since $\bar V^{p\hat z}_j=\bar V^0_j-p\delta_{jz}$, Lemma~\ref{lem:phase-mult}(i)(ii) and the form of $q_H$ in Lemma~\ref{lem:multiplier}(iii) give
\[
q_{H(p\hat z)}[U_p\psi]=\frac c2\sum_j\bigl\|\bigl(\bar V^0_j+p(D_j-\delta_{jz})\bigr)\psi\bigr\|^2+\bigl\|\mathfrak a_\omega\psi+ipG\psi\bigr\|^2=E+p\,\mathsf c_1+\frac{p^2}2\,B,
\]
where $\mathsf c_1:=c\sum_j\mathrm{Re}\langle\bar V^0_j\psi,(D_j-\delta_{jz})\psi\rangle-2\,\mathrm{Im}\langle\mathfrak a_\omega\psi,G\psi\rangle$ and $B:=c\langle|D-e_z|^2\rangle+\langle\Gamma_{\hat X_z}\rangle$; for the quadratic term we used that the $D_j$ are bounded, self-adjoint and commute, and $\|G\psi\|^2=\frac12\langle\Gamma_{\hat X_z}\rangle$. Since $\|U_p\psi\|=1$, the variational principle and $E(p\hat z)\ge E$ (Theorem~\ref{thm:facts}) give $E\le E(p\hat z)\le E+p\,\mathsf c_1+\frac{p^2}2B$ for every $p\in\R$. A polynomial $p\,\mathsf c_1+\frac{p^2}2B$ that is nonnegative on $\R$ has $\mathsf c_1=0$, hence
\[
E(p\hat z)-E\le\frac{p^2}2\,B ,
\]
and Definition~\ref{def:masses} gives $h\ge c/B$. As $\Gamma_{\hat X_z}\le\Gamma_{\hat X}$, this is \eqref{eq:consumer}. The last claim holds because $|D-e_z|$ is the norm of the $z$-th row of $\Rm$.
\end{proof}

No sum rule, no time reversal symmetry and no simplicity of the ground state is used: the term linear in $p$ vanishes because $E(P)\ge E(0)$ for all $P$, by the same principle by which the term linear in $\theta$ in Lemma~\ref{lem:boost} vanishes because $\psi_0$ is a ground state. Bounds of this kind on $E(P)-E(0)$ by trial states are used in \cite{LS20}. Hence $c/h\le C$ follows from an estimator with
\begin{equation}\label{eq:two}
c\,\delta(\hat X)\lesssim1\qquad\text{and}\qquad\langle\Gamma_{\hat X}\rangle\lesssim1 .
\end{equation}
The first condition (almost covariance) carries the large factor $c\asymp\ell^7$; the second (small fluctuation) does not. Our estimator is centred at the core locator $X_c$ of Section~\ref{sec:estimators} and inherits the first property from it (Section~\ref{sec:proof}); the new point is $\langle\Gamma\rangle\to0$.

\subsection{General tools}
The form identities of Setting~\ref{set:full} and Lemmas~\ref{lem:agmon}, \ref{lem:ims} and~\ref{lem:softmax} hold directly in the full model by Lemma~\ref{lem:multiplier}. Statements that use the Schr\"odinger representation of the field (Lemma~\ref{lem:seam}, Lemma~\ref{lem:Xin} and Theorem~\ref{thm:gauge}) use it only for finitely many canonical pairs, and this representation exists in the full model (Lemma~\ref{lem:Xin}). No finite-mode approximation of the model is used.

For a real function $F$ of the configuration, $\Gamma_F=|\nabla_qF|^2$ (with the normalization of the modes), and in the full-space model $\nabla_xF$ denotes the gradient in the electron coordinate.

\begin{setting}[Full-space model and collective coordinates]\label{set:full}
Consider $H_{\mathrm{full}}=\frac c2(p_x+eA(x))^2+\Hf$ on $L^2(\R^3_x)\otimes\F$, and denote by $\transl_a$ the translation of the field by $a$, $(\transl_aA)(y)=A(y-a)$; thus $\transl_{-x}A$ is the field seen from an electron at $x$, and $\Dm=\partial_a\hat X(\transl_aA)|_{a=0}$. For an estimator $\hat X$ (a function of the configuration) we read $\hat X(\transl_{-x}A)$ as the position of the bubble seen from the electron at $x$, and set $X_{\mathrm{abs}}:=x+\hat X(\transl_{-x}A)$ (absolute position) and $X_{\mathrm{rel}}:=\hat X(\transl_{-x}A)$ (relative position). The chain rule gives $\nabla_xX_{\mathrm{abs}}=\Rm(\transl_{-x}A)$ and $\nabla_xX_{\mathrm{rel}}=-\Dm(\transl_{-x}A)$. If $H_{\mathrm{full}}\ge V$ in the sense of forms for a multiplication operator $V$ in $(x,q)$ that is invariant under joint translations, then $H(0)\ge\bar V$ in the fiber, where $\bar V$ is $V$ written in relative coordinates: the direct integral decomposition (Theorem~\ref{thm:facts}) gives $q_{H(P)}[\Phi]\ge\langle\Phi,\bar V\Phi\rangle$ for each $\Phi$ and almost every $P$ (for a fixed $\Phi$ in the common form domain of the $H(P)$, Lemma~\ref{lem:multiplier}(iii), test the form inequality on the wave packets $\int^\oplus g(P)\Phi\,\dd P$ with $g\in L^\infty$ of compact support; this gives $\int|g|^2(q_{H(P)}[\Phi]-\langle\Phi,\bar V\Phi\rangle)\dd P\ge0$ for all $g$, hence the inequality at $\Phi$ for almost every $P$; here $V=V_1-\mathcal W$ with $V_1$ bounded below and $\mathcal W\ge0$, as in all applications ($\mathcal W=0$ in Appendix~\ref{app:core}; in Section~\ref{sec:shell} $V_1$ contains the Hardy margins, which are bounded below by $-B_-$ but not above, and $\mathcal W$ is the non-negative quadratic form of Step~4 there), and the inequality is read as $q_{H(P)}[\Phi]+\langle\Phi,\bar{\mathcal W}\Phi\rangle\ge\langle\Phi,\bar V_1\Phi\rangle$; it is trivial if $\langle\Phi,\bar{\mathcal W}\Phi\rangle=\infty$, and otherwise all terms for the wave packets are integrals over $P$ by Tonelli's theorem, so no continuity of $\Phi\mapsto\langle\Phi,\bar V\Phi\rangle$ is needed), and since $P\mapsto q_{H(P)}[\Psi]$ is a polynomial for each $\Psi\in Q(H(0))$ (Lemma~\ref{lem:multiplier}(iii)), the inequality holds for every $P$, in particular for $P=0$. Non-invariant $V$ are first averaged over translations of the partition lattices used to build them: if the construction on the lattices translated by $t$ gives $H_{\mathrm{full}}\ge V_t$ and is the joint translate of the construction for $t=0$ (all ingredients --- cells, candidates, external regions, $\Ebox$ of the boxes, $r_{\mathrm e}$, $T_\Sigma$, $c^{(r)}$ --- move rigidly with the lattices), then $V_{t+a}(x+a,\transl_aA)=V_t(x,A)$; if moreover $V_t$ is periodic in $t$ with the period $n$ of the lattices, then $V^{\mathrm{av}}:=n^{-3}\int_{[0,n)^3}V_t\dd t$ satisfies $H_{\mathrm{full}}\ge V^{\mathrm{av}}$ and $V^{\mathrm{av}}(x+a,\transl_aA)=V^{\mathrm{av}}(x,A)$ for all $a$ (substitute $t\mapsto t-a$ and use the periodicity).
\end{setting}

\subsubsection{Agmon estimate with deficit}
This is a version of the exponential decay estimates of Agmon \cite{Ag82} in the field configuration. Let $H$ be the fiber Hamiltonian with ground state $\psi$, and $f$ a bounded real function of the configuration in $\mathscr C$. Applying Lemma~\ref{lem:multiplier}(iii) to $\mathrm e^f\in\mathscr C$ (Lemma~\ref{lem:multiplier}(vi)) gives $q_H[\mathrm e^f\psi]-E\|\mathrm e^f\psi\|^2=\frac12\langle \mathrm e^{2f}\GAg_f\rangle$ with the \emph{Agmon metric}
\begin{equation}\label{eq:agmonmetric}
\GAg_f:=c\,|Df|^2+\Gamma_f,
\end{equation}
where $Df$ is the translation derivative of $f$.

\begin{lemma}[Agmon estimate with deficit]\label{lem:agmon}
Assume $H\ge V-c_0$ in the sense of forms, with $V$ a multiplication operator in the configuration and $V_-$ integrable with respect to $\psi$; let $0\le f\le F_{\max}$, and suppose that on $\mathcal Z:=\{f>0\}\cup\{\GAg_f>0\}$
\[
V-c_0-E-\tfrac12\GAg_f\ge \mAg-\Lexc
\]
for some $\mAg>0$ and some function $\Lexc\ge0$. Then for all $F>0$
\[
\mAg\,\mathrm e^{2F}P_\psi(f\ge F)\le\langle(E+c_0-V)_+1_{\mathcal Z^c}\rangle+\mathrm e^{2F_{\max}}\langle\Lexc1_{\mathcal Z}\rangle,
\]
where $P_\psi$ is the distribution of the configuration in $\psi$.
\end{lemma}
\begin{proof}
From $H\ge V-c_0$ and \eqref{eq:agmonmetric}, $\langle \mathrm e^{2f}(V-c_0-E-\frac12\GAg_f)\rangle\le0$. On $\mathcal Z^c$ we have $f=0=\GAg_f$ and the integrand is $\ge-(E+c_0-V)_+$; on $\mathcal Z$ it is $\ge \mathrm e^{2f}(\mAg-\Lexc)$. Hence $\mAg\langle \mathrm e^{2f}1_{\mathcal Z}\rangle\le\langle(E+c_0-V)_+1_{\mathcal Z^c}\rangle+\langle \mathrm e^{2f}\Lexc1_{\mathcal Z}\rangle$. Finally $\{f\ge F\}\subset\mathcal Z$, $\mathrm e^{2f}\ge \mathrm e^{2F}$ there, and $\mathrm e^{2f}\le \mathrm e^{2F_{\max}}$.
\end{proof}

The point is that the deficit is paid in expectation rather than by cutting it off: terms in which the large factor $c$ multiplies the translation derivative of $f$ are harmless as long as they live on rare sets.

\begin{lemma}[Gaussian tails of quadratic field functionals]\label{lem:tail}
Let $h_1,h_2,\dots\in\mathfrak h$ be $J$-real with $\sum\|h_m\|^2<\infty$, let $W:=\sum_m\Phi(h_m)^2$, and let $\Gamma$ be the covariance matrix of the $\Phi(h_m)$ in the free vacuum. For the ground state $\psi$ and $0\le s$ with $2s\|\Gamma\|<1$, $\langle \mathrm e^{sW}\rangle\le\det(1-2s\Gamma)^{-1/2}$; consequently $\langle W\rangle\le\tr\Gamma$ and
\[
P_\psi(W>b^2)\le\exp\Bigl(-\frac{b^2-2\tr\Gamma}{4\|\Gamma\|}\Bigr).
\]
\end{lemma}
Covariance matrices are denoted $\Gamma$, $\Gamma_r$, $\Gamma_\Sigma$, $\Gamma_{r,j}$; the carr\'e du champ $\Gamma_F$ always carries a function as subscript. The proof (domination of the ground state field by the free vacuum in the sense of sub-Gaussian moment generating functions) is given in Appendix~\ref{app:cited}.

\begin{corollary}[Quadratic terms paid by the deficit]\label{cor:agmonquad}
In the setting of Lemma~\ref{lem:agmon} let $V=V_0-\sum_r\kappa_r\sum_ja_{r,j}W_{r,j}$ with $\kappa_r>0$, weights $a_{r,j}\ge0$ that may depend on the configuration, with $\sum_ja_{r,j}\le1$ and $a_{r,j}\le\bar a_{r,j}$ for constants $\bar a_{r,j}$ (for constant weights $\bar a_{r,j}=a_{r,j}$), and $W_{r,j}$ as in Lemma~\ref{lem:tail} with covariances $\Gamma_{r,j}$. Assume $\sum_r\kappa_rb_r^2\le \mAg$ and $V_0-c_0-E-\frac12\GAg_f\ge2\mAg-\mathcal L_0$ on $\mathcal Z$ for some $\mathcal L_0\ge0$. Then
\begin{multline*}
P_\psi(f\ge F)\le\frac{\mathrm e^{-2F}}\mAg\Bigl(\langle(E+c_0-V_0)_+\rangle+\sum_r\kappa_r\sum_j\bar a_{r,j}\tr\Gamma_{r,j}\Bigr)+\frac{\mathrm e^{2(F_{\max}-F)}}\mAg\langle\mathcal L_01_{\mathcal Z}\rangle\\
+\frac{\mathrm e^{2(F_{\max}-F)}}\mAg\sum_r\kappa_r\sum_j\bar a_{r,j}\,4\|\Gamma_{r,j}\|\,\mathrm e^{-(b_r^2-2\tr\Gamma_{r,j})/(4\|\Gamma_{r,j}\|)}.
\end{multline*}
\end{corollary}
\begin{proof}
Using $W\le b^2+(W-b^2)_+$, $\sum_ja_{r,j}\le1$ and $a_{r,j}\le\bar a_{r,j}$, on $\mathcal Z$ we get $V-c_0-E-\frac12\GAg_f\ge \mAg-\Lexc$ with $\Lexc:=\mathcal L_0+\sum_r\kappa_r\sum_j\bar a_{r,j}(W_{r,j}-b_r^2)_+$. Apply Lemma~\ref{lem:agmon}, $(E+c_0-V)_+\le(E+c_0-V_0)_++\sum\kappa_r\bar a_{r,j}W_{r,j}$ (sums with constant coefficients, so that the expectations below apply term by term), $\langle W\rangle\le\tr\Gamma$, and $\langle(W-b^2)_+\rangle=\int_{b^2}^\infty P(W>t)\dd t\le4\|\Gamma\|\mathrm e^{-(b^2-2\tr\Gamma)/(4\|\Gamma\|)}$ (Lemma~\ref{lem:tail}).
\end{proof}

\subsubsection{Localization in collective coordinates}
\begin{lemma}[IMS localization in collective coordinates]\label{lem:ims}
(Cf.\ the IMS localization formula \cite[Sect.~3.1]{CFKS87}.) In Setting~\ref{set:full} let $\hat X$ be $C^2$ with bounded derivatives. Let $\sum_j\chi_j^2=1$ be a partition subordinate to a lattice of cubes of side $n_b$ with $\sum_j|\nabla\chi_j|^2\le C_\chi/n_b^2$, and $\eta_0^2+\eta_1^2=1$ with $\eta_0=1$ on $[0,\rrel]$, $\eta_0=0$ on $[2\rrel,\infty)$, $|\eta_i'|^2\le C_\eta/\rrel^2$. Put
\[
\varphi_j:=\chi_j(X_{\mathrm{abs}})\,\eta_0(|X_{\mathrm{rel}}|),\qquad\varphi_\infty:=\eta_1(|X_{\mathrm{rel}}|).
\]
Then, in the sense of forms,
\[
H_{\mathrm{full}}=\sum_k\varphi_kH_{\mathrm{full}}\varphi_k-\frac c2\sum_k|\nabla_x\varphi_k|^2-\frac12\sum_k\Gamma_{\varphi_k},
\]
and pointwise
\[
\frac c2\sum_k|\nabla_x\varphi_k|^2\le\frac{cC_\chi}{n_b^2}\|\Rm\|^2\,1_{\{|X_{\mathrm{rel}}|\le2\rrel\}}+\frac{2cC_\eta}{\rrel^2}\|\Dm\|^2\,1_{\{\rrel\le|X_{\mathrm{rel}}|\le2\rrel\}}.
\]
If $\varphi_j\ne0$, then $X_{\mathrm{abs}}\in\supp\chi_j$ and $x\in\supp\chi_j+\overline{B(0,2\rrel)}$. The partition is covariant under joint translations of electron and field.
\end{lemma}
\begin{proof}
For $\sum_k\varphi_k^2=1$ and each of the two kinetic terms $T$ one has $\sum_k\varphi_kT\varphi_k=T+\frac12\sum_k[[\varphi_k,T],\varphi_k]$; the electronic term gives $\frac c2\sum|\nabla_x\varphi_k|^2$, the field term $\frac12\sum\Gamma_{\varphi_k}$ (multiplication operators in the configuration commute with $eA(x)$). The pointwise bound follows from the chain rule (Setting~\ref{set:full}) and $|\nabla(\chi\eta)|^2\le2|\nabla\chi|^2\eta^2+2\chi^2|\nabla\eta|^2$; the derivatives of $\eta_0,\eta_1$ are supported in $\{\rrel\le|X_{\mathrm{rel}}|\le2\rrel\}$. The support statements are immediate.
\end{proof}

The kinetic loss (the terms with $c$) is small where the covariance defect is small, and is large only on the shell $\{\rrel\le|X_{\mathrm{rel}}|\le2\rrel\}$, which will be a rare event.

\subsubsection{Selection by functions of the field alone}
\begin{lemma}[Softmax partition in the field]\label{lem:softmax}
Let $g_1,\dots,g_n$ be $C^1$ functions of the configuration only, $\beta_{\mathrm{sel}}>0$, $p_k:=\mathrm e^{-\beta_{\mathrm{sel}} g_k}/\sum_je^{-\beta_{\mathrm{sel}} g_j}$ and $\zeta_k:=\sqrt{p_k}$. Then $H_{\mathrm{full}}=\sum_k\zeta_kH_{\mathrm{full}}\zeta_k-\frac12\sum_k\Gamma_{\zeta_k}$ in the sense of forms (no electronic term appears), and pointwise
\[
\sum_k\Gamma_{\zeta_k}\le\frac{\beta_{\mathrm{sel}}^2}4\sum_kp_k\Gamma_{g_k},\qquad\sum_{k:\,g_k>\min_jg_j+\delta}p_k\le n\,\mathrm e^{-\beta_{\mathrm{sel}}\delta}.
\]
\end{lemma}
\begin{proof}
The $\zeta_k$ do not depend on $x$ and commute with the electronic kinetic term; the field term gives the first identity. Since $\nabla_q\zeta_k=-\frac{\beta_{\mathrm{sel}}}2\zeta_k(\nabla g_k-\sum_jp_j\nabla g_j)$, we get $\sum_k|\nabla\zeta_k|^2=\frac{\beta_{\mathrm{sel}}^2}4\sum_kp_k|\nabla g_k-\overline{\nabla g}|^2\le\frac{\beta_{\mathrm{sel}}^2}4\sum_kp_k|\nabla g_k|^2$. The last bound follows from $p_k\le \mathrm e^{-\beta_{\mathrm{sel}}(g_k-\min g)}$.
\end{proof}

\subsubsection{Completing the square at the seam}
Let $V_T=\Win\oplus\Wext$ be an orthogonal decomposition of the real transverse fields with $\Wext$ finite-dimensional, with coordinates $q=(q_{\mathrm i},q_{\mathrm e})$, and write
\[
\Hf=X_{\mathrm{in}}+\Hee+\Hie,\qquad\Hee=\tfrac12\pi_{\mathrm e}^2+\tfrac12q_{\mathrm e}\cdot\omega_{\mathrm{ee}}^2q_{\mathrm e},\qquad\Hie=q_{\mathrm i}\cdot\Kie q_{\mathrm e},
\]
with $\omega^2$ the squared dispersion, $\omega_{\mathrm{ee}}^2:=P_{\mathrm{ext}}\omega^2P_{\mathrm{ext}}|_{\Wext}$ (positive definite, since $\langle v,\omega^2v\rangle=\int|k|^2|\hat v|^2\dd k>0$ for $v\ne0$), $\Kie=P_{\mathrm{in}}\omega^2P_{\mathrm{ext}}$, and $X_{\mathrm{in}}$ the part acting on $\Win$. Let $E_0:=E_0(\Hee)=\frac12\tr\omega_{\mathrm{ee}}$. In the representation $\F\cong L^2(\R^{\dim\Wext},\dd q_{\mathrm e})\otimes\mathcal K$ of Lemma~\ref{lem:Xin}, $q_{\mathrm e},\pi_{\mathrm e}$ act in the first factor, and the components of $\Kie^{\mathsf T}q_{\mathrm i}$ are commuting field operators along $\Win$ acting in $\mathcal K$.

\begin{lemma}[Seam]\label{lem:seam}
For every $\delta>0$, in the sense of forms,
\[
\Hie\ge-\delta(\Hee-E_0)-\delta^{-1}T_\Sigma,\qquad T_\Sigma:=\tfrac12\,q_{\mathrm i}\cdot\Kie\omega_{\mathrm{ee}}^{-2}\Kie^{\mathsf T}q_{\mathrm i}.
\]
\end{lemma}
\begin{proof}
We argue on the form core $\mathcal D$ of Lemma~\ref{lem:Xin}; the inequality extends to $Q(\Hf)$ by density, as stated at the end of that proof. Since $\Kie^{\mathsf T}q_{\mathrm i}$ commutes with $\Hee$, it suffices to argue at a fixed value of $\Kie^{\mathsf T}q_{\mathrm i}$ in its joint spectral representation. With $y:=\delta^{-1}\omega_{\mathrm{ee}}^{-2}\Kie^{\mathsf T}q_{\mathrm i}$,
\[
\delta(\Hee-E_0)+q_{\mathrm i}\cdot\Kie q_{\mathrm e}=\delta\Bigl[\tfrac12\pi_{\mathrm e}^2+\tfrac12(q_{\mathrm e}+y)\cdot\omega_{\mathrm{ee}}^2(q_{\mathrm e}+y)-E_0\Bigr]-\delta^{-1}T_\Sigma ,
\]
and the bracket is a shifted harmonic oscillator with ground state energy $E_0$.
\end{proof}
The form $T_\Sigma$ only sees field gradients across the boundary (the seam) between the inner and outer regions; its vacuum expectation is bounded in Lemma~\ref{lem:trace}.

\section{Estimators and good sets}\label{sec:estimators}

The estimator required by Lemma~\ref{lem:consumer} reads the position of the bubble around the electron off the field configuration. We first define local field levels and depression indicators, then the core locator $X_c$ of the deep core of the bubble together with the properties we quote for it (Propositions~\ref{prop:Mprime}--\ref{prop:global}), and finally our shallow-level estimator $\Xw$, centred at $X_c$, and the design parameters. We use functions of the field configuration rather than of the total momentum as trial functions because the boundary of the bubble is a free boundary, whose position can be read off the field configuration; heuristically, trial functions of the momentum do not capture the inertia of its translations.

\subsection{Local levels and depressions}
Divide $\R^3$ into blocks $Q$ of side
\[
s:=\lfloor\log\ell\rfloor ,
\]
a block side admissible in \cite[\SerAnum{prop:configBO}]{SerA}, which holds for all integers $s\ge s_0$. For a block $Q$ let $P'_Q$ be the orthogonal projection in $L^2(Q;\R^3)$ onto the orthogonal complement of $\nabla H^1(Q)$ (as in \cite[\SerAnum{sec:finite-model}]{SerA}; it acts locally in $Q$), and let $u'_Q:=\|P'_Q1_QA\|^2$ be the squared norm of the projected field restricted to $Q$, all components, and let $\sat_\tau(u):=\tau(1-\mathrm e^{-u/\tau})$ be a saturation. With the normalized indicator $K'_b$ of the cube $[-(b+\frac s2),b+\frac s2]^3$ define
\[
g^{(\tau)}_b(Y):=\int K'_b(Y-z)\,\sat_\tau\bigl(u'_{Q(z)}/|Q|\bigr)\dd z,
\]
where $Q(z)$ is the block centred at $z$; equivalently, we average over all translates of the block lattice. The function $g_b$ is the local level of the field near $Y$: it is small where the electron density stiffens the field and of order one in the vacuum. It is covariant, $g_b(Y;\transl_aA)=g_b(Y-a;A)$.

For a threshold $g>0$ let $\tilde\chi_g$ be smooth, equal to $1$ on $[0,\frac34g]$, to $0$ on $[g,\infty)$, vanishing quadratically at the upper end, $|\tilde\chi_g'|^2\le C\tilde\chi_g/g^2$. The \emph{depression indicator at level $g$} is $\tilde\chi_g(g_b(Y))$. We use two levels:
\begin{itemize}
\item the \emph{core level}: threshold $g_s\asymp\ell^{-2}$ (the level of the core of the mean-field bubble), saturation $\tau=\rho_*g_s$, cube side $b_\theta\asymp\ell$ (fixed in (D$'$), Section~\ref{sec:constructive}); the indicator is denoted $d(Y)$;
\item the \emph{level $\kappa$}, $0<\kappa<1$: threshold $g_w:=\ell^{2-\kappa}g_s\asymp\ell^{-\kappa}$, saturation $\tau_w:=c_\tau g_w$, cube side $b:=C_bs$; the indicator is denoted $d_w(Y)$.
\end{itemize}
Core-level depressions form the deep core of the bubble, of volume $\asymp\ell^3$; level-$\kappa$ depressions are much shallower and extend to the periphery of the bubble.

\emph{Lifted denominators.} Our estimators are centres of mass $N/W$ of a weight with total mass $W\ge0$. To keep them bounded we replace the denominator by $r_a(W)$, where for $a>0$
\[
r_a(t):=a\,r(t/a),
\]
where $r\in C^\infty([0,\infty))$ is convex with $r(u)=u$ for $u\ge2$, $r(u)\ge\max(1,u)$ and $0\le r'\le1$ (for instance $r(u):=1+\int_0^u\vartheta$ with $\vartheta$ smooth and non-decreasing, $\vartheta=0$ on $[0,\frac12]$, $\vartheta=1$ on $[\frac32,\infty)$ and $\vartheta(1+t)+\vartheta(1-t)=1$, so that $\int_0^2\vartheta=1$ and $r(u)-u=1-\int_0^u(1-\vartheta)\ge0$). Thus $r_a(W)=W$ when $W\ge2a$, $r_a\ge a$, and $|r_a''|\le C/a$. The threshold $a$ is always chosen at most the mass threshold of the corresponding good set, so the lifting acts only on configurations that are rare anyway; on the good set the estimators are exact centres of mass, and exactly covariant away from the window edges (Proposition~\ref{prop:Mprime}(b)).

\subsection{The core locator}
\begin{definition}[Core locator]\label{def:core}
Let $\mathcal M(Y):=|B_{R_c}|^{-1}\int_{B(Y,R_c)}d$ be the local depression mass, $\zeta$ smooth with $\zeta=0$ on $[0,\Mzero]$ and $\zeta=1$ on $[2\Mzero,\infty)$, $\phi(\mathcal M):=\mathrm e^{\mathcal M}\zeta(\mathcal M)$, and $\chi_W(|Y|/r_W)$ a $C^2$ window of radius $r_W$ around the electron (the origin of the fiber coordinates), equal to $1$ on $|Y|\le r_W-w$, with edge width $w:=R_c$ and $\|\nabla\chi_W\|_\infty\lesssim1/w$, $\|\nabla^2\chi_W\|_\infty\lesssim1/w^2$ (a fixed profile rescaled to the edge shell). Set
\[
W_{\mathrm{ms}}:=\int\chi_W\,\phi(\mathcal M(Y))\dd Y,\qquad X_c:=\frac{\int Y\,\chi_W\,\phi(\mathcal M(Y))\dd Y}{r_{a_c}(W_{\mathrm{ms}})},\qquad a_c:=W_*,
\]
with $\Mzero$, $W_*$, $R_c$ as in Section~\ref{sec:constructive} and $r_W$ as in Proposition~\ref{prop:events}, converted to physical units: Section~\ref{sec:constructive} measures lengths in units of $R_*\asymp\ell$ and writes the physical core radius as $R_cR_*$, while here and in Appendix~\ref{app:core} $R_c\asymp\ell$ and $W_*\asymp\ell^3$ denote the physical quantities.
\end{definition}
$X_c$ is a soft mode of the depression mass: the weight $\phi(\mathcal M)$ vanishes unless a ball of radius $R_c$ around the point contains a substantial core-level depression mass, and it favours the points where this mass is largest.

\begin{definition}[Core good set]\label{def:good}
Let $S_M:=\{W_{\mathrm{ms}}<2W_*\}\cup\{M<2M^w_*\}$ ($W_*,M^w_*$ design constants, $M$ as in Definition~\ref{def:Xw}) and let $\mathcal B_\theta$ be the event of Proposition~\ref{prop:events}. The core good set is $\mathcal G_\theta:=(S_M\cup\mathcal B_\theta)^c$.
\end{definition}

\subsection{Properties of the core locator}
The following statements are derived from the mean-field bound of \cite{SerA} and the tools of Section~\ref{sec:prelim}; they are proved in Appendix~\ref{app:core}, except Proposition~\ref{prop:SM}. The order in which constants are chosen is summarized in Remark~\ref{rem:order}. Their roles in the two conditions \eqref{eq:two} are as follows.
\begin{itemize}
\item The first condition $c\,\delta\lesssim1$ comes from Proposition~\ref{prop:Mprime}(b) on the good set, from Propositions~\ref{prop:SM} and~\ref{prop:events} (the bad sets are rare to every power) and from Proposition~\ref{prop:global} (polynomial bounds on the bad sets).
\item The second condition $\langle\Gamma\rangle\lesssim1$ is not reached by $X_c$: Proposition~\ref{prop:Mprime}(a) gives only $\Gamma_{X_c}\lesssim\ell(\log\lambda)^5$ (Remark~\ref{rem:core-estimator}). It is proved for the level-$\kappa$ estimator in Section~\ref{sec:proof}.
\item Proposition~\ref{prop:MF} is the input from \cite{SerA} used to prove Proposition~\ref{prop:SM}.
\end{itemize}

The proofs in Appendix~\ref{app:core} and Section~\ref{sec:constructive} use only the tools of Sections~\ref{sec:prelim} and~\ref{sec:surgery} and the deterministic Lemma~\ref{lem:Mw}; none of them uses Proposition~\ref{prop:shell} or Proposition~\ref{prop:expect}, so there is no circularity (the order of the constants is Remark~\ref{rem:order}).

\begin{proposition}\label{prop:Mprime}
(a) For all configurations, $\Gamma_{X_c}\lesssim\ell(\log\lambda)^5$.
(b) On $\mathcal G_\theta$, $\Rm(X_c)=0$.
\end{proposition}

\begin{remark}\label{rem:Mw}
On $\mathcal G_\theta$ one has $M\ge2M^w_*$ with $M^w_*\asymp\ell^3/b^3$, simply by the definition of $S_M$; it is Proposition~\ref{prop:SM} that makes this a property of typical configurations.
\end{remark}

\begin{proposition}\label{prop:SM}
For every $\kappa_p>0$ there is $K_0$ such that for all window constants of Proposition~\ref{prop:events} at least $K_0$ (that is, $K\ge K_0$ and $K_d\ge K+1$, as in (D5) of Definition~\ref{def:Xw}) there is $\lambda_1$ with $P_\psi(S_M)\le\ell^{-\kappa_p}$ for $\lambda\ge\lambda_1$. (Proved in Section~\ref{sec:constructive} as Proposition~\ref{prop:SMexpl}.)
\end{proposition}

\begin{proposition}\label{prop:events}
Let $r_{\mathrm{near}}=r^{\mathrm{eff}}_{\mathrm{in}}+K\ell\log\lambda\asymp\ell\log\lambda$ be the radius of Appendix~\ref{app:core}, with the window constant $K$, let the window of the core locator satisfy the design inequality
\[
r_W-w-R_c\ge r_{\mathrm{near}}
\]
(we take $r_W:=r_{\mathrm{near}}+w+R_c$), and let $\rrel=K_d\ell\log\lambda>r_{\mathrm{near}}+R_c$. Let $\mathcal B_\theta$ be the event that a core-level depression, a point $Y$ with $g^{(\tau)}_{b_\theta}(Y)\le g_s$, lies at distance between $r_{\mathrm{near}}$ and $r_W+R_c$ from the electron.
\begin{itemize}
\item[(a)] Outside $\mathcal B_\theta$: $|X_{\mathrm{rel}}|\le r_{\mathrm{near}}+R_c<\rrel$, where $X_{\mathrm{rel}}$ is $X_c$ read as relative position (in particular the shell $\rrel\le|X_{\mathrm{rel}}|\le2\rrel$ of Lemma~\ref{lem:ims} lies in $\mathcal B_\theta$); and the event $\Gamma'_R$ that a point $Y$ with $d(Y)>0$ lies at distance between $r_W-w-R_c$ and $r_W+R_c$ from the electron does not occur. Outside $\Gamma'_R\cup S_M$ the covariance defect $\Rm(X_c)$ vanishes.
\item[(b)] $P_\psi(\mathcal B_\theta)\le\mathrm{poly}(\ell)\lambda^{-c_{\mathrm{ev}}K}$, with a degree independent of $K$. Hence for every $\kappa_p$ there is $K_0$ such that for $K\ge K_0$ there is $\lambda_1$ with $P_\psi(\mathcal B_\theta)\le\ell^{-\kappa_p}$ for $\lambda\ge\lambda_1$.
\end{itemize}
\end{proposition}

\begin{proposition}\label{prop:MF}
Let $J_q(\psi)$ be the functional of the electronic wave function of the configuration-wise block Born--Oppenheimer bound \cite[\SerAnum{prop:configBO}]{SerA} at a fixed configuration $q$ (with the convention there for $u'_Q(q)=0$: $h_Q^2/(8u'_Q):=+\infty$ if $h_Q>0$ and $:=0$ if $h_Q=0$, so that $J_q(\psi)\in(-\infty,+\infty]$; the deficits $\Delta_Q$ below are read with the same convention). Then: (a) the Born--Oppenheimer operator inequality $H'_{\mathrm{BO}}\ge\mathcal U(q):=\inf_\psi J_q(\psi)$ holds (multiplication in $q$); (b) the average $\bar J_q(\psi)$ over translates of the block lattice, with the same $\psi$ and the infimum over $\psi$ taken last, gives the fiber bound \eqref{eq:630P} for local functionals $\Xi_0$ of the field around the electron (Appendix~\ref{app:core}); (c) with $\varphi$ the square root of the electron density rescaled by the bubble radius $R_*\asymp\ell$, there are $\varepsilon_1(\ell),\Eloss(\ell)$ with $\varepsilon_1+\Eloss/F\to0$ as $\ell\to\infty$ such that for each lattice translate $t$
\[
(1-\varepsilon_1)\,F\,\frac{\mathcal E_1[\varphi]}{F[1]}+\sum_Q\Delta_Q-\Eloss\le J^{(t)}_q(\psi).
\]
Here $\mathcal E_1[\varphi]=\frac12\|\nabla\varphi\|^2+\|\varphi\|_1$, $F[1]:=\min\{\mathcal E_1[\varphi]:\|\varphi\|_2=1\}=4.9646158\ldots$, $F\asymp\abar\ell^5$ is the mean-field energy of \cite{SerA}, and $\Delta_Q\ge0$ is the deficit of the block $Q$: the arithmetic--geometric deficit of Lemma~\ref{lem:AMGM} if $u'_Q\le u_b$ or $\bar\rho_Q=0$ (for $\bar\rho_Q=0$ it is $h_Q^2/(8u'_Q)$), and the deficit $\Delta^{\mathrm{hi}}_Q$ of Appendix~\ref{app:core} if $u'_Q>u_b$ and $\bar\rho_Q>0$, where $u_b$ is the field ceiling of \cite[\SerAnum{prop:configBO}]{SerA}; in all cases $\Delta_Q$ satisfies the lower bounds (a) and (b) of Lemma~\ref{lem:AMGM} (with $h=h_Q$, $u=u'_Q$, $x=\gamma\bar\rho_Q$, and for (b) when $\eta\le3+2\sqrt3$), as shown in Step~c2 of Appendix~\ref{app:core}; $\Eloss$ contains a bound, independent of $q$, on the ceiling loss $\sum_Qh_Q^2/(2u_b)$, summed over the blocks with $u'_Q>u_b$ that meet the electron box (blocks without electron density need no ceiling loss).
\end{proposition}

\begin{proposition}\label{prop:global}
For all configurations, $\Gamma_{X_c}\le K_G\ell^D$ and $\|\Rm(X_c)\|^2\le K_G\ell^D$, with $D=2$ for the edge width $w=R_c$ of Definition~\ref{def:core} (Remark~\ref{rem:budget}); $D$ depends only on the design, while $K_G$ may depend also on the window constants (the powers of $\log\lambda$ in these bounds are absorbed by one power of $\ell$, Remark~\ref{rem:budget}). The same bounds hold for the level-$\kappa$ estimator of Definition~\ref{def:Xw}.
\end{proposition}

\begin{remark}[The core locator as a trial function]\label{rem:core-estimator}
Inserted directly into Lemma~\ref{lem:consumer}, the core locator gives only $h\gtrsim\lambda^{6/7}$ up to logarithms, since our bound on its fluctuation is $\langle\Gamma_{X_c}\rangle\lesssim\ell(\log\lambda)^5$ (we do not prove a matching lower bound). Our bound is limited by the sensitivity of $X_c$ to the steep wall of the core, which lies inside every window; the level-$\kappa$ estimator below avoids this sensitivity.
\end{remark}

\subsection{The level-$\kappa$ estimator and the design}
\begin{definition}[Level-$\kappa$ estimator and design]\label{def:Xw}
Fix $0<\kappa<1$. For $\rho>0$ let $\chi_\rho$ be radial and of class $C^2$, equal to $1$ on $[0,\rho/2]$, decreasing on $[\rho/2,\rho]$ with $|\chi_\rho'|\le c_\rho/\rho$, and $0$ on $[\rho,\infty)$. Set
\begin{gather*}
M:=\int\chi_\rho(|Y-X_c|)\,d_w(Y)\frac{\dd Y}{b^3},\qquad a_w:=M^w_*,\\
\Xw:=X_c+\frac1{r_{a_w}(M)}\int\chi_\rho(|Y-X_c|)\,(Y-X_c)\,d_w(Y)\frac{\dd Y}{b^3},
\end{gather*}
the centre of the level-$\kappa$ depressions in the window of radius $\rho$ around $X_c$. The \emph{shell event} $\Gamma'_\rho$ is the event that a level-$\kappa$ depression lies in the shell where $\chi_\rho$ varies: there is $Y$ with $\rho/2\le|Y-X_c|\le\rho$ and $g^{(\tau_w)}_b(Y)<g_w$. The design is:
\begin{itemize}
\item[(D1)] cube side $b=C_bs$ with $s=\lfloor\log\ell\rfloor$;
\item[(D2)] the cell side $n_b$ of Section~\ref{sec:shell} is an integer multiple of the candidate spacing $\rho_\delta:=\theta_\delta g_wb/3$;
\item[(D3)] candidate lattice points in Section~\ref{sec:shell} are determined by the localization cell of the electron only, not by the electron position nor by the estimator;
\item[(D4)] $c_\tau>(1+3\theta_\delta)/(1-\mathrm e^{-1})$;
\item[(D5)] $\rrel:=K_d\,\ell\log\lambda$ with $K_d\ge K+1$ for the window constant $K$ of Proposition~\ref{prop:events} (then $\rrel\ge r_{\mathrm{near}}-r^{\mathrm{eff}}_{\mathrm{in}}+\ell\log\lambda>r_{\mathrm{near}}+R_c$ for large $\lambda$, since $r^{\mathrm{eff}}_{\mathrm{in}}+R_c<\ell\log\lambda$ eventually, so the hypothesis on $\rrel$ in Proposition~\ref{prop:events} holds; throughout, ``window constants at least $K_0$'' means $K\ge K_0$ and $K_d\ge K+1$), $n_b:=\rho_\delta\lceil\rrel/\rho_\delta\rceil$ (so that (D2) holds and $\rrel\le n_b\le\rrel+\rho_\delta$), and
\[
\rho:=4\Bigl(\DEl+\tfrac{\sqrt3}2(2b+2s+2D_\eta+\rho_\delta)+\diam\supp\chi_j+2\rrel\Bigr)\asymp\ell\log\lambda,
\]
where $\DEl:=\ell^{1/2}$ is the surgery distance of Corollary~\ref{cor:far}, $D_\eta$ the rim width of Lemma~\ref{lem:hardy} (a constant), and $\chi_j$ the cell functions of Lemma~\ref{lem:ims} ($\diam\supp\chi_j\asymp n_b$).
\end{itemize}
\end{definition}

The shell event $\Gamma'_\rho$ is essential: off it the fluctuation of the window centre $X_c$ does not propagate to $\Xw$ at all (Lemma~\ref{lem:Mw}), and Proposition~\ref{prop:shell} shows that it is rare to every power. Condition (D1) lets the cube grow logarithmically, which makes distant shallow depressions rare to every power (Section~\ref{sec:discussion}). Condition (D3) prevents the sharpness of the selection of candidates from multiplying the sensitivity of the position of the electron or of $X_c$ (Lemma~\ref{lem:softmax} and Step~2 of the proof of Proposition~\ref{prop:shell}). The remaining conditions are bookkeeping for the proof of Proposition~\ref{prop:shell}: (D2) makes the average over the translations of the cells compatible with the candidate lattice (Step~4); (D4) makes the level margin $q_g$ of Lemma~\ref{lem:level} positive up to the level $g_c=(1+3\theta_\delta)g_w$; and (D5) chooses the window radius so that a depression at distance $\rho/4$ from $X_c$ is at distance at least $\DEl$ from every electron box of the localization (Step~5), where the surgery of Corollary~\ref{cor:far} costs only super-polynomially small terms.

\begin{remark}[Covariance]\label{rem:cov}
Write $\Xw=X_c+\mathsf Y(X_c)$, where $\mathsf Y(x)$ depends on the field only through the covariant indicator $d_w$. Under a translation of the field by $a$, one checks $\mathsf Y_{\transl_aA}(x)=\mathsf Y_A(x-a)$, so $\Xw(\transl_aA)-a=y+\mathsf Y_A(y)$ with $y=X_c(\transl_aA)-a$. Differentiating at $a=0$ gives $\Rm(\Xw)=(I+\nabla_x\mathsf Y)\Rm(X_c)$. In particular $\Rm(\Xw)=0$ on $\mathcal G_\theta$ by Proposition~\ref{prop:Mprime}(b).
\end{remark}

\section{Hardy margin and the external field}\label{sec:surgery}

In Section~\ref{sec:shell} we show that configurations with a level-$\kappa$ depression in the shell $[\rho/2,\rho]$ cost an energy margin $\mu_w\asymp b^3\ell^\kappa$. A depression is a region of low local field level; if the field near it is split off into a finite-dimensional ``external'' subspace, the kinetic energy of the external field provides a margin through a Hardy-type inequality (Section~\ref{ssec:hardy}). The difficulty is that the external field is also coupled to the electron, with the large coefficient $c$. We take the external subspace inside the divergence-free fields supported near the depression (the Leray subspace). Then the external vector potential at the position of the electron is exactly a super-polynomially small tail of the band function, and the coupling costs nothing beyond the distance $\ell^{1/2}$ (Theorem~\ref{thm:gauge} and Corollary~\ref{cor:far}).

\subsection{Box model and external subspace}
Confine the electron to a bounded box $\mathcal C_{\mathrm{el}}$ with Dirichlet boundary conditions and keep the transverse field on $\R^3$ with band $|k|\le\pi$:
\[
\Hbox:=\frac c2\bigl(p+eA(x)\bigr)^2+\Hf,\qquad\Ebox:=\inf\sigma(\Hbox).
\]
Extending Dirichlet states by zero gives $E\le\Ebox$.

\begin{setting}[External subspace]\label{set:box}
Let $S$ be a radial $C_c^\infty$ function with $0\le S\le1$, $S=1$ near $0$ and $\supp S\subset\{|k|\le\pi\}$. Let $\Pi$ be the transverse projection restricted to the band, and $\mathcal E\subset\R^3$ a bounded region. Let
\[
L^2_\sigma(\mathcal E):=\bigl\{g\in L^2(\R^3;\mathbb C^3):\ g=0\text{ a.e.\ outside }\mathcal E,\ \mathrm{div}\,g=0\text{ in }\mathcal D'(\R^3)\bigr\}
\]
be the Leray subspace of $\mathcal E$ (divergence-free in $\mathcal E$ with vanishing normal component on $\partial\mathcal E$), $P_\sigma$ the orthogonal projection onto it, and
\[
X:=SP_\sigma S,\qquad\Wext:=1_{[t_0,1]}(X)V_T,\qquad\Win:=\Wext^\perp,
\]
with $t_0\in(0,1)$ and $V_T$ the real transverse fields. With $\Pi_\infty$ the transverse projection without band, $L^2_\sigma(\mathcal E)$ is the intersection of the ranges of $\Pi_\infty$ and of $1_{\mathcal E}$, so $P_\sigma=\Pi_\infty\wedge1_{\mathcal E}\le\Pi_\infty1_{\mathcal E}\Pi_\infty$ (for projections $P\wedge Q\le PQP$), and since $S\Pi_\infty=S\Pi$ we get $0\le X\le S\Pi1_{\mathcal E}\Pi S\le1$; $X$ is real, maps $V_T$ into itself, and is covariant under translations of $\mathcal E$. $\Wext$ is finite-dimensional (Lemma~\ref{lem:trace}). Let $v_m$ be a real orthonormal eigenbasis of $X$ on $\Wext$ ($Xv_m=\mu_mv_m$, $\mu_m\ge t_0$), with coordinates $q_m$ and conjugate momenta $\pi_m$; write $A(x)=A_{\mathrm{in}}(x)+A_{\mathrm e}(x)$, $A_{\mathrm e}(x)=\sum_mv_m(x)q_m$. The field energy is normal ordered, $\Hf\ge0$, and we decompose it as $\Hf=X_{\mathrm{in}}+\Hee+\Hie$ with $\Hee$, $\Hie$ as in Lemma~\ref{lem:seam} (no zero-point subtraction in $\Hee$) and $X_{\mathrm{in}}$ the remainder; $E_0:=E_0(\Hee)$. We use
\[
\Sch(w):=(2\pi)^{-3}\int S(k)\mathrm e^{ik\cdot w}\dd k,\qquad\mathcal N_S(w):=(2\pi)^{-3}\int\frac{S(k)}{|k|^2}\mathrm e^{ik\cdot w}\dd k .
\]
$\Sch$ is a Schwartz function, and $\mathcal N_S=\frac1{4\pi|w|}+\varrho$ with $\varrho$ and all its derivatives decaying super-polynomially (Newton's theorem and smoothness of $S$); the same holds for every power $S^m$.
\end{setting}
The smoothness of $S$ is a choice made in the proof: the cutoff of the model remains sharp, $S$ only serves to define $\Wext$, and the modes near the edge of the band stay in $\Win$.

\begin{lemma}\label{lem:Xin}
The canonical pairs $(q_m,\pi_m)$, $v_m\in\Wext$, commute with all field coordinates and momenta along $\Win$, and the Fock space is unitarily equivalent to $L^2(\R^{\dim\Wext},\dd q_{\mathrm e})\otimes\mathcal K$, with $q_m,\pi_m$ acting in the first factor in the Schr\"odinger representation and the fields along $\Win$ in $\mathcal K$. In this representation $X_{\mathrm{in}}=1\otimes\tilde X_{\mathrm{in}}$ acts in $\mathcal K$ only, in the following sense: there is a dense subspace $\mathcal D_{\mathcal K}\subset\mathcal K$ such that $\mathcal D:=\mathcal S(\R^{\dim\Wext})\otimes_{\mathrm{alg}}\mathcal D_{\mathcal K}$ is a form core of $\Hf$ and $X_{\mathrm{in}}(e\otimes\xi,e'\otimes\xi')=\langle e,e'\rangle\,\tilde X_{\mathrm{in}}(\xi,\xi')$ for a quadratic form $\tilde X_{\mathrm{in}}$ on $\mathcal D_{\mathcal K}$. Moreover $X_{\mathrm{in}}\ge-E_0$ as forms on $Q(\Hf)$.
\end{lemma}
\begin{proof}
Since $[q\cdot u,\pi\cdot w]=i\langle u,w\rangle$ and $\Win\perp\Wext$, the pairs $(q_m,\pi_m)$ commute with the fields along $\Win$, and the unitary equivalence is von Neumann's uniqueness theorem for finitely many canonical pairs. We make the representation explicit. By the end of the proof, $q_m=\Phi(g_m)$ and $\pi_m=\Phi(g'_m)$ with $g_m,g'_m\in D(\omega)\cap D(\omega^{-1/2})$. Let $\h_1:=\mathrm{span}_{\C}\{g_m,g'_m\}$, of complex dimension $M'\le2M$, $M:=\dim\Wext$, and $\h_0:=\h\ominus\h_1$. By the exponential law $\F=\F(\h_1)\otimes\F(\h_0)$, and $\F(\h_1)\cong L^2(\R^{M'})$ in the Schr\"odinger representation of $M'$ modes, with $\mathcal S(\R^{M'})$ the $C^\infty$ vectors of the number operator. In the real symplectic space $(\h_1,\mathrm{Im}\langle\cdot,\cdot\rangle)$ the real span $\mathfrak s$ of the $g_m,g'_m$ is non-degenerate; a linear symplectic map of $\h_1$ onto $\R^{2M}\oplus\R^{2(M'-M)}$ taking $\mathfrak s$ onto the first summand is implemented by a metaplectic unitary $L^2(\R^{M'})\to L^2(\R^M)\otimes L^2(\R^{M'-M})$ that maps $\mathcal S$ onto $\mathcal S$. Put $\mathcal K:=L^2(\R^{M'-M})\otimes\F(\h_0)$. As $\h_0\perp\h_1\supset\mathfrak s$, $\h_0$ is symplectically orthogonal to $\mathfrak s$, so the symplectic complement of $\mathfrak s$ is $\mathfrak s^\sigma=(\mathfrak s^\sigma\cap\h_1)\oplus\h_0$, and the symplectic map above takes $\mathfrak s^\sigma\cap\h_1$ onto the second summand. A field along $\Win$, i.e.\ $q(u)=\Phi(\omega^{-1/2}u/\sqrt2)$ or $\pi(u)=\Phi(i\omega^{1/2}u/\sqrt2)$ with $u\perp\Wext$ (including the band-limited distributions $u$ with $\omega^{-1/2}u\in\h$ that give the components of $A_{\mathrm{in}}(x)$), has its one-particle vector in $\mathfrak s^\sigma$, since $[q(u),\pi_m]=i\langle u,v_m\rangle=0$, $[\pi(u),q_m]=-i\langle u,v_m\rangle=0$, and fields of real vectors commute among themselves. Hence the fields along $\Win$, in particular $A_{\mathrm{in}}(x)$ and the components of $\Kie^{\mathsf T}q_{\mathrm i}$, act in $\mathcal K$. For a subspace $\mathfrak k\subset\h$ let $\F_{\mathrm{fin}}(\mathfrak k)$ denote the algebraic finite-particle vectors over $\mathfrak k$, i.e.\ the linear span of the vectors $a^*(f_1)\cdots a^*(f_j)\Omega$, $j\ge0$, $f_1,\dots,f_j\in\mathfrak k$ (no closure is taken in the $j$-particle sectors). Let $\mathcal D_{\mathcal K}:=\mathcal S(\R^{M'-M})\otimes_{\mathrm{alg}}\F_{\mathrm{fin}}(\h_0\cap D(\omega))$ and $\mathcal D:=\mathcal S(\R^M)\otimes_{\mathrm{alg}}\mathcal D_{\mathcal K}$. Since $\h_1\subset D(\omega)$, $D(\omega)=\h_1\oplus(\h_0\cap D(\omega))$, so $\F_{\mathrm{fin}}(D(\omega))$, a core of $\Hf$, lies in $\mathcal S(\R^{M'})\otimes_{\mathrm{alg}}\F_{\mathrm{fin}}(\h_0\cap D(\omega))$: write $a^*(f)=a^*(f_1)\otimes1+1\otimes a^*(f_0)$ for $f=f_1+f_0$, and note that polynomials in the creation operators of $\h_1$ applied to the vacuum are polynomials times a Gaussian in the Schr\"odinger representation. On this space $\Hf$ is a polynomial of degree two in the $M'$ canonical pairs, plus $\dG(P_0\omega P_0)$ in the second factor, plus the terms $a^*(P_0\omega e)a(e)+\mathrm{h.c.}$, $e\in\h_1$, so its form norm is continuous in the Schwartz topology of the first factor, in which $\mathcal S(\R^M)\otimes_{\mathrm{alg}}\mathcal S(\R^{M'-M})$ is dense. Hence $\mathcal D$ is a form core of $\Hf$, $q_m,\pi_m$ leave $\mathcal D$ invariant, and on $\mathcal D$ each of $\Hf$, $\Hee$, $\Hie$ is a polynomial of degree at most two in $q_{\mathrm e},\pi_{\mathrm e}$ whose coefficients are forms on $\mathcal D_{\mathcal K}$. On $\mathcal D$, $i[\Hf,q_m]=\pi_m=i[\Hee,q_m]$, $[\Hie,q_m]=0$ and $i[\Hf,\pi_m]=-\langle v_m,\omega^2q\rangle=i[\Hee+\Hie,\pi_m]$, so $X_{\mathrm{in}}=\Hf-\Hee-\Hie$ commutes with all $q_m,\pi_m$ as a form on $\mathcal D$. For fixed $\xi,\xi'\in\mathcal D_{\mathcal K}$, the form $(e,e')\mapsto X_{\mathrm{in}}(e\otimes\xi,e'\otimes\xi')$ on $\mathcal S(\R^M)$ is given by a differential operator of order at most two with polynomial coefficients that commutes with all $x_m$ and $\partial_m$, hence by a multiple $\tilde X_{\mathrm{in}}(\xi,\xi')$ of the identity. For $\xi\in\mathcal D_{\mathcal K}$ let $\Psi:=\Omega_{\mathrm{ee}}\otimes\xi\in\mathcal D$ ($\Omega_{\mathrm{ee}}$ is a Gaussian). Then $\langle\Psi,\Hee\Psi\rangle=E_0\|\xi\|^2$; $\langle\Psi,\Hie\Psi\rangle=0$, since $\Hie$ is linear in $q_{\mathrm e}$ and $\langle\Omega_{\mathrm{ee}},q_m\Omega_{\mathrm{ee}}\rangle=0$; and $\langle\Psi,\Hf\Psi\rangle\ge0$, since the field energy is normal ordered. Hence $\tilde X_{\mathrm{in}}[\xi]\ge-E_0\|\xi\|^2$. Every $\Psi\in\mathcal D$ is a finite sum $\sum_je_j\otimes\xi_j$ with orthonormal $e_j\in\mathcal S(\R^M)$, and $X_{\mathrm{in}}[\Psi]=\sum_j\tilde X_{\mathrm{in}}[\xi_j]\ge-E_0\|\Psi\|^2$. Since $X_{\mathrm{in}}$ is bounded with respect to $\Hf+1$ (see below) and $\mathcal D$ is a form core, $X_{\mathrm{in}}\ge-E_0$ on $Q(\Hf)$. The same density argument extends the inequality of Lemma~\ref{lem:seam}, whose proof is carried out on $\mathcal D$, to $Q(\Hf)$. The $v_m$ are band-limited, but this alone does not give $\int|\hat v_m|^2|k|^{-2}\dd k<\infty$. By Theorem~\ref{thm:gauge}(0) (Step~1 of its proof, which does not use this lemma), $v_m=Sf_m$ with $f_m=P_\sigma Sv_m/\mu_m$ supported in the bounded set $\mathcal E$ and $\|f_m\|=\mu_m^{-1/2}$, so $\|\hat v_m\|_\infty\le\|f_m\|_{L^1}\le\vol(\mathcal E)^{1/2}\mu_m^{-1/2}$ and $\int_{|k|\le\pi}|\hat v_m|^2|k|^{-2}\dd k\le4\pi^2t_0^{-1}\vol(\mathcal E)$. Hence $q_m$, $\pi_m$, $\Hee$ and $\Hie$ are form bounded with respect to $\Hf+1$, so $X_{\mathrm{in}}$ is a form on $Q(\Hf)$. No finite-mode approximation is used.
\end{proof}

\begin{lemma}[Point variance of the external field]\label{lem:607N}
Let $\kappa_S$ be the kernel of $S^2/|k|$. Then $|\kappa_S(w)|\le\min(\frac14,\frac{1+V_S}{2\pi^2|w|^2})$ with $V_S:=\int_0^\pi|(S^2)'|$; hence $N_{\mathcal E}:=\sup_{y\in\mathcal E}\int_{\mathcal E}|\kappa_S(y-y')|\dd y'\le\frac{2(1+V_S)}\pi\diam\mathcal E$, and $T_{\mathcal E}:=\omega^{-1/2}S\Pi1_{\mathcal E}$ satisfies $\|T_{\mathcal E}\|^2\le N_{\mathcal E}$.
\end{lemma}
The kernel bound is obtained by integration by parts in Fourier space and the norm bound by Schur's test; see Appendix~\ref{app:cited}.

\subsection{Hardy inequality for the external field and the margin}\label{ssec:hardy}
For a block $Q$ put $Y_{\mathrm e}:=P'_Q1_Q|_{\Wext}$ and $g'_{\mathrm e}:=\tr Y_{\mathrm e}^*Y_{\mathrm e}-2\|Y_{\mathrm e}^*Y_{\mathrm e}\|$.

\begin{lemma}[Hardy inequality and localized zero point]\label{lem:hardy}
(i) If $g'_{\mathrm e}>0$, then $1/u'_Q\le\frac4{g_{\mathrm e}'^2}\int_Q|E_{\mathrm e}|^2$ in the sense of forms, where $E_{\mathrm e}$ is the momentum (electric field) conjugate to the external field. (ii) For a block at distance $D_Q$ from the boundary of $\mathcal E$, $\tr Y_{\mathrm e}^*Y_{\mathrm e}\ge\rho_{\mathrm m}|Q|(\langle p(S^2)\rangle-C_pD_Q^{-3/2})-C_D|\partial Q|$, with the boundary constant $C_D$ of \cite[\SerAnum{cor:hardy}]{SerA} (part (a) there, proved on $\R^3$), where $\rho_{\mathrm m}=\pi/3$ is the mode density and $p$ is a polynomial with $p\le1_{[t_0,1]}$ on $[0,1]$. In particular, for a suitable constant rim width $D_\eta$ and $s$ large enough, all blocks of $\mathcal E$ at distance $\ge D_\eta$ from its boundary satisfy $g'_{\mathrm e}\ge\kappa_{\mathrm e}\rho_{\mathrm m}|Q|$ with a constant $\kappa_{\mathrm e}>0$. (iii) If $H=\frac12\pi\cdot\pi+\frac12q\cdot\Kosc q$ with $\Kosc>0$, $0\le M\le1$, $T_M:=\frac12\pi\cdot M\pi$ and $0\le\varepsilon_{\mathrm H}\le1$, then $H-E_0(H)\ge\varepsilon_{\mathrm H}(T_M-(1+\varepsilon_{\mathrm H})\langle T_M\rangle_0)$, where $\langle\cdot\rangle_0$ is the expectation in the ground state of $H$.
\end{lemma}
The proofs are given in Appendix~\ref{app:cited}. Part (i) is a Hardy inequality in the external coordinates, invariant under shifts by internal components; part (iii) localizes the kinetic energy in the ground state of a harmonic oscillator. In part (ii) the restriction to divergence-free fields creates a boundary layer at $\partial\mathcal E$; it decays like $D_Q^{-3}$, as does the boundary layer created by the cut $1_{\mathcal E}$ itself, and is paid by the same rim width.

\begin{corollary}[Hardy margin]\label{cor:margin}
Let $Q_j$ be the blocks of a cube in $\mathcal E$ outside the rim, with weights $a_j\ge0$, $a_{\max}:=\max a_j$, and $M:=\sum_j(a_j/a_{\max})P_{\Wext}1_{Q_j}P_{\Wext}$. Then
\[
\Hee-E_0\ge\mathcal V(A):=\varepsilon_{\mathrm H}\Bigl(\frac{\kappa_{\mathrm e}^2\rho_{\mathrm m}^2|Q|^2}8\sum_j\frac{a_j/a_{\max}}{u'_{Q_j}}-(1+\varepsilon_{\mathrm H})\langle T_M\rangle_0\Bigr),\qquad\langle T_M\rangle_0\le\frac{e_*}2\sum_j\frac{a_j}{a_{\max}}|Q|,
\]
where $e_*$ is the local density of the vacuum electric energy of the external field.
\end{corollary}
\begin{proof}
Apply Lemma~\ref{lem:hardy}(iii) to $\Hee$ and $M$, and bound $T_M=\frac12\sum_j(a_j/a_{\max})\int_{Q_j}|E_{\mathrm e}|^2$ from below by (i) and (ii): $\frac12\int_{Q_j}|E_{\mathrm e}|^2\ge g_{\mathrm e}'^2/(8u'_{Q_j})\ge\kappa_{\mathrm e}^2\rho_{\mathrm m}^2|Q|^2/(8u'_{Q_j})$. The vacuum expectation is a sum of local densities.
\end{proof}

The margin is a weighted sum of local convex costs $1/u'_{Q_j}$. Bounding it from below through the average level by Markov's inequality loses too much at shallow levels; we use Cauchy--Schwarz instead.

\begin{lemma}[Level margin]\label{lem:level}
(a) Let $\nu$ be a finite positive measure on a family of blocks, $\nu_{\mathrm{tot}}>0$ its total mass, $u_Q:=u'_Q/|Q|>0$, $\tau>0$, and let $g:=\nu_{\mathrm{tot}}^{-1}\int\sat_\tau(u_Q)\dd\nu(Q)$ be the level. If $g<(1-\mathrm e^{-1})\tau$, then with $q_g:=1-g/((1-\mathrm e^{-1})\tau)$
\[
\int\frac{\dd\nu(Q)}{u_Q}\ge(1-\mathrm e^{-1})\,q_g^2\,\frac{\nu_{\mathrm{tot}}}g .
\]
(b) In Corollary~\ref{cor:margin} let $a_\Sigma:=\sum_ja_j$, $\tau:=c_\tau g_w$, and assume that the level $g:=\sum_ja_j\sat_\tau(u_{Q_j})/a_\Sigma$ satisfies $g\le g_c:=(1+3\theta_\delta)g_w$. If $c_\tau>(1+3\theta_\delta)/(1-\mathrm e^{-1})$ (design (D4)), then $q_g>0$ and
\[
\mathcal V(A)\ge\mu_w(g):=\varepsilon_{\mathrm H}\frac{a_\Sigma}{a_{\max}}|Q|\Bigl(\frac{(1-\mathrm e^{-1})\kappa_{\mathrm e}^2\rho_{\mathrm m}^2q_g^2}{8g}-\frac{(1+\varepsilon_{\mathrm H})e_*}2\Bigr).
\]
The function $\mu_w$ is decreasing in $g$; for kernel weights with $a_\Sigma/a_{\max}\asymp b^3/|Q|$ one has $\mu_w(g_c)\asymp b^3/g_w\asymp b^3\ell^\kappa$. From now on $\mu_w$ without argument denotes this margin $\mu_w(g_c)$ at the threshold; the surgery constants of Section~\ref{sec:shell} ($\eta$ and the Agmon margin $\mAg$) and the conditions on $\DEl$ in Corollary~\ref{cor:far} are taken with it. For all configurations $\mathcal V(A)\ge-B_-$, $B_-:=\varepsilon_{\mathrm H}(1+\varepsilon_{\mathrm H})\frac{e_*}2\frac{a_\Sigma}{a_{\max}}|Q|$.
\end{lemma}
\begin{proof}
(a) On the set of large blocks $\mathcal B:=\{Q:u_Q\ge\tau\}$ we have $\sat_\tau(u_Q)\ge(1-\mathrm e^{-1})\tau$, so $\nu(\mathcal B)\le g\nu_{\mathrm{tot}}/((1-\mathrm e^{-1})\tau)=(1-q_g)\nu_{\mathrm{tot}}$. On the complement $\mathcal S$, concavity of $\sat_\tau$ gives $\sat_\tau(u)\ge(1-\mathrm e^{-1})u$, hence $\int_{\mathcal S}u_Q\dd\nu\le g\nu_{\mathrm{tot}}/(1-\mathrm e^{-1})$, and $\nu(\mathcal S)\ge q_g\nu_{\mathrm{tot}}$. By Cauchy--Schwarz
\[
\int_{\mathcal S}\frac{\dd\nu}{u_Q}\ge\frac{\nu(\mathcal S)^2}{\int_{\mathcal S}u_Q\dd\nu}\ge(1-\mathrm e^{-1})\frac{q_g^2\nu_{\mathrm{tot}}}{g}.
\]
(b) Apply (a) to $\nu:=\sum_ja_j\delta_{Q_j}$ and insert the result into Corollary~\ref{cor:margin} using $1/u'_{Q_j}=1/(|Q|u_{Q_j})$. Monotonicity holds since $q_g$ and $1/g$ decrease in $g$; $q_g>0$ for $g\le g_c$ is (D4). The last claim holds because the first term of $\mathcal V$ is non-negative.
\end{proof}
This is a deterministic inequality for each configuration; no probabilistic event is involved.

\begin{lemma}[Dimension and seam trace]\label{lem:trace}
In Setting~\ref{set:box}, $\dim\Wext\le\frac\pi{3t_0}\vol(\mathcal E)$, and $T_\Sigma$ of Lemma~\ref{lem:seam} satisfies
\[
\langle T_\Sigma\rangle_\vac\le\Bigl(\frac\pi2+\frac{\pi\diam\mathcal E}{t_0^2}\Bigr)\frac\pi{3t_0}\vol(\mathcal E).
\]
\end{lemma}
\begin{proof}[Proof of the first claim]
$\dim\Wext\le t_0^{-1}\tr X$ and $X\le S\Pi1_{\mathcal E}\Pi S$, so
\[
\tr X\le\tr(1_{\mathcal E}\Pi S^2\Pi1_{\mathcal E})=2\vol(\mathcal E)(2\pi)^{-3}\int S^2\dd k\le2\vol(\mathcal E)\frac{(4\pi/3)\pi^3}{8\pi^3}=\frac\pi3\vol(\mathcal E).
\] The bound on the vacuum trace (each eigenvector's $\omega^{-1}$-expectation is controlled by the kernel of Lemma~\ref{lem:607N}) is proved in Appendix~\ref{app:cited}.
\end{proof}

\subsection{Splitting off the external field}
The external field couples to the electron through $\frac c2(p+eA)^2$. Paying for this coupling with a fraction $\eta$ of the electron's kinetic energy by the Schwarz inequality leaves a term $\frac{c\abar}2\eta^{-1}|A_{\mathrm e}(x)|^2$. Since $\eta$ must be as small as the ratio of the margin to $\Ebox\asymp\ell^5$, the coefficient is of order $\ell^{12-\kappa}/b^3$. For an external subspace built from $1_{\mathcal E}\Pi$ instead of $P_\sigma$, the cut $1_{\mathcal E}$ creates a divergence at $\partial\mathcal E$, which the transverse projection returns as a non-local gradient field with a dipole tail; $|A_{\mathrm e}(x)|^2$ then decays only like $\dist(x,\mathcal E)^{-6}$, and the depression would have to be at a polynomially large distance (of order $\ell^{2-\kappa/3}b^{-1/3}$) from the electron, which is too large for the window. For the Leray subspace the source of the tail is absent.

With $f_m:=P_\sigma Sv_m/\mu_m\in L^2_\sigma(\mathcal E)$ define
\[
r_m:=\Sch*f_m,\qquad r_{\mathrm e}(x):=\sum_mr_m(x)q_m,\qquad\sigma_{\Sch}(x):=\int_{\mathcal E}|\Sch(x-z)|^2\dd z .
\]

\begin{theorem}[Splitting off the external field]\label{thm:gauge}
In Setting~\ref{set:box}:
\begin{itemize}
\item[(0)] $v_m=r_m$, i.e.\ $A_{\mathrm e}=r_{\mathrm e}$; the $r_m$ are real $C^\infty$ functions with bounded derivatives of all orders.
\item[(i)] (Surgery inequality) For all $\eta\in(0,1)$, as forms on $Q(\Hbox)$,
\[
\Hbox-\Ebox\ge(\Hee-E_0)+\Hie-\kappa_\eta|r_{\mathrm e}(x)|^2-\eta\Ebox-c^{(r)},
\]
with $\kappa_\eta:=\frac{c\abar}2(\eta^{-1}-1)$ and $c^{(r)}:=\frac{c\abar}2\sup_{x\in\mathcal C_{\mathrm{el}}}\langle\Omega_{\mathrm{ee}},|r_{\mathrm e}(x)|^2\Omega_{\mathrm{ee}}\rangle$, $\Omega_{\mathrm{ee}}$ the ground state of $\Hee$.
\item[(ii)] (Bounds) (a) $\sum_m|r_m(x)|^2\le\frac3{t_0}\sigma_{\Sch}(x)$. (b) Each component of $r_{\mathrm e}(x)$ is a $J$-real field $\Phi(h_l)$ with $h_l\in\Wext$, and its free-vacuum covariance satisfies $\tr\Gamma_r\le\frac{3N_{\mathcal E}}{2t_0^2}\sigma_{\Sch}(x)$. (c) $c^{(r)}\le\frac{3c\abar\diam\mathcal E}{4\sqrt2t_0^2}\sup_{x\in\mathcal C_{\mathrm{el}}}\sigma_{\Sch}(x)$.
\end{itemize}
\end{theorem}
The remaining coupling is a quadratic term proportional to the tail $\sigma_{\Sch}$, which is super-polynomially small when $\mathcal C_{\mathrm{el}}$ is far from $\mathcal E$; no gauge transformation and no constant term of order one appear.

\begin{proof}
\emph{Step 1: (0).} $\mu_mv_m=Xv_m=S(P_\sigma Sv_m)$, so $v_m=Sf_m$. Since $S$ acts as a Fourier multiplier, $v_m=Sf_m=\Sch*f_m=r_m$. As $\mathcal E$ is bounded, $f_m\in L^1\cap L^2$ and $\hat f_m$ is bounded; smoothness and boundedness of all derivatives follow from $S|k|^n|\hat f_m|\in L^1$, and reality from the reality of $\Sch$ and of $f_m$ ($X$ is real).

\emph{Step 2: (i).} By (0), $A=A_{\mathrm{in}}+r_{\mathrm e}$. Applying $(a+b)^2\ge(1-\eta)a^2-(\eta^{-1}-1)b^2$ componentwise to $a=p+eA_{\mathrm{in}}(x)$, $b=e\,r_{\mathrm e}(x)$ (it follows from $(\sqrt\eta a+b/\sqrt\eta)^2\ge0$ also for non-commuting $a,b$),
\[
\Hbox\ge Y_\eta+\Hee+\Hie-\kappa_\eta|r_{\mathrm e}(x)|^2,\qquad Y_\eta:=(1-\eta)\frac c2\bigl(p+eA_{\mathrm{in}}\bigr)^2+X_{\mathrm{in}} .
\]
In the representation of Lemma~\ref{lem:Xin}, $A_{\mathrm{in}}(x)$ is a field along $\Win$, so $Y_\eta$ acts in $L^2(\mathcal C_{\mathrm{el}})\otimes\mathcal K$ only in the sense of Lemma~\ref{lem:Xin}: on $\mathcal D_{\mathrm{box}}:=C_c^\infty(\mathcal C_{\mathrm{el}}^\circ)\otimes_{\mathrm{alg}}\mathcal D$, a form core of $\Hbox$ (for the fixed cutoff $Q(\Hbox)=Q(-\Delta_D\otimes1+1\otimes\Hf)$ with equivalent norms, and the algebraic tensor product of form cores of $-\Delta_D$ and of $\Hf$ is a form core of $-\Delta_D\otimes1+1\otimes\Hf$), $Y_\eta(e\otimes\varphi,e'\otimes\varphi')=\langle e,e'\rangle\tilde Y_\eta(\varphi,\varphi')$ for $e,e'\in\mathcal S(\R^{\dim\Wext})$ and $\varphi,\varphi'\in C_c^\infty(\mathcal C_{\mathrm{el}}^\circ)\otimes_{\mathrm{alg}}\mathcal D_{\mathcal K}$, by the argument of that proof. Let $e_\eta$ be the infimum of the form $Y_\eta$ over normalized vectors of $Q(\Hbox)$. Then $\Hbox-\Ebox\ge(\Hee-E_0)+\Hie-\kappa_\eta|r_{\mathrm e}|^2-c'_\eta$ with $c'_\eta:=\Ebox-E_0-e_\eta$. For the bound on $c'_\eta$ take the trial state $\Psi=\Omega_{\mathrm{ee}}\otimes\varphi$ with normalized $\varphi\in C_c^\infty(\mathcal C_{\mathrm{el}}^\circ)\otimes_{\mathrm{alg}}\mathcal D_{\mathcal K}$, so that $\Psi\in\mathcal D_{\mathrm{box}}$. The kinetic cross term and $\Hie$ are linear in $q_{\mathrm e}$ and have zero expectation, since $\langle\Omega_{\mathrm{ee}},q_m\Omega_{\mathrm{ee}}\rangle=0$, so
\[
\Ebox\le\langle\varphi,Y_0\varphi\rangle+E_0+c^{(r)} ,
\]
that is, $\tilde Y_0[\varphi]=Y_0[\Omega_{\mathrm{ee}}\otimes\varphi]\ge\Ebox-E_0-c^{(r)}$ for these $\varphi$. Writing $\Psi\in\mathcal D_{\mathrm{box}}$ as $\sum_je_j\otimes\varphi_j$ with orthonormal $e_j$ gives $Y_0[\Psi]=\sum_j\tilde Y_0[\varphi_j]\ge(\Ebox-E_0-c^{(r)})\|\Psi\|^2$, and since $Y_0$ is bounded with respect to the form norm of $\Hbox$ and $\mathcal D_{\mathrm{box}}$ is a form core, $e_0\ge\Ebox-E_0-c^{(r)}$. On the other hand $Y_\eta\ge(1-\eta)Y_0+\eta X_{\mathrm{in}}$ and Lemma~\ref{lem:Xin} give $e_\eta\ge(1-\eta)e_0-\eta E_0$. Hence $c'_\eta\le\Ebox-(1-\eta)(E_0+e_0)\le\eta\Ebox+(1-\eta)c^{(r)}$, which proves (i). The forms are those of the full model, in the representation of Lemma~\ref{lem:Xin}.

\emph{Step 3: (a).} Let $B:=P_\sigma S$, so $X=B^*B$ and $f_m=Bv_m/\mu_m$; $\{Bv_m/\sqrt{\mu_m}\}$ is an orthonormal system in $L^2_\sigma(\mathcal E)$. For $g\in L^2(\R^3;\R^3)$, Bessel's inequality gives $\sum_m|\langle g,f_m\rangle|^2=\sum_m\mu_m^{-1}|\langle P_\sigma g,Bv_m/\sqrt{\mu_m}\rangle|^2\le t_0^{-1}\|P_\sigma g\|^2\le t_0^{-1}\|1_{\mathcal E}g\|^2$. Since $\Sch$ is real and even, $r_m(x)\cdot e_l=\langle g_l,f_m\rangle$ with $g_l:=1_{\mathcal E}\Sch(x-\cdot)e_l$, $l=1,2,3$.

\emph{Step 4: (b).} Let $h_l:=\sum_m\langle g_l,f_m\rangle v_m$ with $g_l$ as in Step~3. Since $v_m=B^*(Bv_m)/\mu_m$, $h_l=B^*u_l$ with $u_l:=\sum_m\mu_m^{-2}\langle g_l,Bv_m\rangle Bv_m\in L^2_\sigma(\mathcal E)$ and $\|u_l\|^2\le t_0^{-2}\|g_l\|^2$. In the free vacuum $\mathrm{Var}\,\Phi(h)=\frac12\|\omega^{-1/2}h\|^2$, and for $u\in L^2_\sigma(\mathcal E)$ we have $B^*u=Su=S\Pi1_{\mathcal E}u$, so $\omega^{-1/2}B^*u=T_{\mathcal E}u$ (Lemma~\ref{lem:607N}). Summing over $l$ gives (b); $J$-reality holds because $X$ is a real operator.

\emph{Step 5: (c).} $\langle\Omega_{\mathrm{ee}},\Phi(h)^2\Omega_{\mathrm{ee}}\rangle=\frac12\langle h,\omega_{\mathrm{ee}}^{-1}h\rangle\le\frac12\|h\|\langle h,\omega_{\mathrm{ee}}^{-2}h\rangle^{1/2}$. By the Schur complement, $\omega_{\mathrm{ee}}^{-2}\le P_{\mathrm{ext}}\omega^{-2}P_{\mathrm{ext}}$ on $\Wext$; moreover $\langle B^*u,\omega^{-2}B^*u\rangle\le\langle1_{\mathcal E}u,\omega^{-2}1_{\mathcal E}u\rangle\le\|1_{\mathcal E}\omega^{-2}1_{\mathcal E}\|\,\|u\|^2$ (as $S^2\Pi\le1$ commutes with $\omega$), and the kernel $1/(4\pi|w|)$ of $\omega^{-2}$ and Schur's test give $\|1_{\mathcal E}\omega^{-2}1_{\mathcal E}\|\le(\diam\mathcal E)^2/2$. Together with Step~4 ($h_l=B^*u_l$, $\|h_l\|\le\|u_l\|\le\|g_l\|/t_0$) this gives $\langle\Omega_{\mathrm{ee}},\Phi(h_l)^2\Omega_{\mathrm{ee}}\rangle\le\frac12\cdot\frac{\|g_l\|}{t_0}\cdot\frac{\diam\mathcal E}{\sqrt2}\frac{\|g_l\|}{t_0}$, and summing over $l$ with $\sum_l\|g_l\|^2\le3\sigma_{\Sch}(x)$ (as in Step~3) gives (c): $c^{(r)}\le\frac{c\abar}2\cdot\frac{3\diam\mathcal E}{2\sqrt2\,t_0^2}\sup_x\sigma_{\Sch}(x)$.
\end{proof}

\begin{corollary}[Surgery far from the electron]\label{cor:far}
Let $\mathcal E$ be a cube of side $2(b+s)+2D_\eta$, $\mathcal C_{\mathrm{el}}$ an electron box, and $\dist(\mathcal C_{\mathrm{el}},\mathcal E)\ge D\ge1$. Apply Theorem~\ref{thm:gauge} with $\eta:=\mu/(16\Ebox)$, $0<\mu\le\Ebox$. Then
\[
\kappa_\eta\le\frac{8c\abar\Ebox}\mu,\qquad\tr\Gamma_r,\ \frac{c^{(r)}}{c\abar}\lesssim\frac{\diam\mathcal E}{t_0^2}\vol(\mathcal E)\sup_{|w|\ge D}|\Sch(w)|^2 .
\]
Since $\Sch$ is a Schwartz function, for $D=\ell^{1/2}$ the second quantity is smaller than any given power of $\ell$ once $\ell$ is large. We fix $\DEl:=\ell^{1/2}$. With $\mu=\mu_w$, the conditions $\kappa_\eta b_r^2\le\min(\mu_w/8,\mAg)$ and $c^{(r)}\le\mu_w/8$ of Section~\ref{sec:shell} then hold for $D=\DEl$ and $\ell\ge\ell_1$ ($\mAg=\mu_w/8$ the Agmon margin of the proof of Proposition~\ref{prop:shell}; the threshold $b_r^2$ is a sum of $\tr\Gamma_r$ and a multiple of $\|\Gamma_r\|\log\ell$ as in Lemma~\ref{lem:tail}, the multiple being fixed in Step~8 of that proof before $\ell_1$, and $\kappa_\eta$ is bounded by a power of $\ell$). Since $\DEl\ll\rrel\asymp\ell\log\lambda$, the order of the window radius $\rho$ is unchanged.
\end{corollary}
\begin{proof}
By Theorem~\ref{thm:gauge}(ii), $\sigma_{\Sch}(x)\le\vol(\mathcal E)\sup_{|w|\ge D}|\Sch|^2$ for $x\in\mathcal C_{\mathrm{el}}$, and $N_{\mathcal E}\lesssim\diam\mathcal E$ by Lemma~\ref{lem:607N}.
\end{proof}

\section{Shallow depressions in the window shell are rare}\label{sec:shell}

This section proves the central probabilistic ingredient of Theorem~\ref{thm:main}. It is needed because the fluctuation of $\Xw$ inherits that of the core locator unless the shell of the window is free of shallow depressions (Lemma~\ref{lem:Mw}), so that $\langle\Gamma_{\Xw}\rangle\to0$ requires that the shell event $\Gamma'_\rho$ of Definition~\ref{def:Xw} be rare to every power. The difficulty is that a shallow depression has only the small margin $\mu_w\asymp b^3\ell^\kappa$, while every way of locating it relative to the electron or to $X_c$ brings in the kinetic coefficient $c\asymp\ell^7$ or the fluctuation of $X_c$; the proof avoids both by selecting candidates on a lattice fixed by the localization cell and by a surgery that decouples the field near the candidate from the electron.

\begin{proposition}\label{prop:shell}
Fix $\alpha>0$ and $0<\kappa<1$, the design (D1)--(D5) of Definition~\ref{def:Xw}, and the external subspace of Setting~\ref{set:box} with $t_0=\frac12$ and $S\in C_c^\infty$. For every $\kappa_p>0$ there is $K_0$ such that, whenever the window constants ($K_d$ and the window constant $K$ of Proposition~\ref{prop:events}) are at least $K_0$, in the sense $K\ge K_0$ and $K_d\ge K+1$ of (D5), there is $\lambda_1$ such that for $\lambda\ge\lambda_1$
\[
P_\psi(\Gamma'_\rho)\le\ell^{-\kappa_p}.
\]
\end{proposition}

We first explain the idea. On $\Gamma'_\rho$ there is a level-$\kappa$ depression at distance at least $\rho/2$ from the core estimator $X_c$. Splitting off the field near a small cube containing it gives the margin $\mu_w\asymp b^3\ell^\kappa$ of Lemma~\ref{lem:level}, and the Agmon estimate (Lemma~\ref{lem:agmon}) turns this margin into the bound $\ell^{-\kappa_p}$. The issue is how to choose the cube. Choosing it relative to the electron brings in the large coefficient $c$ of the electron's kinetic energy; choosing it relative to $X_c$ multiplies the sharpness of the selection by the fluctuation of $X_c$. We therefore first localize the electron and the bubble in cells of side $n_b$ by the collective-coordinate IMS formula (Lemma~\ref{lem:ims}). Within each cell, candidates are points of an absolute lattice determined by the cell alone, and the selection is done by functions of the field alone (Lemma~\ref{lem:softmax}). The surgery is Theorem~\ref{thm:gauge}, and the coupling to the electron is negligible beyond the distance $\DEl$. The Agmon function, which is not used for the selection, is a soft minimum of the local level plus a trapezoidal radial profile around $X_c$.

\begin{proof}
Let $\mAg:=\mu_w/8$ with the margin $\mu_w=\mu_w(g_c)$, $g_c=(1+3\theta_\delta)g_w$, of Lemma~\ref{lem:level}(b), $F_{\max}:=\kappa_p\log\ell$ and $a:=1+\theta_\delta$.

\emph{Step 0: the Agmon function.} Let $\Upsilon$ be the radial profile that vanishes on $[\rho/2,\rho]$, is linear with slope $4ag_w/\rho$ from $0$ to $ag_w$ on $[\rho/4,\rho/2]$ (inwards) and on $[\rho,5\rho/4]$ (outwards), rises with the same slope to $2ag_w$ beyond these intervals (reached at $r=0$ and at $r=3\rho/2$), and is constant for $r\ge3\rho/2$. Thus $|\Upsilon'|\le4ag_w/\rho$, $\Upsilon(r)<ag_w$ only for $\rho/4<r<5\rho/4$, and $\Upsilon=2ag_w$ for $r\ge3\rho/2$. Let $R_0:=r_W+2\rho$ and $B_0:=B(0,R_0)$ in relative coordinates. Since $|X_c|\le r_W$ for all configurations (Appendix~\ref{app:core}), $B(X_c,3\rho/2)\subset B_0$ and $\Upsilon(|Y-X_c|)=2ag_w$ on $\partial B_0$. With $\mathsf h(Y):=g^{(\tau_w)}_b(Y)+\Upsilon(|Y-X_c|)$ and $\beta':=\ell^{\kappa+1}$ set
\[
\tilde G_w:=-\frac1{\beta'}\log\Bigl(\frac1{|B_0|}\int_{B_0}\mathrm e^{-\beta'\mathsf h(Y)}\dd Y\Bigr),
\]
and let $p\propto\mathrm e^{-\beta'\mathsf h}$ be the Gibbs probability density on $B_0$. This is the integral soft minimum of Step~2 of the proof of Lemma~\ref{lem:edge}, with the shell, the slope length, the level and the electron replaced by $[\rho/2,\rho]$, $\rho/4$, $g_w$ and $X_c$: the two Agmon arguments of Part~II have the same form. The function $\mathsf h$ is Lipschitz with the constant $L_{\mathsf h}:=\sqrt3\,\tau_w/(2b+s)+4ag_w/\rho$, independent of the configuration, since $0\le\sat_{\tau_w}\le\tau_w$ is averaged over cubes of side $2b+s$. We use four properties of $\tilde G_w$.
\begin{itemize}
\item[(a)] $\min_{B_0}\mathsf h\le\tilde G_w$; and if $\beta'>1/\rho$ and $\mathsf h(Y)<g$ at some $Y$ with $|Y-X_c|\le\rho$, then $\tilde G_w\le g+\Delta_{\beta'}$ with $\Delta_{\beta'}:=\beta'^{-1}(L_{\mathsf h}+3\log(\beta'R_0))$. Indeed $B(Y,1/\beta')\subset B_0$, $\mathsf h<g+L_{\mathsf h}/\beta'$ there, and $|B_0|/|B(Y,1/\beta')|=(\beta'R_0)^3$.
\item[(b)] $\Gamma_{\tilde G_w}\le2\bar C_g+2(4ag_w/\rho)^2\Gamma_{X_c}$. Indeed $B_0$ does not depend on the field, so the field gradient is the Gibbs average $\langle\nabla g_b(Y)\rangle_p-\langle\Upsilon'(|Y-X_c|)u\rangle_p^{\mathsf T}\nabla X_c$, $u:=(Y-X_c)/|Y-X_c|$; by Jensen's inequality and the contraction property of $P_T$ the first term has squared norm at most $\sup_Y\|v_Y\|^2=\bar C_g$ (with $v_Y$ as in Section~\ref{sec:proof}). The sharpness $\beta'$ does not enter.
\item[(c)] The translation derivative is $D\tilde G_w=\langle\nabla\Upsilon(|Y-X_c|)\rangle_p^{\mathsf T}\Rm(X_c)+\varphi$, where, by $g_b(Y;\transl_aA)=g_b(Y-a;A)$ and the substitution $Y\mapsto Y+a$, the flux
\[
\varphi=\frac1{\beta'}\,\frac{\int_{\partial B_0}\mathrm e^{-\beta'\mathsf h}\,n\dd S}{\int_{B_0}\mathrm e^{-\beta'\mathsf h}\dd Y}
\]
comes from the translation of the domain. Since $\int_{B_0}\mathrm e^{-\beta'\mathsf h}=|B_0|\mathrm e^{-\beta'\tilde G_w}$ and $\mathsf h\ge2ag_w$ on $\partial B_0$, on $\{\tilde G_w<ag_w\}$
\[
|\varphi|\le\frac{|\partial B_0|}{\beta'|B_0|}\mathrm e^{-\beta'(2ag_w-\tilde G_w)}\le\frac3{\beta'R_0}\,\mathrm e^{-\beta'ag_w}.
\]
\item[(d)] On $\{\tilde G_w<ag_w\}$ every minimizer $Y_0$ of $\mathsf h$ over $B_0$ satisfies $\rho/4<|Y_0-X_c|<5\rho/4$ and $g^{(\tau_w)}_b(Y_0)<ag_w$, since $\Upsilon(|Y_0-X_c|)\le \mathsf h(Y_0)\le\tilde G_w$ by (a) and $g_b\ge0$.
\end{itemize}
For $\ell\ge\ell_1$ we have $\beta'>1/\rho$ and $\Delta_{\beta'}\lesssim\ell^{-1-\kappa}\log\ell\le\theta_\delta g_w/4$. Let $F\in C^1$ be non-increasing with $F=F_{\max}$ on $[0,(1+\theta_\delta/4)g_w]$, $F=0$ on $[ag_w,\infty)$ and $|F'|\le2F_{\max}/(\theta_\delta g_w)$, and $f:=F(\tilde G_w)$. On $\Gamma'_\rho$ the profile vanishes at the depression, so (a) with $g=g_w$ gives $\Gamma'_\rho\subset\{\tilde G_w\le g_w+\Delta_{\beta'}\}\subset\{f=F_{\max}\}$; and as $F=F'=0$ on $[ag_w,\infty)$, $\mathcal Z:=\{f>0\}\cup\{\GAg_f>0\}\subset\{\tilde G_w<ag_w\}$.

The function $f$ belongs to $\mathscr C$: $\tilde G_w$ is bounded, depends on the field only in a bounded region ($B_0$ enlarged by the support of the kernel) and through $X_c$, and is Fr\'echet $C^1$ with bounded gradient by (b) and Proposition~\ref{prop:global}, so (C1) follows from Lemma~\ref{lem:multiplier}(iv). Condition (C2) asks for a first-order Taylor expansion in the translation parameter, uniform in the configuration, and is checked directly: $a\mapsto X_c(\transl_aA)-a$ is $C^2$ with derivatives bounded uniformly in the configuration (Section~\ref{sec:prelim}), the translation of the domain contributes the flux, whose integrand $\mathrm e^{-\beta'\mathsf h}$ is Lipschitz with a constant independent of the configuration (for a Lipschitz function $u$ and a ball $B$, $\bigl|\int_{B+a}u-\int_Bu-a\cdot\int_{\partial B}u\,\nu\,\dd S\bigr|\le\frac12\mathrm{Lip}(u)\,|\partial B|\,|a|^2$, since the gradient in $a$ of $\int_{B+a}u=\int_Bu(\cdot+a)$ is $\int_{\partial B}u(\cdot+a)\,\nu\,\dd S$), and the profile term is continuous in $L^1$; these remainders stay uniform in the configuration after division by $\int_{B_0}\mathrm e^{-\beta'\mathsf h}\dd Y\ge|B_0|\,\mathrm e^{-\beta'(\tau_w+2ag_w)}$, since $0\le \mathsf h\le\tau_w+2ag_w$. (The kinks of $\Upsilon$ and the hard ball, together with the merely Lipschitz dependence of $g_b(Y)$ on $Y$ (the kernel $K'_b$ is an indicator), prevent a direct use of Lemma~\ref{lem:multiplier}(v); the same argument applies to the soft minimum of Lemma~\ref{lem:edge}.) Lemma~\ref{lem:multiplier}(vi) gives $f,\mathrm e^f\in\mathscr C$.

\emph{Step 1: localization in cells.} Apply Lemma~\ref{lem:ims} in the full-space model with $\hat X=X_c$ and the cell side $n_b\ge\rrel$ of (D5). On cell $j$ the absolute position of the bubble lies in $\supp\chi_j$ and the electron in $\mathcal C_e^j:=\supp\chi_j+\overline{B(0,2\rrel)}$ (side $\le n_b+4\rrel$). The localization errors are:
\begin{itemize}
\item the field term $\frac12\sum_k\Gamma_{\varphi_k}\lesssim\Gamma_{X_c}/n_b^2\lesssim\ell^{-1}(\log\lambda)^3$ by Proposition~\ref{prop:Mprime}(a), a fraction $\ell^{-1-\kappa}$polylog of the margin (term (g) of Table~\ref{tab:terms});
\item the kinetic term $\frac{cC_\chi}{n_b^2}\|\Rm(X_c)\|^2$, which vanishes on $\mathcal G_\theta$ by Proposition~\ref{prop:Mprime}(b);
\item everything else, which is put into the deficit (term (h) of Table~\ref{tab:terms}): the kinetic loss on bad sets and on the shell $\{\rrel\le|X_{\mathrm{rel}}|\le2\rrel\}$, and the missing margin on the cell $\varphi_\infty$, where we only use $H-E\ge0$.
\end{itemize}

\emph{Step 2: selection among absolute candidates.} For each cell $j$ let
\[
\mathcal K_j:=\bigl\{Y_k\in\rho_\delta\Z^3:\ \dist(\mathcal E_k,\mathcal C_e^j)\ge\DEl,\ |Y_k-c_j|\le R_{\mathrm{out}}\bigr\},
\]
where $\mathcal E_k$ is the cube of side $2(b+s)+2D_\eta$ centred at $Y_k$ (so that all blocks entering $g^{(\tau_w)}_b(Y_k)$ lie in $\mathcal E_k$ at distance $\ge D_\eta$ from its boundary), $c_j$ is the centre of the cell, and $R_{\mathrm{out}}:=5\rho/4+\diam\supp\chi_j+\rho_\delta\asymp\ell\log\lambda$. By (D3), $\mathcal K_j$ depends only on $j$. The functions $g_k:=g^{(\tau_w)}_b(Y_k)$ depend on the field only, so Lemma~\ref{lem:softmax} applies inside the cell: the $\zeta_k$ commute with $\varphi_j$ and
\[
\varphi_jH\varphi_j=\sum_k\zeta_k\varphi_jH\varphi_j\zeta_k-\frac12\varphi_j^2\sum_k\Gamma_{\zeta_k},\qquad\sum_k\Gamma_{\zeta_k}\le\frac{\beta_{\mathrm{sel}}^2}4\max_k\Gamma_{g_k}\le\frac{\beta_{\mathrm{sel}}^2}4\cdot\frac{2\tau_w}{\mathrm e\,b^3},
\]
the last bound by $\Gamma_{g_k}=\|P_Tv_{Y_k}\|^2\le\|v_{Y_k}\|^2\le\bar C_g$ for the unprojected gradients $v_Y$ of Section~\ref{sec:proof}, whose bound uses $|\partial \sat_\tau(u)|^2\le4u\mathrm e^{-2u/\tau}\le2\tau/\mathrm e$. With $\beta_{\mathrm{sel}}:=\log(|\mathcal K_j|/\eta_p)/(\theta_\delta g_w)$ the ratio of this loss to the margin is $(\log|\mathcal K_j|)^2/(\theta_\delta^2b^6)\to0$ by (D1) (term (e) of Table~\ref{tab:terms}). No electronic double commutator appears, since $\nabla_x\zeta_k=0$.

\emph{Step 3: surgery for each cell and candidate.} On states localized by $\varphi_j\zeta_k$ the electron lies in $\mathcal C_e^j$, so $H-E\ge\Hbox(\mathcal C_e^j)-\Ebox(\mathcal C_e^j)$. Apply Theorem~\ref{thm:gauge} with external region $\mathcal E_k$ and $\eta:=\mu_w/(16\Ebox)$, Lemma~\ref{lem:seam} to the seam, and Corollary~\ref{cor:margin} and Lemma~\ref{lem:level} to the external field. The last two are applied with the kernel weights of the cube at $Y_k$, for each translate of the block lattice, and with $a_{\max}:=|Q|\sup K'_b$ for all translates. Since $\mathcal V$ of Corollary~\ref{cor:margin} is linear in the weights, its average over the translates is the same expression for the averaged measure $\nu(\dd z)=K'_b(Y_k-z)\dd z$ on block centres, whose level is exactly $g_k$; Lemma~\ref{lem:level}(a) applied to $\nu$ shows that the average is at least $\mu_w(g_k)$ (with $a_\Sigma$ the total mass of $\nu$) whenever $g_k\le g_c$, and it is at least $-B_-$ always. For a constant $\delta\in(0,\frac12]$,
\[
H-E\ge(1-\delta)\mathcal V_{jk}(A)-\delta^{-1}T_{\Sigma,jk}-\kappa_\eta|r_{\mathrm e,jk}(x)|^2-\frac{\mu_w}{16}-c^{(r)}_{jk},
\]
with $\mathcal V_{jk}\ge\mu_w(g_k)$ when $g_k\le g_c$ and $\mathcal V_{jk}\ge-B_-$ always. By Corollary~\ref{cor:far} (with $\DEl=\ell^{1/2}$ and $\ell\ge\ell_1$), $c^{(r)}_{jk}\le\mu_w/8$. The forms $T_\Sigma$ and $|r_{\mathrm e}|^2$ are quadratic in the field as in Lemma~\ref{lem:tail}; by Lemma~\ref{lem:trace}, $\tr\Gamma_\Sigma\lesssim b^4$, and by Theorem~\ref{thm:gauge}(ii)(b) and Corollary~\ref{cor:far}, $\tr\Gamma_r$ is super-polynomially small.

\emph{Step 4: multiplicative lower bound.} Combining Steps 1--3, define
\[
V:=\sum_j\varphi_j^2\sum_kp_k\Bigl((1-\delta)\mathcal V_{jk}-\frac{\mu_w}{16}-c^{(r)}_{jk}\Bigr)-\frac12\sum_j\varphi_j^2\sum_k\Gamma_{\zeta_k}-\frac12\sum_k\Gamma_{\varphi_k}-\frac c2\sum_k|\nabla_x\varphi_k|^2
\]
and $\mathcal W:=\sum_j\varphi_j^2\sum_kp_k(\kappa_\eta|r_{\mathrm e,jk}|^2+\delta^{-1}T_{\Sigma,jk})$. Then $H_{\mathrm{full}}-E\ge V-\mathcal W$ in the sense of forms, and the weights $\varphi_j^2p_k$ sum to at most one. Averaging the cell partition over translations of period $n_b$ (compatible with the candidate lattice by (D2)) makes $V$ and $\mathcal W$ translation invariant, and Setting~\ref{set:full} gives $H(0)-E\ge\bar V-\bar{\mathcal W}$ in the fiber.

\emph{Step 5: margin on $\mathcal Z$.} On $\mathcal Z$ we have $\tilde G_w<ag_w$, so by Step~0(d) there is $Y_0$ with $\rho/4<|Y_0-X_c|<5\rho/4$ and $g^{(\tau_w)}_b(Y_0)<(1+\theta_\delta)g_w$. On a cell $j$ with $\varphi_j\ne0$ the bubble lies in $\supp\chi_j$ and the electron in $\mathcal C_e^j$, so (in absolute coordinates)
\[
\dist(Y_0,\mathcal C_e^j)\ge\frac\rho4-(\diam\supp\chi_j+2\rrel)=\DEl+\tfrac{\sqrt3}2(2b+2s+2D_\eta+\rho_\delta).
\]
The nearest lattice point $Y_k$ (at distance $\le\frac{\sqrt3}2\rho_\delta$) therefore has $\dist(\mathcal E_k,\mathcal C_e^j)\ge\DEl$, and $Y_k\in\mathcal K_j$ since $|Y_k-c_j|\le5\rho/4+\diam\supp\chi_j+\rho_\delta=R_{\mathrm{out}}$. By the Lipschitz continuity of $g_b$ and $\rho_\delta=\theta_\delta g_wb/3$, $g_k\le(1+2\theta_\delta)g_w$ for $\ell\ge\ell_1$, so $\min_kg_k\le(1+2\theta_\delta)g_w$. By Lemma~\ref{lem:softmax}, candidates with $g_k>\min g+\theta_\delta g_w$ carry total weight at most $\eta_p$, and all other candidates have $g_k\le g_c=(1+3\theta_\delta)g_w$; with the monotonicity in Lemma~\ref{lem:level} and $\mathcal V\ge-B_-$,
\[
\sum_kp_k\mathcal V_{jk}\ge\mu_w-\eta_p(\mu_w+B_-).
\]
Subtracting the losses of Steps 1--2 and the constants of Step 3, and using Step 6 below for $\GAg_f$, we obtain for $\ell\ge\ell_1$, on $\mathcal Z\cap\mathcal G_\theta\cap\{|X_{\mathrm{rel}}|\le\rrel\}$,
\[
\bar V-c_0-\tfrac12\GAg_f\ge4\mAg ,
\]
where $c_0$ collects constants produced by the localizations and averaging; here $\eta_p,\delta$ are constants with $(1-\delta)(1-\eta_p)>11/16$ (the constants of Step~3 cost at most $3\mu_w/16$) and $B_-/\mu_w\asymp g_w\to0$.

\emph{Step 6: Agmon metric.} By Step~0(c), $Df=F'(\tilde G_w)\bigl(\langle\nabla\Upsilon\rangle_p^{\mathsf T}\Rm(X_c)+\varphi\bigr)$, and $F'\ne0$ only where $\tilde G_w<ag_w$. On $\mathcal G_\theta$, $\Rm(X_c)=0$ by Proposition~\ref{prop:Mprime}(b), so
\[
c|Df|^2\le c\Bigl(\frac{2F_{\max}}{\theta_\delta g_w}\Bigr)^2\frac9{\beta'^2R_0^2}\,\mathrm e^{-2\beta'ag_w},
\]
which is super-polynomially small since $\beta'g_w\asymp\ell$. Elsewhere, since $|\langle\nabla\Upsilon\rangle_p|\le4ag_w/\rho$, $c|Df|^2$ is a polynomial in $\ell$ (Proposition~\ref{prop:global}) carried by the bad sets. For the carr\'e du champ, Step~0(b) gives
\[
\Gamma_f\le F'(\tilde G_w)^2\,\Gamma_{\tilde G_w}\le\Bigl(\frac{2F_{\max}}{\theta_\delta g_w}\Bigr)^2\Bigl(2\bar C_g+2\Bigl(\frac{4ag_w}\rho\Bigr)^2\Gamma_{X_c}\Bigr)\lesssim F_{\max}^2\Bigl[\frac1{g_wb^3}+\frac{\ell(\log\lambda)^5}{\rho^2}\Bigr],
\]
using $\bar C_g\le2\tau_w/(\mathrm e\,b^3)$ (Section~\ref{sec:proof}) and Proposition~\ref{prop:Mprime}(a). Relative to the margin $\mu_w\asymp b^3/g_w$ this is $\lesssim\kappa_p^2\log^2\ell/(\theta_\delta^2b^6)+\ell^{-1-\kappa}$polylog$\,\to0$ by (D1) and $\rho\asymp\ell\log\lambda$ (term (i) of Table~\ref{tab:terms}). The sharpness $\beta'$ of the soft minimum does not enter $\Gamma_f$.

\emph{Step 7: the Agmon bound and the seam threshold.} Apply Corollary~\ref{cor:agmonquad} with $2\mAg$ in place of $\mAg$ to the quadratic terms $\mathcal W$ (weights $\varphi_j^2p_k$, coefficients $\kappa_\eta$ and $\delta^{-1}$, thresholds $b_r^2,b_\Sigma^2$), with $\mathcal L_0$ the shortfall of the bound of Step~5 on the part of $\mathcal Z$ where it is not established ($\mathcal G_\theta^c$, the shell, the cell $\varphi_\infty$). The thresholds are chosen with $\kappa_\eta b_r^2\le \mAg$ (Corollary~\ref{cor:far}) and $\delta^{-1}b_\Sigma^2\le \mAg$, where $b_\Sigma^2:=C(b^4+b^3(2F_{\max}+K_{\rm t}\log\ell))$ with a tail constant $K_{\rm t}$ (not related to the window constant $K$ of Proposition~\ref{prop:events}); here $\tr\Gamma_\Sigma\lesssim b^4$ by Lemma~\ref{lem:trace}, and $\|\Gamma_\Sigma\|\lesssim\diam\mathcal E\asymp b$ by Step~2 of its proof in Appendix~\ref{app:cited}, so that the Gaussian tails of Lemma~\ref{lem:tail} are at most any power of $\ell$. The ratio $\delta^{-1}b_\Sigma^2/\mu_w\lesssim(b+(\kappa_p+K_{\rm t})\log\ell)/\ell^\kappa\to0$ (from $\mu_w\asymp b^3\ell^\kappa$ and $F_{\max}\lesssim\kappa_p\log\ell$). \emph{This is where $\kappa>0$ is used} (term (c) of Table~\ref{tab:terms}). The seam costs $\delta^{-1}b_\Sigma^2\asymp b^4+b^3\log\ell$, while the margin of one shallow depression (Lemma~\ref{lem:level}) is only $\mu_w\asymp b^3\ell^\kappa$. For $\kappa=0$ the margin would be of the order $b^3$ of the volume of the cube, and the extra factor $b+\log\ell$ of the seam could not be paid; any fixed $\kappa>0$ beats these polylogarithms. The shortfall $\mathcal L_0$ and the global terms are bounded by $\mathrm{poly}(\ell)\,1_{S_M\cup\mathcal B_\theta}$ (Proposition~\ref{prop:global}), so their expectations are at most $\ell^{-K'}$ for any $K'$ by Propositions~\ref{prop:SM} and~\ref{prop:events}. The Gaussian tails are at most any power of $\ell$ by Lemma~\ref{lem:tail}. With $F=F_{\max}$,
\[
P_\psi(\Gamma'_\rho)\le P_\psi(f\ge F_{\max})\le\frac{\ell^{-2\kappa_p}}{2\mAg}\langle(c_0-\bar V)_+\rangle+\frac1{2\mAg}\langle\mathcal L_01_{\mathcal Z}\rangle+(\text{tails}).
\]
(Lemma~\ref{lem:agmon} is applied with $E+\bar V-\bar{\mathcal W}$ in place of $V$.)

\emph{Step 8: constant majorants and the choice of constants.} The weights $\varphi_j^2p_k$ depend on the configuration, so Corollary~\ref{cor:agmonquad} is used with constant majorants: inside the average over the translations of the cell lattice (constant weights) we take $\bar a=1$ for every pair $(j,k)$ with $k\in\mathcal K_j$ and $0\in\supp\chi_j+\overline{B(0,2\rrel)}$ in relative coordinates (a bounded number of cells). There are at most $N_{\mathrm{cand}}\le C(R_{\mathrm{out}}/\rho_\delta)^3$ such pairs, a polynomial in $\ell$ whose degree does not depend on $\kappa_p$. Hence the trace terms, at most $N_{\mathrm{cand}}(\kappa_\eta\sup\tr\Gamma_r+\delta^{-1}\sup\tr\Gamma_\Sigma)$, join $\langle(c_0-\bar V)_+\rangle$ below, and the Gaussian tails acquire the factor $N_{\mathrm{cand}}$, which is absorbed by taking the multiples of $\log\ell$ in the thresholds $b_r^2,b_\Sigma^2$ larger than its degree. The quantity $\langle(c_0-\bar V)_+\rangle$ is bounded by $B_-$, the constants of Step~3 and the localization errors, a polynomial in $\ell$ whose degree does not depend on $\kappa_p$ (the window constants grow with $\kappa_p$ but do not enter as powers of $\ell$). Since $\ell^{-\kappa_p}$ decreases in $\kappa_p$, it suffices to treat $\kappa_p$ larger than this degree. Choosing $K'$ larger than $\kappa_p$ plus this degree, and the window constants accordingly, the right-hand side is at most $\ell^{-\kappa_p}$ for $\lambda\ge\lambda_1$. The auxiliary quantities $\beta'$, $R_0$ and $F$ of Step~0 are explicit functions of $\ell$ and of constants chosen before, so the order of Remark~\ref{rem:order} is respected.
\end{proof}

\begin{remark}[What fixes the window]\label{rem:window}
In the formula for $\rho$ only the relative localization length $\rrel$ is of order $\ell\log\lambda$; $\DEl=\ell^{1/2}$, and $b$, $D_\eta$, $\rho_\delta$ are at most polylogarithmic. With the conventional Schwarz inequality instead of Theorem~\ref{thm:gauge}, a distance of order $\ell^{2-\kappa/3}b^{-1/3}$ would take the place of $\DEl$, and the bound $\ell^{\kappa-3}\rho^2$ of Lemma~\ref{lem:Mw} would not tend to zero. Reducing this distance to $\ell^{1/2}$ is what makes Proposition~\ref{prop:shell} compatible with Theorem~\ref{thm:main}.
\end{remark}

\section{Fluctuation of the estimator and proof of Theorem~\ref{thm:main}}\label{sec:proof}

\emph{Structure of the proof.} Theorem~\ref{thm:main} follows from Lemma~\ref{lem:consumer} applied to $\Xw$: the first condition of \eqref{eq:two} holds because $\Rm(\Xw)=0$ on the good set (Remark~\ref{rem:cov} and Proposition~\ref{prop:Mprime}(b)) and the bad sets are rare to every power and carry only polynomial bounds (Propositions~\ref{prop:SM}, \ref{prop:events}, \ref{prop:global}); the second is Lemma~\ref{lem:Mw} together with Proposition~\ref{prop:shell} (shallow depressions in the shell are rare). The constants enter as in Table~\ref{tab:constants}; the order in which they are chosen is Remark~\ref{rem:order}.
\begin{center}\small
\captionof{table}{Constants of Part~II and their roles.}\label{tab:constants}
\begin{tabular}{p{0.2\textwidth}p{0.3\textwidth}p{0.4\textwidth}}
constant & fixed in & role\\\hline
$\kappa$ & Definition~\ref{def:Xw} & level $g_w\asymp\ell^{-\kappa}$ of the shallow depressions; $1-\kappa$ is the margin in Lemma~\ref{lem:Mw} (its limit $\kappa\to0$ gives the exponent $8/7$ of Theorem~\ref{thm:main}), and $\kappa>0$ is needed for the seam threshold in Step~7 of the proof of Proposition~\ref{prop:shell}\\
$C_b$ & (D1) & cube side $b=C_bs$ of the local levels\\
$c_\tau$, $\theta_\delta$ & (D4), Lemma~\ref{lem:level} & saturation $\tau_w=c_\tau g_w$ (Definition~\ref{def:Xw}) and the level margin $g\le g_c=(1+3\theta_\delta)g_w$\\
$\eta_p$, $\theta_\delta$ & Lemma~\ref{lem:softmax}, (D2) & selection of candidates and their spacing $\rho_\delta$\\
$\delta$ & Lemma~\ref{lem:seam} & seam completion in Proposition~\ref{prop:shell}\\
$\kappa_p$, $K$, $K_d$ & Propositions~\ref{prop:SM}, \ref{prop:events}, (D5) & bad sets rare to the power $\ell^{-\kappa_p}$; window lengths $K\ell\log\lambda$ and $\rrel$\\
$W_*$, $M^w_*$; $a_c$, $a_w$ & (D$'$), Definitions~\ref{def:core}, \ref{def:Xw} & mass thresholds of the good set; lifting thresholds $a_c=W_*$, $a_w=M^w_*$\\
$\rho$ & (D5) & radius of the window of $\Xw$\\
$t_*$, $R_c$, $\sigma$ & Proposition~\ref{prop:massloc}, Lemma~\ref{lem:design} & mass localization at energy $(1+t_*)F$, core radius ($R_*$ units), smooth thresholds $\Sigma_k$\\
\end{tabular}
\end{center}

\subsection{A deterministic fluctuation bound}
The field-gradient of $\Xw$ has two sources: the motion of the shallow indicator $d_w$ inside the window, and the motion of the centre $X_c$ of the window. The next lemma bounds the first deterministically by the inverse of the depression mass $M$ in the window and shows that the second vanishes unless a shallow depression lies in the shell of the window (the shell event $\Gamma'_\rho$ of Definition~\ref{def:Xw}). The first bound needs no probabilistic input: the contribution of the shallow-level indicator is bounded by the inverse of its total mass $M$ in the window, whatever the number of shallow depressions.

The functional $g_b(Y)=g^{(\tau_w)}_b(Y)$ is defined through $u'_Q(A)=\|P'_Q1_QA\|^2$ for every square integrable vector field $A$; let $v_Y$ be its gradient in $L^2(\R^3;\R^3)$, which is supported in the cube of side $2(b+s)$ around $Y$. The gradient in the field configuration is $\nabla g_b(Y)=P_Tv_Y$, where $P_T$ is the orthogonal projection onto the band-limited transverse fields. Since $P_T$ is not local, the kernel $\langle\nabla g_b(Y),\nabla g_b(Y')\rangle$ need not vanish for distant $Y,Y'$; we therefore work with the unprojected gradients. Let $\bar C_g:=\sup_Y\|v_Y\|^2$ and define the overlap constant
\[
K_{\mathrm S}:=\sup_Y\int\bigl|\langle v_Y,v_{Y'}\rangle\bigr|\frac{\dd Y'}{b^3\,\bar C_g}.
\]
This kernel vanishes unless $|Y-Y'|_\infty<2(b+s)$, so $K_{\mathrm S}\le(4(b+s))^3/b^3$. As $P_T$ is a contraction, Schur's test gives $\|\int a\,\nabla g_b(Y)\dd Y/b^3\|^2\le\|\int a\,v_Y\dd Y/b^3\|^2\le K_{\mathrm S}\bar C_g\int|a|^2\dd Y/b^3$ for vector-valued $a$.

\begin{lemma}\label{lem:Mw}
Let $X_c$ be any bounded field functional in $\mathscr C$, and $\chi_\rho,M,\Xw$ and the shell event $\Gamma'_\rho$ as in Definition~\ref{def:Xw}. Assume $|\tilde\chi_{g_w}'|^2\le C\tilde\chi_{g_w}/g_w^2$ and $\bar C_g\le2\tau_w/(\mathrm e\,b^3)$ with $\tau_w=c_\tau g_w$ (the derivative of the saturation gives $\|v_z\|^2\le2\tau_w/(\mathrm e\,s^3)$ for the unprojected gradient $v_z$ of $\sat_{\tau_w}(u'_{Q(z)}/|Q|)$, which is supported in the block $Q(z)$; Schur's test for the unprojected kernel $\langle v_z,v_{z'}\rangle$, which vanishes for $|z-z'|_\infty\ge s$, gives $\|v_Y\|^2\le(2s)^3\sup K'_b\cdot2\tau_w/(\mathrm e\,s^3)\le2\tau_w/(\mathrm e\,b^3)$ when $\sup K'_b\le(2b)^{-3}$, e.g.\ for $K'_b$ the normalized indicator of its cube). On configurations with $M\ge\max(M_*,2a_w)$ outside $\Gamma'_\rho$, the part of the field gradient of $\Xw$ through $X_c$ vanishes and
\begin{equation}\label{eq:Mw}
\Gamma_{\Xw}\le\frac{2C\,c_\tau}{\mathrm e}\,K_{\mathrm S}\,\frac{(2\rho)^2}{b^3\,g_w\,M_*}.
\end{equation}
\end{lemma}

\begin{proof}
On $\{M\ge2a_w\}$ we have $r_{a_w}(M)=M$ and $r_{a_w}'(M)=1$, so the derivatives below are those of $\Xw=X_c+Z/M$ with $Z:=\int\chi_\rho(|Y-X_c|)(Y-X_c)d_w(Y)\dd Y/b^3$. Since $|Z|\le\rho M$, $|\Xw-X_c|\le\rho$, and $|Y-\Xw|\le2\rho$ on the support of $\chi_\rho$. Split the derivative $\partial$ with respect to the configuration into (a) the part through $d_w$ and (b) the part through $X_c$.

(a) With $X_c$ fixed, the derivatives of $Z$ and $M$ combine to
\[
\partial_{(a)}\Xw=\frac1M\int\chi_\rho\,(Y-\Xw)\,\tilde\chi'(g_b(Y))\,\partial g_b(Y)\frac{\dd Y}{b^3}.
\]
By the contraction property of $P_T$ and Schur's test for the kernel $\langle v_Y,v_{Y'}\rangle$ (see the definition of $K_{\mathrm S}$), $\|\int a\,\partial g_b\dd Y/b^3\|^2\le K_{\mathrm S}\bar C_g\int|a|^2\dd Y/b^3$ for vector-valued $a$. Here $|a|^2\le\chi_\rho(2\rho)^2C\,d_w/(g_w^2M^2)$, since $0\le\chi_\rho\le1$ and $d_w=\tilde\chi_{g_w}(g_b)$; hence, by the definition of $M$, $\int|a|^2\dd Y/b^3\le(2\rho)^2C/(g_w^2M)\le(2\rho)^2C/(g_w^2M_*)$, and with $\bar C_g\le2c_\tau g_w/(\mathrm e\,b^3)$ this gives $\Gamma_{(a)}\le\frac{2Cc_\tau}{\mathrm e}K_{\mathrm S}\frac{(2\rho)^2}{b^3g_wM_*}$.

(b) With $d_w$ fixed, $\partial_{X_c}\chi_\rho(|Y-X_c|)=-\chi_\rho'(|Y-X_c|)u$ with $u:=(Y-X_c)/|Y-X_c|$, and $\partial_{(b)}\Xw=\mathsf M\,\partial X_c$, where
\[
\mathsf M:=-\frac1M\int\chi_\rho'\,(Y-\Xw)\,u^{\mathsf T}d_w\frac{\dd Y}{b^3}
\]
(the $\chi_\rho'$-terms from $Z$ and from $M$ combine through $Y-X_c$ and $\Xw-X_c$, and the terms $\pm I$ from $Z$ and from $X_c$ cancel). The support of $\chi_\rho'$ is the shell $\rho/2\le|Y-X_c|\le\rho$. Outside $\Gamma'_\rho$ every $Y$ in this shell has $g_b(Y)\ge g_w$, hence $d_w(Y)=0$ since $\tilde\chi_{g_w}=0$ on $[g_w,\infty)$. Thus the integrand vanishes identically, $\mathsf M=0$ and $\partial_{(b)}\Xw=0$, so $\partial\Xw=\partial_{(a)}\Xw$ and $\Gamma_{\Xw}=\Gamma_{(a)}$; no doubling is needed.
\end{proof}

Without the exclusion of $\Gamma'_\rho$ the depression mass in the shell is controlled only by the volume of the shell: the coefficient $\mathsf M$ is then of order $|B_\rho|/(b^3M_*)$, the term $\Gamma_{X_c}$ survives, and the bound degrades to $\langle\Gamma\rangle\lesssim\ell$ up to logarithms (Remark~\ref{rem:core-estimator}). This is why Proposition~\ref{prop:shell} is needed.

The matrix $\mathsf M$ equals $I+\nabla_x\mathsf Y$ of Remark~\ref{rem:cov}. Hence
\begin{equation}\label{eq:Rw}
\Rm(\Xw)=\mathsf M\,\Rm(X_c)\qquad(M\ge\max(M_*,2a_w)),
\end{equation}
and in particular $\Rm(\Xw)=0$ on $\mathcal G_\theta$ (Remark~\ref{rem:cov}).

\subsection{Expectations}
\begin{proposition}\label{prop:expect}
Assume the design of Definition~\ref{def:Xw} with $0<\kappa<1$. There is $\lambda_1=\lambda_1(\kappa)$ such that for $\lambda\ge\lambda_1$, with implicit constants independent of $\lambda$ (Remark~\ref{rem:order}),
\[
\langle\Gamma_{\Xw}\rangle\lesssim\ell^{\kappa-1}(\log\lambda)^2,\qquad c\,\delta(\Xw)\lesssim\ell^{-1}.
\]
\end{proposition}
\begin{proof}
Let $\mathcal G':=\mathcal G_\theta\cap(\Gamma'_\rho)^c$. Remark~\ref{rem:Mw} gives $M\ge2M^w_*=2a_w$ on $\mathcal G'$ with $M^w_*\asymp\ell^3/b^3$. Apply Lemma~\ref{lem:Mw} with $M_*=M^w_*$. As $g_w\asymp\ell^{-\kappa}$, $K_{\mathrm S}=O(1)$ and $\rho\asymp\ell\log\lambda$, the bound \eqref{eq:Mw} is
\[
\asymp\frac{\rho^2\ell^\kappa}{b^3(\ell^3/b^3)}=\ell^{\kappa-3}\rho^2\asymp\ell^{\kappa-1}(\log\lambda)^2 .
\]
On $\mathcal G'^c$ we use $\Gamma_{\Xw}\le K_G\ell^D$ (Proposition~\ref{prop:global}) and $P(\mathcal G'^c)\le P(S_M)+P(\mathcal B_\theta)+P(\Gamma'_\rho)\le3\ell^{-\kappa_p}$ (Propositions~\ref{prop:SM}, \ref{prop:events}, \ref{prop:shell}, applied with common window constants at least the largest of their thresholds $K_0$) with $\kappa_p>D+1$. The contribution of $\mathcal G'^c$ is $O(\ell^{-1})$, hence at most $\ell^{\kappa-1}(\log\lambda)^2$ for large $\lambda$. This proves the first bound.

For the second, $\delta\le\langle\|\Rm(\Xw)\|^2\rangle$ by Lemma~\ref{lem:consumer}. On $\mathcal G_\theta$, $\Rm(\Xw)=0$ by Remark~\ref{rem:cov}; Proposition~\ref{prop:shell} is not needed here. On $\mathcal G_\theta^c$, $c\|\Rm\|^2\le K_G\ell^{7+D}$ and the probability is at most $2\ell^{-\kappa_p}$ with $\kappa_p>D+8$.
\end{proof}

\subsection{Proof of Theorem~\ref{thm:main}}
The estimator $\hat X:=\Xw$ belongs to $\mathscr C$ (Section~\ref{sec:prelim}). By Lemma~\ref{lem:consumer} and Proposition~\ref{prop:expect},
\[
\frac ch\le c\,\delta+\langle\Gamma_{\hat X}\rangle\le C_\kappa\,\ell^{\kappa-1}(\log\lambda)^2\qquad(\lambda\ge\lambda_1(\kappa)),
\]
where we used $\ell^{-1}\le\ell^{\kappa-1}$ and $\log\lambda\ge1$ (enlarging $\lambda_1$ if necessary), and $C_\kappa$, $\lambda_1(\kappa)$ do not depend on $\lambda$ (Remark~\ref{rem:order}). Since $\kappa<1$, the right-hand side is bounded (in fact it tends to zero), which already gives the linear bound $h\ge\lambda/C$ for large $\lambda$. For the full statement fix $\epsilon>0$ and put $\kappa:=\min(7\epsilon/2,\frac12)$; every $0<\kappa<1$ is admissible in Definition~\ref{def:Xw} and Propositions~\ref{prop:shell} and~\ref{prop:expect}. With $c=\lambda/\pi$ and $\ell=(\lambda/(\pi\abar))^{1/7}$,
\[
h(\lambda)\ge\frac{\lambda\,\ell^{1-\kappa}}{\pi C_\kappa(\log\lambda)^2}=\frac{(\pi\abar)^{-(1-\kappa)/7}}{\pi C_\kappa}\,\frac{\lambda^{8/7-\kappa/7}}{(\log\lambda)^2}\qquad(\lambda\ge\lambda_1(\kappa)).
\]
As $\kappa/7\le\epsilon/2$, the right-hand side is at least $\lambda^{8/7-\epsilon}$ once $\lambda^{\epsilon/2}(\log\lambda)^{-2}$ exceeds a constant depending only on $\alpha$ and $\epsilon$; this defines $\lambda_\epsilon\ge\lambda_1(\kappa)$. In particular $h(\lambda)/\lambda\to\infty$, so $h(\lambda)\ge\lambda$ for $\lambda\ge\lambda_*$ with some $\lambda_*\ge1$, and for $\lambda<\lambda_*$, $h\ge1\ge\lambda/\lambda_*$. Hence $\iPF(\alpha)\ge\lambda_*^{-1}>0$. \qed

Corollary~\ref{cor:LL} was proved in Section~\ref{sec:model}.

\begin{remark}[Order of choice of constants]\label{rem:order}
Fix $\alpha$ and $\kappa$, and choose the design constants ($C_b$, $c_\tau$, $\theta_\delta$, $\eta_p$, the constants of the surgery of Section~\ref{sec:surgery}, $\delta$). The constants of the core locator are chosen in the order of (D$'$) and Lemma~\ref{lem:design} (thresholds, then the auxiliary scale $L$, then $t_*$, then the side constant $K_e$ of the electron box, then $R_c,\Mzero,W_*,M^w_*$), and the constants $\eta_S,K_n$ of the window-edge estimate (Appendix~\ref{app:core}) after them. These determine the degree $D$ of Proposition~\ref{prop:global} and the polynomial degrees in Proposition~\ref{prop:shell}, hence an upper bound for the required power $\kappa_p^*$. Next choose the window constants ($K$ of Proposition~\ref{prop:events} and $K_d$) for $\kappa_p^*$; this is possible for every $\kappa_p^*$ because the exponents are proportional to the window lengths. The window constants enlarge the constant in $\rho$, the constant in Proposition~\ref{prop:expect} (which is $\propto K_d^2$) and $\lambda_1$, but do not enter as powers of $\ell$; in particular they do not change the polynomial degrees fixed before them. Throughout Part~II, $A\lesssim B$ means $A\le CB$ with $C$ depending only on $\alpha$, $\kappa$, the design constants and the window constants, and not on $\lambda$ (equivalently on $\ell$); $A\asymp B$ means $A\lesssim B\lesssim A$. Every dependence on $\lambda$ is displayed as a power of $\ell$ or of $\log\lambda$.
\end{remark}

Appendix~\ref{app:terms} records the growing factors and the probability estimates used to absorb them.

\begin{remark}[Exponent budget]\label{rem:budget}
The following coarse choices realize the order of Remark~\ref{rem:order}; none is optimized. With the edge width $w=R_c$ of Definition~\ref{def:core}, Proposition~\ref{prop:global} holds with $D=2$: the bounds of Appendix~\ref{app:core} give $\|\Rm(\Xw)\|^2\lesssim(\log\lambda)^{14}$ and $\Gamma_{\Xw}\lesssim\ell(\log\lambda)^{11}$, and $\log\lambda=7\log\ell+O(1)$. In Table~\ref{tab:budget}, $C$ may depend on all constants, including the window constants, and $\lambda_1$ is chosen last. The column ``required power'' lists the target powers $\kappa_p^*$, $p_{\mathrm{bad}}$; in the column ``probability bound'', $\kappa_p$ is the auxiliary exponent $F_{\max}=\kappa_p\log\ell$ of the Agmon function in the proof of the respective proposition, chosen in terms of the target power. The factor $C\ell^7$ collects the terms of Step~8 of the proof of Proposition~\ref{prop:shell} (the trace terms of the $N_{\mathrm{cand}}$ candidate pairs and $\langle(c_0-\bar V)_+\rangle$, divided by $2\mAg$), and $\Pi''$ is the prefactor defined in Lemma~\ref{lem:edge} (with the candidate number $N'$ and the margin $\mAg\asymp\ell^5$ of that lemma).
\begin{center}\small
\captionof{table}{Exponent budget for the bad events (Remark~\ref{rem:budget}).}\label{tab:budget}
\begin{tabular}{p{0.15\textwidth}p{0.17\textwidth}p{0.27\textwidth}p{0.17\textwidth}p{0.14\textwidth}}
event & polynomial factor & probability bound & required power & fixed before\\\hline
$\mathcal G'^c$ (Prop.~\ref{prop:expect}) & $K_G\ell^{D}$, $cK_G\ell^{D}$ & each event $\le\ell^{-\kappa_p^*}$ & $\kappa_p^*:=D+9=11$ & design, (D$'$), $\eta_S$, $K_n$, $w$\\
$\Gamma'_\rho$ (Prop.~\ref{prop:shell}) & $C\ell^{7}$ (Steps 7--8) & $C\ell^{7-2\kappa_p}+C\ell^{7}P_\psi(S_M\cup\mathcal B_\theta)+{}$tails & $p_{\mathrm{bad}}:=\kappa_p^*+8$ for $S_M$, $\mathcal B_\theta$ & $\kappa_p^*$, $F_{\max}$, $K_{\rm t}$\\
$S_M$ (Prop.~\ref{prop:SMexpl}) & $C\ell$ & $2\ell^{-2\kappa_p}+C\ell\,P_\psi(\Gamma'_R)$ & $p_{\mathrm{bad}}+2$ for $\Gamma'_R\subset\mathcal B_\theta$ & $p_{\mathrm{bad}}$\\
$\mathcal B_\theta$ (Prop.~\ref{prop:events}) & $\Pi''\le C\ell^{6}$ & $C\ell^{6}\lambda^{-c_{\mathrm{ev}}K}$ & $7c_{\mathrm{ev}}K\ge p_{\mathrm{bad}}+9$ & $b_{r'}^2$\\
\end{tabular}
\end{center}
The thresholds $b_{r'}^2$ of Lemma~\ref{lem:edge}, the tail constant $K_{\rm t}$ and $F_{\max}=\kappa_p\log\ell$ depend on the target powers, but they enter only through ratios to the margins that tend to zero for $\lambda\ge\lambda_1$, and they change neither $D$ nor $c_{\mathrm{ev}}$. The rate is $c_{\mathrm{ev}}=2\frac{a-1}a\sqrt{2\eta c_1/\abar}$ for a lower bound $\mu'_1\ge c_1\ell^5$ with $c_1$ depending only on the design and on $\eta_S,K_n$ (all chosen before $K$); it holds for $\lambda\ge\lambda_1(K)$ since $S'_{\mathrm{part}}/\mu(g_s)\lesssim\ell^{-2}\log^2(K\ell)$. Hence $c_{\mathrm{ev}}$ does not depend on $K$, and $K\ge(p_{\mathrm{bad}}+9)/(7c_{\mathrm{ev}})$, $K_d=K+1$ is an admissible choice of the window constants.
\end{remark}

\section{Localization of the depression mass; proof of Proposition~\ref{prop:SM}}\label{sec:constructive}

Sections~\ref{sec:shell}--\ref{sec:proof} use the core good set only through the \emph{mass} of the depression set: the hypothesis $W_{\mathrm{ms}}\ge W_*$ of Proposition~\ref{prop:Mprime}(a), the threshold $W_{\mathrm{ms}}\ge2W_*$ of the lifted denominator behind Proposition~\ref{prop:Mprime}(b), and $M\ge2M^w_*$ in Remark~\ref{rem:Mw}. The position of the bubble is needed only up to the coarse scale $\rho/2\asymp\ell\log\lambda$. In this section we show that almost minimizing configurations carry a depression of mass $\asymp R_*^3$ near the region where the electron density concentrates, and deduce Proposition~\ref{prop:SM}.

The difficulty is that the input from \cite{SerA} (Proposition~\ref{prop:MF}) controls the field only block by block, through the deficits $\Delta_Q$, while it is the electron density, not the field, that concentrates. The transfer from the concentration of the density to the mass of the field depression (Section~\ref{ssec:transfer}) is the core of the argument. No shape information is needed, so we do not use concentration compactness \cite{Li84}: the electron is already confined to a box of side $O(\ell)$, which plays the role of the concentration cube. All statements are elementary inequalities, and the numerical conditions are met by choosing the constants in a fixed order of limits (Lemma~\ref{lem:design}). We do not track the size of $\lambda_1$.

\subsection{Lemmas for the mean-field functional}
Both terms of $\mathcal E_1[u]=\frac12\|\nabla u\|^2+\|u\|_1$ (Proposition~\ref{prop:MF}) are nonnegative, so $\mathcal E_1[\varphi]\le(1+t)F[1]$ implies $\|\nabla\varphi\|^2\le2(1+t)F[1]$ and $\|\varphi\|_1\le(1+t)F[1]$.

\begin{lemma}[Transfer to cubes]\label{lem:cubes}
Divide $\R^3$ into cubes $B$ of side $L$, and let $Q_c$ be the block of side $s/R_*$ (in units of $R_*$) centred at $c\in\R^3$. Let $\bar\varphi_B:=|B|^{-1}\int_B\varphi$, $\bar\rho_Q:=|Q|^{-1}\int_Q\varphi^2$, $\Tgrad :=\|\nabla\varphi\|^2$, and $0<\sqrt{s_1}<\bar a$. (i) $\sum_{B:\bar\varphi_B<\bar a}\int_B\varphi^2\le(L/\pi)^2\Tgrad +\bar a\|\varphi\|_1$. (ii) The centres $c$ with $\bar\rho_{Q_c}<s_1$ and $Q_c\subset B$ for a cube $B$ with $\bar\varphi_B\ge \bar a$ form a set of volume at most $(L/\pi)^2\Tgrad /(\bar a-\sqrt{s_1})^2$. (iii) For measurable $A$, $\int_A\varphi^2\le S_3^{-1}\Tgrad |A|^{2/3}$ with $S_3=3(\pi/2)^{4/3}$.
\end{lemma}
\begin{proof}
We use the Neumann Poincar\'e inequality $\int_B|\varphi-\bar\varphi_B|^2\le(L/\pi)^2\int_B|\nabla\varphi|^2$. (i) $\int_B\varphi^2=\int_B(\varphi-\bar\varphi_B)^2+\bar\varphi_B\int_B\varphi$. (ii) For such $c$, by Cauchy--Schwarz $\bar\varphi_{Q_c}\le\sqrt{\bar\rho_{Q_c}}<\sqrt{s_1}$, so $\int_{Q_c}(\varphi-\bar\varphi_B)^2\ge|Q|(\bar\varphi_B-\bar\varphi_{Q_c})^2\ge|Q|(\bar a-\sqrt{s_1})^2$. Integrate over these $c$: each $y\in B$ lies in $Q_c$ for a set of $c$ of volume $|Q|$, so by Fubini the integral of the left-hand side is at most $|Q|\int_B(\varphi-\bar\varphi_B)^2$; sum over $B$. (iii) H\"older's inequality and Sobolev's inequality with the sharp constant \cite[Theorem~8.3]{LL01}.
\end{proof}

\begin{lemma}[Block deficit]\label{lem:AMGM}
For $h,u,x>0$, $\frac{h^2}{8u}+xu=h\sqrt{x/2}+\bigl(\frac h{\sqrt{8u}}-\sqrt{xu}\bigr)^2$. With a threshold $u_s$ and $x_s:=h^2/(8u_s^2)$: (a) if $u\le u_s$ and $x\le x_s/4$, the deficit is at least $h^2/(32u_s)$; (b) if $u\ge u_s$ and $x\ge(1+\eta)x_s$, it is at least $\frac{h^2}{8u_s}(\sqrt{1+\eta}-1)^2$.
\end{lemma}
\begin{proof}
The identity is expansion. (a) $h/\sqrt{8u}\ge h/\sqrt{8u_s}$ and $\sqrt{xu}\le\sqrt{x_su_s/4}=\frac12h/\sqrt{8u_s}$. (b) $h/\sqrt{8u}=u_s\sqrt{x_s/u}$, so the difference is $\sqrt{x_s}(\sqrt{(1+\eta)u}-u_s/\sqrt u)$, increasing in $u$, equal to $\sqrt{x_su_s}(\sqrt{1+\eta}-1)$ at $u=u_s$.
\end{proof}
In Proposition~\ref{prop:MF}(c), $x$ is proportional to the electron density of the block and $u$ is the field level of the block. On the low branch ($u'_Q$ below the field ceiling $u_b$, or $\bar\rho_Q=0$) the deficit $\Delta_Q$ is the deficit of Lemma~\ref{lem:AMGM}; on the high branch it is the deficit $\Delta^{\mathrm{hi}}_Q$ of Appendix~\ref{app:core}, which is not of this form but satisfies the same two lower bounds (a) and (b) (for (b) when $\eta\le3+2\sqrt3$). Lemma~\ref{lem:avg} below is written with a single block constant $h$: on $\mathbb T_L$ all integer cubes have the same $h_Q$ (proof of \cite[\SerAnum{cor:hardy}(c)]{SerA}), and by \cite[\SerAnum{prop:configBO}(iv)]{SerA} so do the translated lattices. If the $h_Q$ differed, it would suffice to define the density threshold of (a) and the constants $e'_b$, $C^{(a)}_\Delta$ with $\min_Qh_Q$ and the density threshold of (b) with $\max_Qh_Q$; these are comparable by \cite[\SerAnum{prop:configBO}(ii)]{SerA}.

\subsection{Transfer to depressions}\label{ssec:transfer}
Measure lengths in units of the bubble radius $R_*\asymp\ell$, and let $\varphi(y):=R_*^{3/2}|\psi|(R_*y)$ for an electronic wave function $\psi$. Blocks are taken at continuous positions ($Q_c$ centred at $c$). For block thresholds $u'_s<u_s$ let $D^-:=\{c:u'_{Q_c}\le u'_s\}$ and $D^+:=\{c:u'_{Q_c}\le u_s\}$ be the sets of block centres at which the block itself is below the threshold, with corresponding density thresholds $\sigma_s$ and $\sigma'_s:=\kappa_s\sigma_s$, $\kappa_s=(u_s/u'_s)^2$; fractions are averages over block centres, i.e.\ over all translates of the block lattice.

\begin{lemma}[Averaged deficits]\label{lem:avg}
Assume Proposition~\ref{prop:MF}(c) for each translate $t$ and $\varepsilon_1+\Eloss/F\le t_*/4$, $t_*\le1$. If $\bar J_q(\psi):=\mathrm{avg}_tJ^{(t)}_q(\psi)\le(1+t_*)F$ for one $\psi$, then $\mathcal E_1[\varphi]\le(1+2t_*)F[1]$ and $\mathrm{avg}_t\sum_{Q\in\mathcal P_t}\Delta_Q\le2t_*F$. In particular, the set of centres $c$ with $\bar\rho_{Q_c}\ge2\sigma'_s$ and $c\notin D^-$ has volume at most $C^{(b)}_\Delta t_*$, and the set of centres with $\bar\rho_{Q_c}\le\sigma_s/4$ and $c\in D^+$ has volume at most $C^{(a)}_\Delta t_*$, where $C^{(b)}_\Delta:=2F|Q|/e'_b$ with $e'_b:=\frac{h^2}{8u'_s}(\sqrt2-1)^2$ and $C^{(a)}_\Delta:=64Fu_s|Q|/h^2$; these are independent of $\ell$ (volumes in units of $R_*^3$).
\end{lemma}
\begin{proof}
Since $\varphi$ does not depend on $t$, averaging the inequality of Proposition~\ref{prop:MF}(c) gives $(1-\varepsilon_1)F\mathcal E_1[\varphi]/F[1]+\mathrm{avg}_t\sum\Delta_Q-\Eloss\le(1+t_*)F$; with $\mathcal E_1\ge F[1]$ and $\Delta\ge0$ both claims follow. Further $\mathrm{avg}_t\sum_{Q\in\mathcal P_t}\Delta_Q=s^{-3}\int\Delta_{Q_c}\dd c$, and the bounds (b) (threshold $u'_s$, $\eta=1$) and (a) (threshold $u_s$) of Lemma~\ref{lem:AMGM}, valid on both branches by Proposition~\ref{prop:MF}(c) ($\eta=1\le3+2\sqrt3$), give the volume bounds. That $F/e'_b$ and $Fu_s/h^2$ are of order $R_*^3/|Q|$ follows from the mean-field scaling of \cite{SerA}.
\end{proof}

We quote two properties relating the core indicator $d$ to the thresholds (Lemma~\ref{lem:T1T2}): (T1) there is a design constant $p_d>0$ such that $d=1$ at points where the kernel-weighted fraction of $(D^-)^c$ is at most $p_d$; (T2) there is $q_d>0$ such that at points with $d(y)>0$ the kernel-weighted fraction of $D^+$ is at least $q_d$. For the level-$\kappa$ indicator we prove the analogue.

\begin{lemma}[Level-$\kappa$ indicator]\label{lem:levelw}
Let $\theta_wg_w$ ($\theta_w=\frac34$) be the end of the plateau of $\tilde\chi_{g_w}$ and assume $u'_s/|Q|\le\theta_wg_w/2$. If $d_w(y)<1$, the kernel-weighted fraction of $(D^-)^c$ at $y$ is at least $f_w:=\theta_w/(2c_\tau)$.
\end{lemma}
\begin{proof}
$\sat_\tau(u)\le\min(u,\tau)$. On blocks $Q_c$ with $c\in D^-$, $\sat_{\tau_w}(u'_Q/|Q|)\le u'_s/|Q|\le\theta_wg_w/2$; elsewhere $\sat_{\tau_w}\le\tau_w=c_\tau g_w$. Thus, with $\mathrm{frac}^w(y)$ the fraction, $g^{(\tau_w)}_b(y)\le\theta_wg_w/2+\mathrm{frac}^w(y)c_\tau g_w$. If $\mathrm{frac}^w(y)<f_w$ then $g^{(\tau_w)}_b(y)<\theta_wg_w$ and $d_w(y)=1$.
\end{proof}
The assumption $u'_s/|Q|\le\theta_wg_w/2$ holds for large $\ell$, since $u'_s/|Q|\lesssim g_s$ and $g_s/g_w=\ell^{\kappa-2}$.

\subsection{The volume of the depression and the choice of constants}
The thresholds $u'_s,u_s,\sigma_s$ (and with them $\kappa_s$, $p_d,q_d,f_w,C^{(a)}_\Delta,C^{(b)}_\Delta$) are fixed first. Put $s_1:=2\sigma'_s$, $\bar a:=2\sqrt{s_1}$, and require
\[
\text{(L0)}\qquad \bar a\,F[1]<\tfrac13 ,
\]
which holds for $\sigma_s$ small. The auxiliary scale $L$ (Lemma~\ref{lem:cubes}) is a device of the proof; it enters the estimators only through the constants $b_\theta$, $R_c$, $\Mzero$, $W_*$, $M^w_*$ fixed below. The core cube side is $b_\theta:=c_bLR_*/(2\sqrt3)$, with $c_b$ such that $(1-c_b)^3\ge\frac12$; then $r_b:=\sqrt3(b_\theta+s)/R_*\to c_bL/2$ as $\ell\to\infty$. Let $\Tgradmax:=2(1+2t_*)F[1]$, so that $\Tgrad\le\Tgradmax$ and $\|\varphi\|_1\le\frac12\Tgradmax$ under the conclusion of Lemma~\ref{lem:avg}, let $r_w:=\sqrt3(b+s)/R_*$, and
\[
\mu_{\mathrm{hi}}:=1-\bigl[(L/\pi)^2+\tfrac12\bar a\bigr]\Tgradmax,\quad V_{\mathrm{hi}}:=\Bigl(\frac{\mu_{\mathrm{hi}}S_3}{\Tgradmax}\Bigr)^{3/2},\quad N_H:=\frac{(L/\pi)^2\Tgradmax}{s_1}+C^{(b)}_\Delta t_*,
\]
\[
V_d:=\Bigl(1-\frac{2r_b}L\Bigr)^3V_{\mathrm{hi}}-\frac{N_H}{p_d},\qquad V^w_d:=\Bigl(1-\frac{2r_w}L\Bigr)^3V_{\mathrm{hi}}-\frac{N_H}{f_w},\qquad V_{\mathrm{far}}:=\frac{C^{(a)}_\Delta t_*}{q_d}.
\]
Their limits as $t_*\to0$ and $s/R_*\to0$ are, with $\mu_0(L):=1-(L/\pi)^2\,2F[1]-F[1]\,\bar a$,
\begin{gather*}
V_d^{(0)}(L):=(1-c_b)^3\Bigl(\frac{\mu_0(L)S_3}{2F[1]}\Bigr)^{3/2}-\frac{(L/\pi)^2\,2F[1]}{s_1p_d},\\
V^{w,(0)}_d(L):=\Bigl(\frac{\mu_0(L)S_3}{2F[1]}\Bigr)^{3/2}-\frac{(L/\pi)^2\,2F[1]}{s_1f_w}.
\end{gather*}

\begin{lemma}[Choice of the auxiliary constants]\label{lem:design}
Assume (L0). There is $L_0>0$ such that for $0<L\le L_0$
\[
V_\flat:=\tfrac12V_d^{(0)}(L)>0,\qquad V^w_\flat:=\tfrac12V^{w,(0)}_d(L)>0,\qquad \mu_0(L)\ge\tfrac12 .
\]
Fix such $L$. Then there is $\bar t(L)>0$, and for $0<t_*\le\bar t(L)$ there is $\ell_0$ such that for $\ell\ge\ell_0$: $\mu_{\mathrm{hi}}>0$, $V_d\ge V_\flat$, $V^w_d\ge V^w_\flat$, $V_{\mathrm{far}}\le V_\flat/8$ and $r_b\le c_bL$.
\end{lemma}
\begin{proof}
By (L0), $\mu_0(L)\ge\frac23-(L/\pi)^22F[1]$, and $(1-c_b)^3\ge\frac12$; hence $V_d^{(0)}(L)\ge\frac12(\frac12S_3/(2F[1]))^{3/2}-CL^2/(s_1p_d)>0$ for small $L$; likewise $V^{w,(0)}_d(L)>0$ and $\mu_0(L)\ge\frac12$ for small $L$. All quantities are continuous in $(t_*,s/R_*)$ at $(0,0)$: $\Tgradmax\to2F[1]$, $C^{(b)}_\Delta t_*\to0$, $r_w\to0$, $2r_b/L\to c_b$, and $V_{\mathrm{far}}\to0$. Choose $t_*$ small, then $\ell$ large.
\end{proof}

\medskip\noindent\textbf{Design (D$'$).} With the thresholds, $L$ and $t_*$ fixed as in Lemma~\ref{lem:design}, let $\kappa_\pm$ be constants with $\kappa_-\ell\le R_*\le\kappa_+\ell$, choose $K_e\ge4\kappa_+\sqrt{C_{\mathrm{IMS}}/(t_*F[1])}$ (this fixes the side $n_e:=2s\lceil K_e\ell/(2s)\rceil$ of the electron box in Appendix~\ref{app:core}) and put $\bar K_e:=2K_e/\kappa_-$, so that $n_e/R_*\le(K_e\ell+2s)/(\kappa_-\ell)\le\bar K_e$ once $2s\le K_e\ell$, i.e.\ for large $\ell$. Let $\mathsf C''$ be the cube centred at the electron position with side $2\bar K_e+2(1+c_b)L$; it contains the rescaled electron box $\mathcal C_e/R_*$ with the margin $(1+c_b)L$ (which contains the margins $L$ and $r_b$ used in the proof, as $r_b\le c_bL$). Let $D_{\mathsf C}:=\diam\mathsf C''=\sqrt3\,(2\bar K_e+2(1+c_b)L)$, and set $R_c:=D_{\mathsf C}$, $\Mzero:=V_\flat/(4|B_{R_c}|)$, $W_*:=\frac18\mathrm e^{4\Mzero}|B_{R_c/2}|$, $M^w_*:=(R_*/b)^3V^w_\flat/8$ (the integral defining $M$ carries the factor $b^{-3}$, so $M^w_*\asymp\ell^3/b^3$), and the switching levels $a_c:=W_*$, $a_w:=M^w_*$ of the denominators in Definitions~\ref{def:core} and~\ref{def:Xw}. Here $D_{\mathsf C}$ and $R_c$ are lengths in units of $R_*$, and $V_\flat$, $V^w_\flat$, $|B_{R_c}|$, $W_*$ are volumes in units of $R_*^3$; in these units all these constants are independent of $\ell$, and the number $M^w_*$ carries the factor $(R_*/b)^3$. In absolute units (Definition~\ref{def:core}, Appendix~\ref{app:core}) the radius is $R_cR_*\asymp\ell$ and $W_*\asymp\ell^3$; Proposition~\ref{prop:massloc} writes the factor $R_*$ explicitly next to the absolute radii $r_W,w,\rho$. The constants $\eta_S,K_n$ of Lemma~\ref{lem:edge} and the window constants of Proposition~\ref{prop:events} are chosen after $L$ (they depend on $b_\theta/R_*$), and $\ell$ (equivalently $\lambda$) last. In this order no constant is needed before it is chosen.

\subsection{Localization of the depression mass}
\begin{proposition}[Mass localization]\label{prop:massloc}
Fix the design (D$'$). There is $\ell_*$ such that for $\ell\ge\ell_*$ the following holds. Let $q$ be a configuration, $x\in\mathcal C_e$ an electron position and $\psi\in H^1_0(\mathcal C_e)$ normalized, where $\mathcal C_e$ is the electron box of Proposition~\ref{prop:MF}(b) and the blocks entering $\bar J_q$ are those of $\mathcal C_e^+\supset B(x,r_W+R_cR_*+2(b_\theta+s))$ (Appendix~\ref{app:core}; the margin $2(b_\theta+s)$ contains the support of the core kernel). Assume $\bar J_q(\psi)\le(1+t_*)F$. Then, with $X_c$, $W_{\mathrm{ms}}$, $M$ computed from the field seen from $x$:
\begin{itemize}
\item[(a)] $W_{\mathrm{ms}}\ge8W_*$;
\item[(b)] $|X_c|\le(R_c+D_{\mathsf C}/2)R_*$;
\item[(c)] if $\rho/2\ge(R_c+D_{\mathsf C})R_*$, then $M\ge8M^w_*$ for $M$ of Definition~\ref{def:Xw}.
\end{itemize}
\end{proposition}
The conditions defining $\ell_*$ are: $\ell\ge\ell_0$ of Lemma~\ref{lem:design}; $\varepsilon_1+\Eloss/F\le t_*/4$ (possible since $\varepsilon_1+\Eloss/F\to0$); $u'_s/|Q|\le\theta_wg_w/2$; $n_e/R_*\le\bar K_e$; and $r_W-w\ge(D_{\mathsf C}+R_c)R_*$ (this puts the relevant part of the window where $\chi_W=1$; it holds for large $\ell$ since $r_W\asymp\ell\log\lambda$ and $R_*\asymp\ell$).
\begin{proof}
\emph{Step 0.} Lemma~\ref{lem:avg}: $\mathcal E_1[\varphi]\le(1+2t_*)F[1]$ and $\mathrm{avg}_t\sum\Delta_Q\le2t_*F$. \emph{Step 1.} Since $\psi\in H^1_0(\mathcal C_e)$ and $x\in\mathcal C_e$, the support of $\varphi$ (relative to $x$) lies in $\mathcal C_e/R_*\subset[-\bar K_e,\bar K_e]^3$, hence in $\mathsf C''$ at distance at least $(1+c_b)L$ from its boundary; no concentration argument is needed, since the electron is already confined to a box of side $O(\ell)$. \emph{Step 2.} Let $\mathcal H$ be the cubes of side $L$ with $\bar\varphi_B\ge \bar a$; each meets the support of $\varphi$, hence lies in $\mathsf C''$. Lemma~\ref{lem:cubes}(i) and $\|\varphi\|_1\le\frac12\Tgradmax$ give $\int_{\cup\mathcal H}\varphi^2\ge\mu_{\mathrm{hi}}$, and (iii) gives $|\cup\mathcal H|\ge V_{\mathrm{hi}}$. \emph{Step 3.} Centres $c\notin D^-$ with $Q_c\subset B$ for some $B\in\mathcal H$ either have $\bar\rho_{Q_c}<s_1$ (volume $\le(L/\pi)^2\Tgradmax/s_1$ by Lemma~\ref{lem:cubes}(ii)) or carry the first deficit of Lemma~\ref{lem:avg} (volume $\le C^{(b)}_\Delta t_*$); in total at most $N_H$. \emph{Step 4.} At points of $B\in\mathcal H$ at distance $\ge r_b$ from $\partial B$ the core kernel charges only centres $c$ with $Q_c\subset B$, since the blocks it charges lie in the cube of half-side $(b_\theta+s)/R_*\le r_b$ around the point (likewise with $r_w$ for the level-$\kappa$ kernel). By (T1), Fubini (the kernel is normalized) and Markov, points where the fraction exceeds $p_d$ have volume at most $N_H/p_d$; hence $\int_{\mathsf C''}d\ge V_d\ge V_\flat$. Likewise Lemma~\ref{lem:levelw} gives $\int_{\mathsf C''}d_w\ge V^w_d\ge V^w_\flat$. \emph{Step 5.} Only the indicator $d$ on $B(x,r_W+R_cR_*)$ enters (a)--(c); it is computed from blocks inside $B(x,r_W+R_cR_*+2(b_\theta+s))\subset\mathcal C_e^+$. For $y\in B(x,r_W+R_cR_*)\setminus\mathsf C''$ the cube of the core kernel at $y$ has half-side $(b_\theta+s)/R_*<(1+c_b)L$, so all its blocks miss the support of $\varphi$ and have $\bar\rho_Q=0\le\sigma_s/4$; hence the centres in $D^+$ there have volume $\le C^{(a)}_\Delta t_*$ by Lemma~\ref{lem:avg} (blocks without electron density included, since their deficits satisfy bound (a)). By (T2) and Markov, $\int_{B(x,r_W+R_cR_*)\setminus \mathsf C''}d\le V_{\mathrm{far}}\le V_\flat/8$. \emph{Step 6.} Since $\mathsf C''\subset B(0,D_{\mathsf C}/2)$ (relative to $x$) and $R_c=D_{\mathsf C}$, if $|Y|\le R_c/2$ then $B(Y,R_c)\supset \mathsf C''$, so $\mathcal M(Y)\ge V_\flat/|B_{R_c}|=4\Mzero$ and $\phi\ge \mathrm e^{4\Mzero}$; these $Y$ lie where $\chi_W=1$ by the condition on $r_W$, which gives (a). If $|Y|>R_c+D_{\mathsf C}/2$ and $|Y|\le r_W$ then $B(Y,R_c)$ misses $\mathsf C''$ and lies in $B(x,r_W+R_cR_*)$, so $\mathcal M(Y)\le V_{\mathrm{far}}/|B_{R_c}|<\Mzero$ and $\phi=0$. Since $W_{\mathrm{ms}}\ge8W_*\ge2a_c$, the denominator of $X_c$ is $W_{\mathrm{ms}}$, and $X_c$ is a weighted average of points of $B(0,R_c+D_{\mathsf C}/2)$; this is (b). Finally, under the condition of (c) the ball $B(X_c,\rho/2)$, where the window of $M$ equals one, contains $\mathsf C''$, and Step~4 gives $M\ge(R_*/b)^3V^w_\flat=8M^w_*$.
\end{proof}

Since $R_c$ covers the whole bubble, $X_c$ locates the bubble only up to a constant times $R_*$; for Sections~\ref{sec:shell}--\ref{sec:proof} a window $\rho/2\asymp\ell\log\lambda$ suffices.

\begin{proposition}[Mass event]\label{prop:SMexpl}
Under (D$'$), for every $\kappa_p$ and all window constants of Proposition~\ref{prop:events} at least a threshold $K_0(\kappa_p)$ there is $\lambda_1$ such that $P_\psi(S_M)\le\ell^{-\kappa_p}$ for $\lambda\ge\lambda_1$, with $S_M=\{W_{\mathrm{ms}}<2W_*\}\cup\{M<2M^w_*\}$ as in Definition~\ref{def:good}. In particular Proposition~\ref{prop:SM} holds.
\end{proposition}
\begin{proof}
Let $\sigma$ be smooth and increasing, $\sigma=0$ on $[0,1]$, $\sigma=1$ on $[2,\infty)$, and put $\Sigma_k:=\sigma(W_{\mathrm{ms}}/kW_*)\,\sigma(M/kM^w_*)$ for $k=2,4$.

\emph{Step 1 (deterministic).} Suppose $\bar{\mathcal U}(q)\le(1+t_*/2)F-c_{\mathrm{BO}}$. Then some $\psi$ has $\bar J_q(\psi)\le(1+t_*)F$, and Proposition~\ref{prop:massloc} gives $W_{\mathrm{ms}}\ge8W_*$ and, since $\rho/2\ge(R_c+D_{\mathsf C})R_*$ for $\lambda\ge\lambda_1$, $M\ge8M^w_*$; hence $\Sigma_4=1$. Contrapositively, $\Xi_0:=1-\Sigma_4>0$ implies $\bar{\mathcal U}(q)>(1+t_*/2)F-c_{\mathrm{BO}}$.

\emph{Step 2 (operator bound).} Apply \eqref{eq:630P} with $\Xi_0$, which is Lipschitz in the field with a polynomial constant and depends on the field in a ball of radius $\rho_\Xi\lesssim\ell\log\lambda$. Its hypotheses hold by Proposition~\ref{prop:MF}(c) (lower bound $(1-\varepsilon_1)F$ for all $q$) and by Step~1. Hence $H(0)\ge(1-\varepsilon')F+\frac{t_*}2F\,\Xi_0$ with $\varepsilon':=\varepsilon_1+(\Eloss+C_{\mathrm{IMS}}c/n_e^2)/F\le t_*/8$ for large $\ell$ (here $c_{\mathrm{BO}}=\Eloss$, $\varepsilon_1+\Eloss/F\to0$, and $C_{\mathrm{IMS}}c/n_e^2\le t_*F/16$ for all $\ell$ by the choice of $K_e$ in (D$'$), since $F=cF[1]/R_*^2$, Appendix~\ref{app:core}).

\emph{Step 3 (Agmon).} Let $f:=F_{\max}(1-\Sigma_2)$, $F_{\max}=\kappa_p\log\ell$. Then $S_M\subset\{f=F_{\max}\}$, and on $\mathcal Z=\{f>0\}\cup\{\GAg_f>0\}$ we have $\Sigma_2<1$, hence $\Sigma_4=0$ and $\Xi_0=1$. The upper bound $E\le(1+r_{\mathrm{MF}})F$ of \cite{SerA} gives $V-E\ge\frac{t_*}4F$ on $\mathcal Z$ once $\varepsilon'+r_{\mathrm{MF}}\le t_*/4$. The derivatives of $f$ are non-zero only where both masses exceed twice their thresholds. There, outside the window-edge event $\Gamma'_R$, the translation derivatives of $W_{\mathrm{ms}}$ and $M$ vanish ($\Rm(X_c)=0$ there, Appendix~\ref{app:core}), so $c|Df|^2=0$. The carr\'e du champ of each mass divided by its threshold squared is at most $\ell^{-1}$ times a power of $\log\lambda$ (Schur's test as in Appendix~\ref{app:core} and Lemma~\ref{lem:Mw}), so $\Gamma_f\ll t_*F$. In detail: with the prefactor $(b_\theta+s)^3\bar C_g\|\tilde\chi'\|^2_\infty\lesssim\ell^{3-5+4}$ of Appendix~\ref{app:core} and the window volume $\lesssim(\ell\log\lambda)^3$, one gets $\Gamma_{W_{\mathrm{ms}}}/W_*^2\lesssim\ell^{2}(\ell\log\lambda)^3/\ell^6=\ell^{-1}(\log\lambda)^3$; for $M$, the part through $d_w$ is $\lesssim\ell^{\kappa-3}(\log\lambda)^3$, and the part through $X_c$ (not removed here, since $\Gamma'_\rho$ is not excluded) is $\lesssim(\rho^2/b^3)^2\,\Gamma_{X_c}/(M^w_*)^2\lesssim(\rho^2/b^3)^2\,\ell(\log\lambda)^5/(\ell^3/b^3)^2\asymp\ell^{-1}(\log\lambda)^9$ by Proposition~\ref{prop:Mprime}(a) and $\rho\asymp\ell\log\lambda$. Hence on $\mathcal Z$, $V-E-\frac12\GAg_f\ge\frac{t_*}8F-\Lexc$ with $\Lexc=\frac12c|Df|^2$ supported on $\Gamma'_R$ and bounded there by $CF(\log\lambda)^{14}$: the translation derivatives are $DW_{\mathrm{ms}}=\int\nabla\chi_W\,\phi\dd Y$ and $DM=\int\chi_\rho'(|Y-X_c|)\,u^{\mathsf T}\Rm(X_c)\,d_w\dd Y/b^3$ ($u:=(Y-X_c)/|Y-X_c|$), so $|DW_{\mathrm{ms}}|/W_*\lesssim r_W^2/\ell^3$ and $|DM|/M^w_*\lesssim\rho^2\|\Rm(X_c)\|/\ell^3$, with $\|\Rm(X_c)\|\lesssim(\log\lambda)^4$ for the edge width $w=R_c$ (Appendix~\ref{app:core}). Lemma~\ref{lem:agmon} with $F=F_{\max}$ gives
\[
P_\psi(S_M)\le \mathrm e^{-2F_{\max}}\frac{\langle(E-V)_+\rangle}{t_*F/8}+\frac{\mathrm{poly}(\ell)}{t_*F/8}P_\psi(\Gamma'_R),
\]
and $\langle(E-V)_+\rangle\le(\varepsilon'+r_{\mathrm{MF}})F$. The first term is at most $2\ell^{-2\kappa_p}$ and the second is at most any power of $\ell$ by Proposition~\ref{prop:events}, whose proof (Appendix~\ref{app:core}) does not use Proposition~\ref{prop:SM}.
\end{proof}

\section{Discussion}\label{sec:discussion}

\subsection{Why the fluctuation tends to zero}
The self-energy asymptotics of \cite{SerA} enter as they stand (Section~\ref{ssec:inputA}); there the difficulty behind the gap between the exponents of \cite{LL99} (the gauge covariant kinetic energy does not directly penalize the local size of a field that is locally close to a gradient) was overcome. The additional task here is to make the fluctuation of the estimator tend to zero. The boost bound (Lemma~\ref{lem:consumer}) needs $\langle\Gamma\rangle=O(1)$ for a linear rate, and $\langle\Gamma\rangle\to0$ for growth faster than linear. An estimator built on the core alone reads the steep wall of the core, and its fluctuation is bounded only by $\ell$ up to logarithms (Proposition~\ref{prop:Mprime}(a)), which would give $h\gtrsim\lambda^{6/7}$ up to logarithms. The shallow-level estimator does not read the wall but is sensitive to shallow depressions in the window shell. The following four estimates control the losses in this construction; in each, a large factor would otherwise cancel against a fixed power of $\ell$.
\begin{itemize}
\item[(i)] The coupling of the external field to the electron, paid by a fraction of the electron's kinetic energy, would force the depression to a polynomially large distance from the electron. Taking the external subspace inside the Leray subspace removes the source of the dipole tail (Theorem~\ref{thm:gauge}); the only remaining input is the smoothness of $S$, a choice of construction and not an assumption on the state.
\item[(ii)] The uniform localization loss $c/n^2$ in the electron coordinate is replaced, by the collective-coordinate IMS formula (Lemma~\ref{lem:ims}), by losses on the covariance defect and on the relative-localization shell; both live on the rare events of Propositions~\ref{prop:SM} and~\ref{prop:events}.
\item[(iii)] Choosing candidates relative to the estimator would multiply the sharpness $1/g_w$ of the selection by the sensitivity of the centre. This is excluded structurally by the absolute candidates (D3).
\item[(iv)] The surgery constant $\eta\Ebox$ is adapted to the margin, and the price $\kappa_\eta$ multiplies only super-polynomially small quantities (Corollary~\ref{cor:far}).
\end{itemize}

The remaining difficulty sits in three places: the statement that the events of Propositions~\ref{prop:SM}, \ref{prop:events} and~\ref{prop:shell} are rare to \emph{every} power (with a fixed power $\ell^{-a}$, the polynomial bound of Proposition~\ref{prop:global} on the bad set would not be compensated unless $a$ exceeds its degree plus $7$); the smoothness of $S$; and the margin $1-\kappa$ together with $\kappa>0$.

\subsection{Which estimates exclude slower growth}
The proof splits into deterministic inequalities (Lemmas~\ref{lem:consumer}, \ref{lem:Mw}, Proposition~\ref{prop:global}) and probabilistic estimates (Propositions~\ref{prop:SM}, \ref{prop:events}, \ref{prop:shell}). Let $\mathcal B$ be the union of the events of Propositions~\ref{prop:SM} and~\ref{prop:events}. On the complement of $\mathcal B\cup\Gamma'_\rho$ the deterministic inequalities bound the contribution to the right-hand side of \eqref{eq:consumer} for $\hat X=\Xw$ by $C\ell^{\kappa-1}(\log\lambda)^2$ (Lemma~\ref{lem:Mw} and Proposition~\ref{prop:Mprime}), while on $\mathcal B\cup\Gamma'_\rho$ only the pointwise polynomial bound $2K_G\ell^{7+D}$ of Proposition~\ref{prop:global} is available. If $h(\lambda_n)\le\lambda_n^{8/7-\epsilon}$ along a sequence $\lambda_n\to\infty$, then $c/h\gtrsim\ell_n^{-1+7\epsilon}$, and choosing $\kappa<7\epsilon$ in the estimator shows that $P_\psi(\mathcal B\cup\Gamma'_\rho)\gtrsim\ell_n^{-(D+8)}$. Thus a growth with an exponent strictly below $8/7$ would force the bad events to have polynomially large probability, and this is what the three probabilistic estimates exclude. For this estimator the bad events are controlled as follows; the physical behaviour named in each item is an interpretation and not part of the proof. (i) A perturbative behaviour $h\simeq\lambda^\gamma$ with $\gamma<6/7$ is excluded already by the core locator, whose input is that the mass event is rare to every power; this rests on the self-energy asymptotics of \cite{SerA} through Proposition~\ref{prop:SMexpl}. (ii) A soft periphery of the bubble with shallow satellite depressions in the window shell at polynomial probability (which would give $6/7\le\gamma<8/7$) is excluded by Proposition~\ref{prop:shell}. Its inputs are the Hardy margin (Lemma~\ref{lem:hardy}), the splitting off of the external field in the Leray subspace (Theorem~\ref{thm:gauge}), the selection and localization tools (Lemmas~\ref{lem:softmax}, \ref{lem:agmon}, \ref{lem:ims}) and the logarithmic growth (D1) of the cube side; with cubes of fixed size the probability of a shallow depression would only be bounded by a fixed power. (iii) A spreading of the relative localization of electron and bubble to $\ell^{1+\epsilon'}$, $\epsilon'>0$, is excluded by Proposition~\ref{prop:events}; through the factor $\rho^2$ of Lemma~\ref{lem:Mw} it would worsen the exponent obtained from this bound for every positive $\epsilon'$, and would prevent the bound from tending to zero for $\epsilon'\ge(1-\kappa)/2$. Physically, an electron outside the bubble pays the diamagnetic energy $\asymp\abar\ell^7$, exceeding the kinetic scale $\ell^5$ inside the bubble, so the choice $\rrel=K_d\ell\log\lambda$ is conservative.

\subsection{What is known about $\iPF$ and $h$}
\emph{Upper bounds at a fixed cutoff.} By Theorem~\ref{thm:main}, $h$ is not bounded uniformly in $\Lambda$. For $0<\alpha<\alpha_0$, where $\alpha_0>0$ is the smallness threshold of the coupling in \cite[Theorem~III.3]{FP10}, the results of Fr\"ohlich and Pizzo at their fixed cutoff value $\lambda_0$ give $h(\lambda_0)<\infty$ and $h(\lambda_0)\to1$ as $\alpha\to0$: in their units ($m=1$, kinetic term $\frac12(p+\alpha^{1/2}A)^2$, sharp cutoff $\Lambda\approx1$), \cite[Theorem~III.3 and Corollary~III.4]{FP10} show that the limit $E_P$ of the infrared-regularized fiber energies $E^\sigma_P$ as $\sigma\to0$ is twice differentiable in $|P|<\frac13$ with second derivative $\Sigma_{|P|}\to1$ as $\alpha\to0$, uniformly in $P$; $E_P$ is the bottom of the spectrum of the unregularized fiber, since the infrared part $A_{<\sigma}$ of the field satisfies $\|A_{<\sigma}\psi\|\le C\sigma^{1/2}\|(\Hf+1)^{1/2}\psi\|$, so that $H_P-H^\sigma_P$ is form-bounded with respect to $H^\sigma_P$ with a relative bound tending to zero and $H^\sigma_P\to H_P$ in the norm resolvent sense; hence $h(\lambda_0)=1/\Sigma_0$; hence $\iPF(\alpha)\le h(\lambda_0)/\lambda_0$ and $\limsup_{\alpha\to0}\iPF(\alpha)\le1/\lambda_0$, with $\alpha_0$ and $\lambda_0$ not explicit. In the infrared-regularized model of Hiroshima and Spohn \cite{HSp05}, the analogous quantity tends to zero as $\alpha\to0$ by their analyticity statement.

\emph{The exponent $8/7$.} In Theorem~\ref{thm:main}, $8/7=1+1/7$ comes from the bound $\ell^{\kappa-3}\rho^2\asymp\ell^{\kappa-1}$ (up to logarithms) of Lemma~\ref{lem:Mw}, that is, from the ratio of the squared radius of the window of $\Xw$ to the volume $\gtrsim\ell^3$ of the shallow depressions in it and from the sensitivity of the shallow level, together with $\ell\asymp\lambda^{1/7}$; the other contributions to the final estimate are $O(\ell^{-1})$ (Proposition~\ref{prop:expect}). Since $\kappa>0$ is needed (Step~7 of the proof of Proposition~\ref{prop:shell}), the present estimates yield every exponent strictly below $8/7$ but do not establish the endpoint.

\subsection{Open problems}
The value of $\iPF$, the true growth exponent of $h$ (Theorem~\ref{thm:main} gives the lower bound $8/7-\epsilon$ for every $\epsilon>0$) and an upper bound on the growth of $h(\lambda)$ as $\lambda\to\infty$ remain open; determining the optimal exponent requires such an upper bound. The relation between the curvature mass and the binding-energy mass of \cite{LL02}, proved equivalent for polaron-type models by Lieb and Seiringer \cite{LS14}, is open for the Pauli--Fierz model. We have not treated the spin-$\frac12$ electron (the Pauli term $e\sigma\cdot B$), external potentials, smooth ultraviolet cutoffs in the model, or lattice models; with spin, the magnetic term changes the bubble and the magnetic floor, and the analysis of \cite{SerA} would have to be revisited.

\appendix
\section{Facts from the literature}\label{app:facts}

The sources state the following facts for bare mass $1$, i.e.\ kinetic term $\frac12(\cdot)^2$. The general case follows from Lemma~\ref{lem:scaling} with $a=1/m$: $U_{1/m}H_{m,\Lambda}(P)U_{1/m}^*=m\,H_{1,\Lambda/m}(P/m)$. The charge $e$ is unchanged, and self-adjointness, the domain, existence and uniqueness of the ground state and the inequality $E(P)\ge E(0)$ transfer between the two sides. Theorem and corollary numbers of \cite{Hi07} refer to the arXiv version.

\begin{fact}[Ground state]\label{fact:gs}
For the sharp cutoff and without spin, $H(0)$ has a ground state for all $e\in\R$ \cite[Thm.~2]{HSi23}; Hypothesis~A of \cite{HSi23} is the sharp cutoff $\rho=(2\pi)^{-3/2}\mathbf 1_{[0,\Lambda]}$, and the polarization vectors may depend measurably on $\hat k$. The ground state is unique up to a phase \cite[Thm.~3.5 and Cor.~3.6]{Hi07} (in the Schr\"odinger representation, $\vartheta\ee^{-tH(0)}\vartheta^{-1}$ is positivity improving). The existence result is restricted to the sharp cutoff.
\end{fact}

\begin{fact}[Diamagnetic inequality]\label{fact:diamagnetic}
For all $P$ and real $e$, $E(P)\ge E(0)$ \cite[Cor.~3.8(1)]{Hi07}, which states $0=E(0,0)\le E(0,e^2)\le E(P,e^2)$. The inequality is not strict and holds for each $\Lambda$.
\end{fact}

\begin{fact}[Domain and self-adjointness]\label{fact:domain}
$H(P)$ is self-adjoint on $D(\Hf)\cap D(\Pf^2)$, and the domain does not depend on $P$ or $e$ \cite[Thm.~2.2; for the sharp cutoff, Cor.~2.5]{Hi07}. $H(P)$ is essentially self-adjoint on every core of $H_{00}(P)=(P-\Pf)^2/2m+\Hf$ (proof of \cite[Thm.~2.2]{Hi07}). The space $\F_{{\rm b}\infty}$ of finite-particle vectors whose components are smooth with compact support is a core of $H_{00}$, since $H_{00}$ acts as a multiplication operator in each particle sector.
\end{fact}

\begin{fact}[Relative bound]\label{fact:rel-bound}
For $f$ with $f/\sqrt\omega\in L^2$,
\begin{equation}\label{eq:rel-bound}
\lVert a^\sharp(f)\Psi\rVert\le\lVert f/\sqrt\omega\rVert\,\lVert\Hf^{1/2}\Psi\rVert+\lVert f\rVert\,\lVert\Psi\rVert
\end{equation}
(see \cite[(2.2)]{Hi07}, where the vector on the left-hand side is misprinted).
\end{fact}

For background we mention that self-adjointness of the Pauli--Fierz Hamiltonian for arbitrary values of the coupling was proved in \cite{Hir02} and its domain was identified in \cite{HH08}, and that functional integral representations of the model go back to \cite{Hi97}.

\section{The pointwise pull-through formula}\label{app:pullthrough}

The pull-through formula is a standard tool in the analysis of Fock space models (see e.g.\ \cite{GLL01}); we give the version used in Section~\ref{sec:partI-proof} with the measure-theoretic details. For $\psi=(\psi^{(n)})_n\in\F$ define
\[
(a_r(k)\psi)^{(n)}(k_1,r_1,\dots,k_n,r_n):=\sqrt{n+1}\,\psi^{(n+1)}(k,r,k_1,r_1,\dots,k_n,r_n).
\]
By Tonelli's theorem, $\sum_r\int\lVert a_r(k)\psi\rVert^2\rho(k)\,d^3k=\ip{\psi}{\dG(\rho)\psi}$ for $\rho\ge0$; if the right-hand side is finite, $a_r(k)\psi$ is a well defined vector in $\F$ for a.e.\ $k$.

\begin{fact}\label{fact:pullthrough}
(i) For all $\delta>0$ and $\psi\in D(\Hf^{1/2})$, $\sum_r\int_{|k|\ge\delta}\lVert a_r(k)\psi\rVert^2\,d^3k\le\delta^{-1}\lVert\Hf^{1/2}\psi\rVert^2$. In particular $k\mapsto a_r(k)\psi_0$ is an a.e.\ defined, locally square integrable $\F$-valued function on $\R^3\setminus\{0\}$, and for $\zeta\in\h_\perp$ vanishing near $0$,
\[
a(\zeta)\psi_0=\int\sum_r\overline{\zeta(k,r)}\,a_r(k)\psi_0\,d^3k
\]
as a Bochner integral; indeed $\int_{|k|\ge\delta}|\zeta|\,\lVert a(k)\psi_0\rVert\le\lVert\zeta\rVert\delta^{-1/2}\lVert\Hf^{1/2}\psi_0\rVert$.

(ii) For a.e.\ $k$ with $0<|k|\le\Lambda$, \eqref{eq:PT} holds: $a_r(k)\psi_0=-\frac{e\varkappa_k}m\Rk\,\varepsilon_r(k)\cdot V\psi_0$; for a.e.\ $|k|>\Lambda$, $a_r(k)\psi_0=0$.
\end{fact}

\begin{proof}
(i) Read $\delta\int_{|k|\ge\delta}a^*(k)a(k)\le\dG(|k|)$ in the $n$-particle sectors. The Bochner representation follows from $\ip{\Phi}{a(\zeta)\psi_0}=\int\bar\zeta\ip{\Phi}{a(k)\psi_0}$ for fixed $\Phi$ and Pettis' theorem.

(ii) The formal relations $a(k)\Pf=(\Pf+k)a(k)$, $a(k)\Hf=(\Hf+|k|)a(k)$ and $a_r(k)A_\mu(0)=A_\mu(0)a_r(k)+\varkappa_k\mathbf 1_{\{|k|\le\Lambda\}}\varepsilon_{r,\mu}(k)$, together with transversality, give
\[
a_r(k)H(0)=\bigl(H(-k)+|k|\bigr)a_r(k)+\frac{e\varkappa_k\mathbf 1_{\{|k|\le\Lambda\}}}m\,\varepsilon_r(k)\cdot V .
\]
To make this rigorous, take bounded $\zeta$ with bounded support vanishing near $0$ (we need $k^2\zeta\in L^2$ so that $a^*(\zeta)\Phi\in D(\Pf^2)$) and $\Phi\in\F_{{\rm b}\infty}$. Rewriting the weak identity $\ip{\Phi}{a(\zeta)(H(0)-E_0)\psi_0}=0$ with the relations above gives
\[
\int\overline{\zeta(k,r)}\,\bigl\langle(H(-k)-E_0+|k|)\Phi,\,a_r(k)\psi_0\bigr\rangle\,d^3k=-\frac em\int\overline{\zeta(k,r)}\,\varkappa_k\mathbf 1_{\{|k|\le\Lambda\}}\ip{\Phi}{\varepsilon_r(k)\cdot V\psi_0}\,d^3k ;
\]
all terms are finite because $\Phi$ has finitely many particles, and integrable in $k$ by (i). Since $\zeta$ is arbitrary, the integrands agree for a.e.\ $k$. Take a countable family in $\F_{{\rm b}\infty}$ that is dense in the graph norm of $H_{00}$, so that the exceptional null set can be chosen independent of $\Phi$. By Fact~\ref{fact:domain} and the closed graph theorem the graph norms of $H(-k)$ and $H_{00}$ are equivalent, so this family is a core of $H(-k)$ for every $k$; boundedness of $\Phi\mapsto\ip{(H(-k)-E_0+|k|)\Phi}{a_r(k)\psi_0}$ then gives $a_r(k)\psi_0\in D(H(-k)^*)=D(H(-k))$ and
\[
(H(-k)-E_0+|k|)\,a_r(k)\psi_0=-\frac{e\varkappa_k\mathbf 1_{\{|k|\le\Lambda\}}}m\,\varepsilon_r(k)\cdot V\psi_0 .
\]
Since $H(-k)-E_0+|k|\ge|k|>0$ by Fact~\ref{fact:diamagnetic}, we may invert; for $|k|>\Lambda$ the right-hand side vanishes, so $a_r(k)\psi_0=0$.
\end{proof}

\section{Closed-form integrals and numerical values}\label{app:integrals}

With $g=1/\Omega_k$ on $[k_1,K]$ we have $kg=\frac{2m}{2m+k}$ and $\beta(k)^2=\frac{2m+k}{4k}$, hence
\[
k^2g\beta=m\sqrt{\frac k{2m+k}},\qquad k^3g^2=\frac{4m^2k}{(2m+k)^2}.
\]
Substituting $k=2mt$ and $\sigma=K/2m$, and enlarging the band to $[0,K]$,
\[
\int_{k_1}^Kk^2g\beta\,dk\le2m^2\mathsf L(\sigma),\qquad\int_{k_1}^Kk^3g^2\,dk\le4m^2\Bigl(\log(1+\sigma)-\frac\sigma{1+\sigma}\Bigr),
\]
with $\mathsf L$ as in Section~\ref{sec:partI-result}. The first gives $\mathcal A\le\frac{|e|}{16\pi m}\cdot2m^2\mathsf L(\sigma)=\mathsf a_K$ in Lemma~\ref{lem:amplitude}(iii). By the first angular integral, $\lVert kf\rVert^2=\int|k|^2|f|^2=\frac1{6\pi^2}\int_{k_1}^Kk^3g^2\,dk\le\mathsf N_K$. Dividing by the limit $mS(0)=\frac{m^2}{3\pi^2}\log(1+\sigma)$ of $mS$ as $k_1\downarrow0$ and using $e^2=4\pi\alpha$ gives $\mathsf a_K^2/(mS(0))=t_K$ and $\mathsf N_K/(mS(0))=q_K$. None of these quantities contains $\Lambda$.

\subsection{Numerical values}\label{app:constants}
The following values were computed with interval arithmetic (\texttt{mpmath.iv}, 40 decimal digits); the script \texttt{constants\_partI.py} is available as an ancillary file. For $K=m$, i.e.\ $\sigma=\frac12$, see Table~\ref{tab:partI-values}.
\begin{center}
\captionof{table}{Constants of Part~I for $K=m$ (interval arithmetic).}\label{tab:partI-values}
\begin{tabular}{lll}
quantity & value (leading digits) & used as\\\hline
$x_m/\alpha=\frac4{3\pi}\log\frac32$ & $0.172084736551$ & \\
$t_m/\alpha$ & $0.0625789538050$ & \\
$q_m$ & $0.355797691749$ & \\
$2x_m/\alpha$ & $0.344169473102$ & $\ge0.3441$\\
$2\sqrt{q_mt_m/\alpha}$ & $0.298432218877$ & $\le0.2985$\\
$2t_m/\alpha$ & $0.125157907610$ & $\le0.1252$
\end{tabular}
\end{center}
The two closed forms above agree with numerical quadrature to within $10^{-40}$ for $\sigma\in\{0.1,0.5,2\}$. The computation checks closed-form arithmetic only; it says nothing about infinite dimensionality or the limit $\Lambda\to\infty$.

\section{Bookkeeping of large factors}\label{app:terms}

A large quantity ($c\asymp\ell^7$, $\Ebox\asymp\ell^5$, $\mathrm e^{2F_{\max}}=\ell^{2\kappa_p}$, or $\kappa_\eta\asymp\ell^{12-\kappa}/b^3$) must never multiply a quantity that is only a fixed negative power of $\ell$, since the powers would then cancel; the mechanisms listed in Section~\ref{sec:discussion} are of this kind. Table~\ref{tab:terms} lists every term of the proof as ``large quantity $\times$ multiplier'', with the lemma bounding it; each multiplier is (i) super-polynomially small, (ii) a power that the design can choose, (iii) the probability of an event that is rare to every power, (iv) a Gaussian tail, or (v) smaller than the margin by a negative power of $\ell$. Only one large factor multiplies a fixed power: $c/\rrel^2$, which appears only on the rare shell event. On the good set the kinetic factor $c$ multiplies the covariance defect, which vanishes there because the denominators are lifted only below the mass thresholds. The power margins actually used are two: the bound of Lemma~\ref{lem:Mw} ($1-\kappa$) and the seam threshold ($\ell^\kappa$ against polylogarithms).

Units: $c\asymp\ell^7$, $\Ebox\asymp\ell^5$, margin $\mu_w\asymp b^3\ell^\kappa$, $\kappa_\eta\asymp\ell^{12-\kappa}/b^3$, $\Gamma_{X_c}\lesssim\ell\,(\log\lambda)^5$, $b=C_b\lfloor\log\ell\rfloor$, $\rrel\asymp\rho\asymp\ell\log\lambda$. ``Any power'' means a probability bounded by $\ell^{-\kappa_p}$ with $\kappa_p$ chosen before the window constants (Remark~\ref{rem:order}). On the good set $\mathcal G_\theta$ the covariance defects of $X_c$ and $\Xw$ vanish (Proposition~\ref{prop:Mprime}(b), \eqref{eq:Rw}), so the kinetic terms $c\|\Rm\|^2$ and $c|Df|^2$ appear only on bad sets.

\begin{center}
\small
\captionof{table}{Large factors of Part~II and where they are bounded.}\label{tab:terms}
\resizebox{\textwidth}{!}{\begin{tabular}{llll}
\hline
& term & large factor $\times$ multiplier & ratio to the margin; where bounded\\
\hline
(a) & $\eta\Ebox$ in the surgery & $\Ebox\times\eta$ (chosen) & $\mu_w/16$; Prop.~\ref{prop:shell}, Step 3\\
(b) & $\kappa_\eta|r_{\mathrm e}|^2$, $c^{(r)}$ & $\ell^{12-\kappa}\times\sup_{|w|\ge\DEl}|\Sch|^2$ & any power via $\DEl$; Cor.~\ref{cor:far}\\
(c) & seam $\delta^{-1}b_\Sigma^2$ & $\delta^{-1}\times(b^4+b^3(2F_{\max}+K_{\rm t}\log\ell))$ & $(b+(\kappa_p+K_{\rm t})\log\ell)/\ell^\kappa$, needs $\kappa>0$; Lem.~\ref{lem:seam}, \ref{lem:trace}\\
(d) & zero point in $\mathcal V$ & $e_*b^3$ & $\ell^{-\kappa}$; Lem.~\ref{lem:level}\\
(e) & softmax selection & $\beta_{\mathrm{sel}}^2\bar C_g$ & $(\log|\mathcal K_j|)^2/b^6$; Lem.~\ref{lem:softmax}\\
(f) & softmax bad weight & $|\mathcal K_j|\mathrm e^{-\beta_{\mathrm{sel}}\theta_\delta g_w}(\mu_w+B_-)$ & $\eta_p$; Lem.~\ref{lem:softmax}\\
(g) & IMS, field part & $\Gamma_{X_c}/n_b^2$ & $\ell^{-1-\kappa}$polylog; Lem.~\ref{lem:ims}\\
(h) & IMS, kinetic, bad sets, shell & $c\cdot\mathrm{poly}\times$ prob.\ & any power; Props.~\ref{prop:SM}, \ref{prop:events}\\
(i) & Agmon $\Gamma_f$ & $F_{\max}^2\times(1/(g_wb^3)+\Gamma_{X_c}/\rho^2)$ & $\kappa_p^2\log^2\ell/(\theta_\delta^2b^6)+\ell^{-1-\kappa}$polylog; Prop.~\ref{prop:shell}, Step 6\\
(j) & Gaussian tails & $\kappa_r\times$ tail & any power via thresholds; Lem.~\ref{lem:tail}, Cor.~\ref{cor:agmonquad}\\
(k) & the bound \eqref{eq:Mw} off $\Gamma'_\rho$ & deterministic $\ell^{\kappa-3}\rho^2$ & $\ell^{\kappa-1}(\log\lambda)^2$, the only use of $1-\kappa$; Lem.~\ref{lem:Mw}\\
(l) & global polynomial on $\mathcal G'^c$ & $\mathrm{poly}\times$ prob.\ & any power; Prop.~\ref{prop:global}\\
\hline
\end{tabular}}
\end{center}

If the probabilities in lines (h) and (l) were only a fixed power $\ell^{-a}$, the bad sets would contribute $\abar K_G\ell^{7+D-a}$ to $c\,\delta$ (Proposition~\ref{prop:global}); this tends to zero only if $a>7+D$, and it is $O(\ell^{-1})$, as required in Proposition~\ref{prop:expect} (there $\kappa_p>D+8$), only if $a\ge D+8$; the quoted propositions give every power because their exponents are proportional to window lengths.

\section{Proofs of the field lemmas}\label{app:cited}

This appendix proves Lemma~\ref{lem:tail} (Gaussian tails), Lemma~\ref{lem:607N} (point variance), Lemma~\ref{lem:hardy} (Hardy inequality and localized zero point) and the second claim of Lemma~\ref{lem:trace}. Fourier transforms are $\hat f(k)=\int f(y)\mathrm e^{-ik\cdot y}\dd y$, so $\|f\|^2=(2\pi)^{-3}\|\hat f\|^2$; $\omega=|k|$, and $\Pi$ is the transverse projection restricted to the band $|k|\le\pi$.

\subsection{Sub-Gaussian domination and Lemma~\ref{lem:tail}}
As in \cite[\SerAnum{sec:model}]{SerA}, $\Phi(f):=a(f)+a^*(f)$, so that $\langle\Omega_\vac,\Phi(f)^2\Omega_\vac\rangle=\|f\|^2$ and the covariance matrix of Lemma~\ref{lem:tail} is $\Gamma_{mn}=\mathrm{Re}\langle h_m,h_n\rangle$; in Theorem~\ref{thm:gauge}(ii)(b) a real configuration-space vector $h$ stands for $\Phi$ of the corresponding one-photon vector, whose squared norm is $\frac12\|\omega^{-1/2}h\|^2$. In the notation of Section~\ref{sec:tools-multiplier}, $\Phi(f)=\sqrt2\,\varphi(f)$. A vector $f\in\mathfrak h$ is \emph{$J$-real} if $Jf=f$ with $J$ of Section~\ref{sec:tools-multiplier}, that is, if it is real-valued in position space. For $J$-real $f,g$ the inner product $\langle f,g\rangle$ is real, hence $\Phi(f)$ and $\Phi(g)$ commute strongly (Weyl relations).

For the semi-relativistic Pauli--Fierz Hamiltonian with an external potential a Gaussian domination of the ground state was proved by Hiroshima by the same Gaussian transformation of the functional integral \cite[Lemma~6.7 and Theorem~6.8]{Hi14} (a Gaussian domination argument of a different kind is used for the polaron effective mass in \cite{Se24}); the following lemma is the version we need, for the fiber at zero total momentum and with the free-vacuum variance. We include its short proof, which inserts the moment generating function directly into the functional integral.

\begin{lemma}[Sub-Gaussian domination]\label{lem:subgauss}
Let $f$ be $J$-real with $\|f\|+\|f/\sqrt\omega\|<\infty$. Then $\langle\psi,\mathrm e^{s\Phi(f)}\psi\rangle\le \mathrm e^{s^2\|f\|^2/2}$ for all $s\in\R$.
\end{lemma}
\begin{proof}
Let $\psi_t:=\mathrm e^{-tH(0)}\Omega_\vac/\|\mathrm e^{-tH(0)}\Omega_\vac\|$ and let $\mu_t$, $\mu$ be the spectral measures of $\Phi(f)$ in $\psi_t$ and in $\psi$. We use the functional integral representation of $\mathrm e^{-tH(0)}$ in terms of Brownian paths and the Euclidean field \cite[Theorem~3.3]{Hi07}, with a function of the field inserted at time $t$ as in \cite[Theorem~4.3, (4.7)]{Hi07} (for the model with an external potential see \cite{Hi97}). Conditionally on the path, the field variable $\phi$ (the Euclidean field at $f$ translated along the path, as the fiber decomposition dictates; the path is substituted into a measurable version of the field) and the phase $\theta$ produced by the minimal coupling are jointly centred Gaussian, with $\mathbb E\phi^2=\|f\|^2$ independent of the path, since the Euclidean covariance is translation invariant; let $\sigma^2:=\mathbb E\theta^2$ and $r_{\phi\theta}:=\mathbb E[\phi\theta]$. The representation gives, for bounded functions $G$ of the field at time zero,
\[
\int G\dd\mu_t=\langle\psi_t,G(\Phi(f))\psi_t\rangle=\frac{\int\mathbb E[G(\phi)\mathrm e^{i\theta}\mid\text{path}]\dd(\text{path})}{\int\mathrm e^{-\sigma^2/2}\dd(\text{path})};
\]
it is stated for $G=\mathrm e^{iu\,\cdot}$ in \cite[Theorem~4.3]{Hi07} with $m=2$, $P_0=P_1=0$, no factors $\mathrm e^{-(s_j-s_{j-1})K}$ and $\Phi_1=\mathrm e^{iuA(f)}$ (this is the second equality in the proof of \cite[Corollary~4.7]{Hi07}, there for general $P$; we use $P=0$), extends to Schwartz functions by Fourier inversion and Fubini, and to bounded continuous $G$ by bounded pointwise approximation and dominated convergence on both sides (the path measure is finite).

Fix $s\in\R$ and let $G_N(x):=\mathrm e^{sx}\min\bigl(1,(N+1-|x|)_+\bigr)$, bounded, continuous and increasing to $\mathrm e^{sx}$. On the left, monotone convergence gives $\int G_N\dd\mu_t\uparrow\int\mathrm e^{sx}\dd\mu_t\in[0,\infty]$. On the right, $|G_N(\phi)\mathrm e^{i\theta}|\le\mathrm e^{s\phi}$ and $\mathbb E[\mathrm e^{s\phi}\mid\text{path}]=\mathrm e^{s^2\|f\|^2/2}$ does not depend on the path, so dominated convergence applies. Since $s\phi+i\theta$ is a complex linear combination of jointly Gaussian variables, $\mathbb E[\mathrm e^{s\phi+i\theta}\mid\text{path}]=\exp\bigl(\tfrac12(s^2\|f\|^2-\sigma^2)+isr_{\phi\theta}\bigr)$, and therefore
\[
\int\mathrm e^{sx}\dd\mu_t=\mathrm e^{s^2\|f\|^2/2}\langle\mathrm e^{isr_{\phi\theta}}\rangle_{\mathrm p}=\mathrm e^{s^2\|f\|^2/2}\langle\cos(sr_{\phi\theta})\rangle_{\mathrm p}\le\mathrm e^{s^2\|f\|^2/2},
\]
where $\langle\cdot\rangle_{\mathrm p}$ is the expectation with respect to the probability measure on paths with density proportional to $\mathrm e^{-\sigma^2/2}$, and the left-hand side is real.

The ground state is unique and has positive overlap with $\Omega_\vac$, by the positivity improving property of $\vartheta \mathrm e^{-tH(0)}\vartheta^{-1}$ \cite[Theorem~3.5 and Corollary~3.6]{Hi07}; hence $\psi_t\to\psi$ in norm as $t\to\infty$, so $\int G\dd\mu_t\to\int G\dd\mu$ for bounded continuous $G$, i.e.\ $\mu_t\Rightarrow\mu$. As $x\mapsto\mathrm e^{sx}$ is non-negative and continuous, $\int\mathrm e^{sx}\dd\mu\le\liminf_t\int\mathrm e^{sx}\dd\mu_t\le\mathrm e^{s^2\|f\|^2/2}$.
\end{proof}

\begin{proof}[Proof of Lemma~\ref{lem:tail}]
\emph{Step 1: removing the infrared condition.} Let $f$ be $J$-real with $\|f\|<\infty$ and $f_\varepsilon:=1_{\{\omega\ge\varepsilon\}}f$. Since $\omega$ is even, $f_\varepsilon$ is $J$-real, $\|f_\varepsilon/\sqrt\omega\|<\infty$, $\|f_\varepsilon\|\le\|f\|$ and $f_\varepsilon\to f$. Lemma~\ref{lem:subgauss} gives $\int \mathrm e^{sx}\dd\mu_\varepsilon\le \mathrm e^{s^2\|f\|^2/2}$ for the spectral measures $\mu_\varepsilon$ of $\Phi(f_\varepsilon)$. Weyl operators are strongly continuous in the one-particle vector, so the characteristic functions converge pointwise and $\mu_\varepsilon\Rightarrow\mu$ (L\'evy); by lower semicontinuity of $\int \mathrm e^{sx}\dd\mu$ under weak convergence, $\langle \mathrm e^{s\Phi(f)}\rangle\le \mathrm e^{s^2\|f\|^2/2}$.

\emph{Step 2: quadratic forms.} For $J$-real $h_1,\dots,h_M$ the $\Phi(h_m)$ commute and have a joint spectral measure $\mu$ in $\psi$; the covariance $\Gamma_{mn}=\mathrm{Re}\langle h_m,h_n\rangle$ is real symmetric. With an auxiliary standard Gaussian vector $\xi\in\R^M$, $\mathrm e^{s|x|^2}=\mathbb E_\xi \mathrm e^{\sqrt{2s}\,\xi\cdot x}$, and by Tonelli and Step~1 applied to $h_\xi:=\sum_m\xi_mh_m$,
\[
\int \mathrm e^{s|x|^2}\dd\mu=\mathbb E_\xi\langle \mathrm e^{\sqrt{2s}\Phi(h_\xi)}\rangle\le\mathbb E_\xi \mathrm e^{s\,\xi\cdot\Gamma\xi}=\det(1-2s\Gamma)^{-1/2}\qquad(2s\|\Gamma\|<1).
\]
Differentiating at $s=0$ gives $\langle W\rangle\le\tr\Gamma$. With $-\log(1-x)\le x/(1-x)$ eigenvalue by eigenvalue and Markov's inequality at $s=1/(4\|\Gamma\|)$, $P_\psi(W>b^2)\le \mathrm e^{-sb^2}\mathrm e^{2s\tr\Gamma}=\exp(-(b^2-2\tr\Gamma)/(4\|\Gamma\|))$.

\emph{Step 3: countably many terms.} If $\sum_m\|h_m\|^2<\infty$, then $W_M:=\sum_{m\le M}\Phi(h_m)^2\uparrow W$, and $\det(1-2s\Gamma_M)^{-1/2}\le\det(1-2s\Gamma)^{-1/2}$ (Fredholm determinants, $\Gamma$ trace class). Monotone convergence gives the claim.
\end{proof}

\subsection{Proof of Lemma~\ref{lem:607N}}
Write $\kappa_S(w)=(2\pi)^{-3}\int S^2|k|^{-1}\mathrm e^{ik\cdot w}\dd k=\frac1{2\pi^2r}\int_0^\pi S(k)^2\sin(kr)\dd k$, $r=|w|$. From $|\sin kr|\le kr$, $|\kappa_S|\le\frac1{2\pi^2}\int_0^\pi k\dd k=\frac14$. Integrating by parts ($S(0)=1$, $S(\pi)=0$), $\int_0^\pi S^2\sin(kr)\dd k=\frac1r+\frac1r\int_0^\pi(S^2)'\cos(kr)\dd k$, whose modulus is at most $(1+V_S)/r$; hence $|\kappa_S|\le(1+V_S)/(2\pi^2r^2)$. For $y\in\mathcal E$, $\int_{\mathcal E}|\kappa_S(y-y')|\dd y'\le\frac{1+V_S}{2\pi^2}\int_{|w|\le\diam\mathcal E}|w|^{-2}\dd w=\frac{2(1+V_S)}\pi\diam\mathcal E$.

For $\varphi\in L^2(\mathcal E;\mathbb C^3)$ (extended by zero), $\varphi\in L^1$ and $\hat\varphi$ is bounded; since $S^2|k|^{-1}\in L^1(\R^3)$ and $\Pi(k)$ is an orthogonal projection for each $k$,
\begin{multline*}
\|T_{\mathcal E}\varphi\|^2=(2\pi)^{-3}\int S^2|k|^{-1}|\Pi\hat\varphi|^2\le(2\pi)^{-3}\int S^2|k|^{-1}|\hat\varphi|^2\\
=\sum_c\iint\overline{\varphi_c(y)}\kappa_S(y-y')\varphi_c(y')\dd y\dd y'\le N_{\mathcal E}\|\varphi\|^2
\end{multline*}
by Fubini and the symmetric Schur test. \qed

\subsection{Proof of Lemma~\ref{lem:hardy}}
\emph{(i), finite-dimensional core.} Let $L:\R^n\to\mathcal H$ be linear into a real Hilbert space, $b\in\mathcal H$, $v(x)=Lx+b$, $u=|v|^2$, and $g:=\tr L^*L-2\|L^*L\|>0$. The vector field $F:=L^*v/u$ satisfies $\mathrm{div}\,F=\tr L^*L/u-2|L^*v|^2/u^2\ge g/u$, because $|L^*v|^2\le\|L^*L\|u$. Since $g>0$, $L$ has at least three non-zero singular values, so the zero set of $v$ is empty or an affine subspace of codimension at least three, where $|F|\lesssim1/\mathrm{dist}$; hence $F$ and $\mathrm{div}F$ are locally integrable and the distributional divergence is the pointwise one. For $\psi\in C_c^\infty(\R^n)$,
\[
g\int\frac{|\psi|^2}u\le\int|\psi|^2\,\mathrm{div}F=-2\,\mathrm{Re}\int\bar\psi\,\nabla\psi\cdot F,\qquad|\nabla\psi\cdot F|=\frac{|\langle L\nabla\psi,v\rangle|}u\le\frac{|L\nabla\psi|}{\sqrt u},
\]
and Cauchy--Schwarz gives $\int|\psi|^2/u\le\frac4{g^2}\int|L\nabla\psi|^2$. The shift $b$ enters nowhere.

\emph{(i), infinite modes.} $G'_{\mathrm e}:=Y_{\mathrm e}^*Y_{\mathrm e}$ is trace class. Let $Y_{\mathrm e}=\sum_j\sigma_jv_j\langle w_j,\cdot\rangle$ be a singular value decomposition with $w_j\in\Wext$, and $x_j:=\phi(w_j)$, $p_j:=\pi(w_j)$. Choose $N$ with $g'_N:=\sum_{j\le N}\sigma_j^2-2\sigma_1^2>0$. The Fock space factorizes as $L^2(\R^N)\otimes\mathcal K$, where the field operators orthogonal to $w_1,\dots,w_N$ (all internal components and the remaining external ones) act in $\mathcal K$ and commute; in their joint spectral representation $\mathcal K=L^2(\Xi,\nu)$ and $u'_Q=|\sum_{j\le N}\sigma_jx_jv_j+b(\xi)|^2$ with $b(\xi)$ independent of $x$. Applying the finite-dimensional inequality fiberwise gives $1/u'_Q\le4g_N'^{-2}\sum_{j\le N}\sigma_j^2p_j^2$ on $C_c^\infty(\R^N)\otimes_{\mathrm{alg}}\mathcal K$, which extends to the form domain by lower semicontinuity. Finally $\sum_{j\le N}\sigma_j^2p_j^2\le\|Y_{\mathrm e}E_{\mathrm e}\|^2=\int_Q|P'_QE_{\mathrm e}|^2\le\int_Q|E_{\mathrm e}|^2$, and $g'_N\uparrow g'_{\mathrm e}$.

\emph{(ii).} We have
\[
\tr G'_{\mathrm e}=\tr(1_QP_{\Wext}1_Q)-\tr(P_{\nabla,Q}1_QP_{\Wext}1_QP_{\nabla,Q}),
\]
and the last term is at most $\tr(P_{\nabla,Q}1_Q\Pi1_QP_{\nabla,Q})\le C_D|\partial Q|$ by the boundary estimate of \cite[\SerAnum{cor:hardy}]{SerA}; also $\|G'_{\mathrm e}\|\le1$. For a polynomial $p=\sum_{j\le J}c_jt^j$ with $p\le1_{[t_0,1]}$ on $[0,1]\supset\sigma(X)$, $p(X)\le P_{\Wext}$ on $V_T$ (the constant term read as $c_0\Pi$), so $\tr(1_QP_{\Wext}1_Q)\ge\tr(1_Qp(X)1_Q)$. With $\Pi_\infty$ as in Setting~\ref{set:box} write
\[
X=S^2\Pi-\Delta-SR'S,\qquad\Delta:=S\Pi1_{\mathcal E^c}\Pi S,\qquad R':=\Pi_\infty1_{\mathcal E}\Pi_\infty-P_\sigma ,
\]
where $0\le R'=\Pi_\infty R'\Pi_\infty\le\Pi_\infty$ since $P_\sigma\le\Pi_\infty1_{\mathcal E}\Pi_\infty$. The telescoping identity $X^j-(S^2\Pi)^j=-\sum_{a+b=j-1}X^a(\Delta+SR'S)(S^2\Pi)^b$ and $|\tr(AB)|\le\|A\|_{\mathrm{HS}}\|B\|_{\mathrm{HS}}$ give, since $\|X\|\le1$, $X=\Pi_\infty X$ and $\|S\Pi1_Q\|_{\mathrm{HS}}^2\le\rho_{\mathrm m}|Q|$,
\[
\bigl|\tr\bigl(1_Q(X^j-S^{2j}\Pi)1_Q\bigr)\bigr|\le j\sqrt{\rho_{\mathrm m}|Q|}\max_{m\le2j-1}\Bigl(\|1_{\mathcal E^c}\Pi S^m1_Q\|_{\mathrm{HS}}+\|R'^{1/2}S^m\Pi1_Q\|_{\mathrm{HS}}\Bigr) .
\]
The kernel $K_m$ of $S^m\Pi$ is the Schwartz kernel of $S^m$ plus $\partial_i\partial_j\mathcal N_{S^m}$, which is $O(|w|^{-3})$ for $|w|\ge1$ (the first moment of the radial function $\check{S^m}$ vanishes); hence $\|1_{\mathcal E^c}\Pi S^m1_Q\|_{\mathrm{HS}}^2\le\frac{4\pi}3C_m^2|Q|D_Q^{-3}$, where $|K_m(w)|\le C_m(1+|w|)^{-3}$.

For the new term let $\varphi_{y,l}:=K_m(\cdot-y)e_l$, a column of the kernel; it lies in the range of $\Pi_\infty$, so $\langle\varphi_{y,l},R'\varphi_{y,l}\rangle=\|1_{\mathcal E}\varphi_{y,l}\|^2-\|P_\sigma1_{\mathcal E}\varphi_{y,l}\|^2=\|(1-P_\sigma)1_{\mathcal E}\varphi_{y,l}\|^2\le\|1_{\mathcal E}\varphi_{y,l}-\psi\|^2$ for every $\psi\in L^2_\sigma(\mathcal E)$, and
\[
\|R'^{1/2}S^m\Pi1_Q\|_{\mathrm{HS}}^2=\int_Q\sum_l\langle\varphi_{y,l},R'\varphi_{y,l}\rangle\dd y .
\]
The symbol of $\varphi_{y,l}$ is $S^m(e_l-\hat k\hat k_l)$, so $\varphi_{y,l}=\mathrm{curl}\,a_{y,l}$ with $a_{y,l}:=\nabla\mathcal N_{S^m}(\cdot-y)\times e_l$, and $|a_{y,l}(w)|\le C'_m(1+|w-y|)^{-2}$ by the decomposition of $\mathcal N_{S^m}$ in Setting~\ref{set:box}. Let $y\in Q$, $d:=\dist(y,\partial\mathcal E)\ge D_Q$, $\chi:=\chi_0(|\cdot-y|/d)$ with $\chi_0=1$ on $[0,\frac12]$ and $\chi_0=0$ on $[1,\infty)$, and $\psi:=\mathrm{curl}(\chi a_{y,l})=\chi\varphi_{y,l}+\nabla\chi\times a_{y,l}$. Then $\psi$ is divergence-free, square integrable and supported in $\bar B(y,d)\subset\bar{\mathcal E}$, so $\psi\in L^2_\sigma(\mathcal E)$, and as $|1_{\mathcal E}-\chi|\le1-\chi$,
\[
\langle\varphi_{y,l},R'\varphi_{y,l}\rangle\le2\int_{|w|>d/2}\frac{C_m^2}{|w|^6}\dd w+2\frac{\|\chi_0'\|_\infty^2}{d^2}\,C_m'^2\Bigl(\frac d2\Bigr)^{-4}\frac{4\pi}3d^3=\Bigl(\frac{64\pi}3C_m^2+\frac{128\pi}3\|\chi_0'\|_\infty^2C_m'^2\Bigr)d^{-3}.
\]
Hence $\|R'^{1/2}S^m\Pi1_Q\|_{\mathrm{HS}}^2\le C''_m|Q|D_Q^{-3}$, of the same form as the first term. The main term is $\tr(1_QS^{2j}\Pi1_Q)=\rho_{\mathrm m}|Q|\langle S^{2j}\rangle$, with $\langle g\rangle:=\int_{|k|\le\pi}g(|k|)\dd^3k/(4\pi^4/3)$. Summing over $j$ gives the bound with $C_p:=\rho_{\mathrm m}^{-1/2}\sum_j|c_j|\,j\max_{m\le2j-1}(\sqrt{4\pi/3}\,C_m+\sqrt{C''_m})$. Approximating $1_{[t_0+\eta,1]}$ from below by polynomials (Weierstrass), $\langle p(S^2)\rangle\ge\langle1\{S^2\ge t_0+\eta\}\rangle-3\eta$; with $D_Q\ge D_\eta$ and $s$ so large that $C_D|\partial Q|+2\le\eta\rho_{\mathrm m}|Q|$, we obtain $g'_{\mathrm e}\ge(\langle1\{S^2\ge t_0+\eta\}\rangle-5\eta)\rho_{\mathrm m}|Q|$. With $\kappa_*:=\langle1\{S^2>t_0\}\rangle>0$ (as $S=1$ near $0$) we have $\langle1\{S^2\ge t_0+\eta\}\rangle\uparrow\kappa_*$ as $\eta\downarrow0$; we set $\kappa_{\mathrm e}:=\kappa_*/2$ and choose $\eta$ so small that $\langle1\{S^2\ge t_0+\eta\}\rangle-5\eta\ge\kappa_*/2$, and then $D_\eta$ accordingly.

\emph{(iii).} Let $\varepsilon_{\mathrm H}<1$ ($\varepsilon_{\mathrm H}=1$ follows by a limit) and $A:=1-\varepsilon_{\mathrm H} M\ge1-\varepsilon_{\mathrm H}>0$. Then $H-\varepsilon_{\mathrm H} T_M=\frac12\pi\cdot A\pi+\frac12q\cdot Kq$; the linear change of variables $q=A^{1/2}x$ shows $\inf\sigma(H-\varepsilon_{\mathrm H} T_M)=\frac12\tr(A^{1/2}KA^{1/2})^{1/2}\ge\frac12\tr(\Kosc ^{1/4}A^{1/2}\Kosc ^{1/4})$. The scalar inequality $\sqrt{1-x}\ge1-\frac x2-\frac{x^2}2$ on $[0,1]$ and $M^2\le M$ give $A^{1/2}\ge1-\frac{\varepsilon_{\mathrm H}}2(1+\varepsilon_{\mathrm H})M$, hence $\inf\sigma(H-\varepsilon_{\mathrm H} T_M)\ge E_0(H)-\frac{\varepsilon_{\mathrm H}}4(1+\varepsilon_{\mathrm H})\tr(\Kosc ^{1/2}M)$. Since $\langle\pi\pi^{\mathsf T}\rangle_0=\frac12\Kosc ^{1/2}$, $\frac14\tr(M\Kosc^{1/2})=\langle T_M\rangle_0$. \qed

\subsection{Proof of the second claim of Lemma~\ref{lem:trace}}
\emph{Step 1.} For $v\in\Wext$, $\langle v,\omega^{-1}v\rangle\le\frac{2\diam\mathcal E}{\pi t_0^2}\|v\|^2$. Indeed $u:=X^{-1}v\in\Wext$ has $\|u\|\le\|v\|/t_0$, and $v=Sy=S\Pi1_{\mathcal E}y$ with $y:=P_\sigma Su\in L^2_\sigma(\mathcal E)$, $\|y\|\le\|v\|/t_0$. As $S^2\Pi\le1$ commutes with $\omega$, $\langle v,\omega^{-1}v\rangle\le\|1_{\mathcal E}\omega_\pi^{-1}1_{\mathcal E}\|\,\|y\|^2$, where $\omega_\pi^{-1}$ has the kernel $\kappa(r)=\frac{1-\cos\pi r}{2\pi^2r^2}\in[0,\bar\kappa(r)]$, $\bar\kappa(r):=\min(\frac14,\frac1{\pi^2r^2})$ decreasing. By Schur's test, rearrangement of the decreasing majorant and the isodiametric inequality ($|\mathcal E|\le|B(0,\diam\mathcal E/2)|$, \cite[Sect.~2.2]{EG15}), the norm is at most $\int_{B(0,\diam\mathcal E/2)}\frac{\dd w}{\pi^2|w|^2}=\frac{2\diam\mathcal E}\pi$.

\emph{Step 2.} In the free vacuum $q_{\mathrm i}$ has covariance $\frac12P_{\mathrm{in}}\omega^{-1}P_{\mathrm{in}}$, so
\[
\langle T_\Sigma\rangle_\vac=\tfrac14\tr_{\Wext}(\omega_{\mathrm{ee}}^{-1}\Kie^{\mathsf T}P_{\mathrm{in}}\omega^{-1}P_{\mathrm{in}}\Kie\omega_{\mathrm{ee}}^{-1})\le\tfrac14\dim\Wext\sup_{\|w\|=1}\langle \Kie u,\omega^{-1}\Kie u\rangle
\]
with $u:=\omega_{\mathrm{ee}}^{-1}w$. Since $u\in\Wext$, $\Kie u=P_{\mathrm{in}}\omega^2u=\omega^2u-\omega_{\mathrm{ee}}^2u$ and $\langle u,\omega^2u\rangle=\langle u,\omega_{\mathrm{ee}}^2u\rangle=\|w\|^2$. Hence
\[
\langle \Kie u,\omega^{-1}\Kie u\rangle\le2\langle u,\omega^3u\rangle+2\langle\omega_{\mathrm{ee}}^2u,\omega^{-1}\omega_{\mathrm{ee}}^2u\rangle\le2\|\omega\|+\frac{4\diam\mathcal E}{\pi t_0^2}\|\omega\|^2
\]
by Step~1 applied to $\omega_{\mathrm{ee}}^2u\in\Wext$, $\|\omega_{\mathrm{ee}}^2u\|=\|\omega_{\mathrm{ee}}w\|\le\|\omega\|$. With $\|\omega\|\le\pi$ and the first claim of Lemma~\ref{lem:trace} the bound follows, without any commutation of $P_{\mathrm{ext}}$ with $\omega$. \qed

\section{Proofs of the propositions of Section~\ref{sec:estimators}}\label{app:core}

Throughout, constants are chosen in the order of Remark~\ref{rem:order}, and the parameters of the core locator are those of the design (D$'$) of Section~\ref{sec:constructive}. For a field functional $G$ we write $\Rm(G)$ for its covariance defect and use the chain rule of Remark~\ref{rem:cov}.

\begin{lemma}[A Lipschitz local term satisfies (W$\pm$)]\label{lem:Wpm}
Let $\mathcal C_e$ be a cube of side $n_e$, $\Xi_0\in[0,1]$ a functional of the field depending only on the field in $B(0,\rho_\Xi)$ with $|\Xi_0(A+B)-\Xi_0(A)|\le \LipXi\sup_{|y|\le\rho_\Xi}|B(y)|$, and $\Xi(x,a):=\Xi_0(\transl_{-x}a)$. For $t',F\ge0$ and $\sigma>0$ let
\[
\tilde h[a]:=\frac c2(p+ea(x))^2+t'F\bigl(1-\Xi(x,a)\bigr)+\sigma\qquad\text{on }L^2(\mathcal C_e)\text{ (Dirichlet)}.
\]
Then $\tilde h[a]\ge0$, its form domain does not depend on $a$, and the hypothesis \cite[\SerAnum{eq:W}]{SerA} holds with $\mathcal C_e$ replaced by the cube $\mathcal C'_e\supset\mathcal C_e+B(0,\rho_\Xi)$ of side $n_e+2\rho_\Xi$ in the norm $\|b\|_{\infty,\mathcal C'_e}$, and with the constant $C'_W=\frac c2e^2+(t'F\LipXi)^2/(4\sigma)$. Consequently \cite[\SerAnum{thm:periodization}(a)]{SerA} applies to $\tilde h$, and since $\inf\sigma(\tilde h^{\bullet})=\inf\sigma(h'^{\bullet})+\sigma$ for $h'=\tilde h-\sigma$ ($\bullet=\infty,L$), for every $\delta\in(0,\frac12]$ and $L\ge L_0(\delta,n_e+2\rho_\Xi,C'_W)$
\[
\inf\sigma(h'^\infty)\ge\frac{1-\delta}{1+\delta}\inf\sigma(h'^L)-\delta-\frac{2\delta}{1+\delta}\sigma.
\]
\end{lemma}

\begin{proof}
Write $T_{\mathrm{el}}[a]=\frac c2(p+ea)^2$ and $V[a]=t'F(1-\Xi(\cdot,a))\in[0,t'F]$. The magnetic Laplacian satisfies (W$\pm$) with $C_W=\frac c2e^2$ and the norm on $\mathcal C_e\subset\mathcal C'_e$ \cite[\SerAnum{sec:periodization}]{SerA}. For $x\in\mathcal C_e$ the functional $\Xi(x,\cdot)$ reads the field only in $\mathcal C'_e$, so $|V[a+b]-V[a]|\le t'F\LipXi\|b\|_{\infty,\mathcal C'_e}\le\delta\sigma+\frac{(t'F\LipXi)^2}{4\delta\sigma}\|b\|^2_{\infty,\mathcal C'_e}$ by the arithmetic-geometric mean inequality. Hence, using $V\ge0$ and $\delta\le1$,
\[
\tilde h[a+b]\ge(1-\delta)T_{\mathrm{el}}[a]-\tfrac{C_W}\delta\|b\|^2+V[a]-\delta\sigma-\tfrac{(t'F\LipXi)^2}{4\delta\sigma}\|b\|^2+\sigma\ge(1-\delta)\tilde h[a]-\tfrac{C'_W}\delta\|b\|^2,
\]
since $(1-\delta)(V[a]+\sigma)\le V[a]+\sigma-\delta\sigma$. Similarly $\tilde h[a+b]\le(1+\delta)T_{\mathrm{el}}[a]+C_W(1+\delta^{-1})\|b\|^2+V[a]+\delta\sigma+\frac{(t'F\LipXi)^2}{4\delta\sigma}\|b\|^2+\sigma\le(1+\delta)\tilde h[a]+C'_W(1+\delta^{-1})\|b\|^2$, since $V[a]+\sigma+\delta\sigma\le(1+\delta)(V[a]+\sigma)$. The form domain is that of the Dirichlet magnetic Laplacian, as $V$ is bounded. The last statement is \cite[\SerAnum{thm:periodization}(a)]{SerA} for $\tilde h$, rewritten for $h'$: $\inf\sigma(h'^\infty)+\sigma\ge\frac{1-\delta}{1+\delta}(\inf\sigma(h'^L)+\sigma)-\delta$.
\end{proof}

\subsection{The configuration-wise mean-field bound (Proposition~\ref{prop:MF})}
\emph{(a).} Let $\mathcal C_e$ be the electron box, of side $n_e:=2s\lceil K_e\ell/(2s)\rceil\in2s\mathbb N$ with the constant $K_e$ of (D$'$), so that $K_e\ell\le n_e<K_e\ell+2s$; hence the IMS loss in \eqref{eq:630P} satisfies $C_{\mathrm{IMS}}c/n_e^2\le t_*F/16$ for all $\ell$ (by $n_e\ge K_e\ell$, $F=cF[1]/R_*^2$, see (c), and $R_*\le\kappa_+\ell$) while $n_e=o(\ell\log\lambda)$. Let $\mathcal C_e^+:=\mathcal C_e+[-r_+,r_+]^3$ be the concentric box of (b) and $\mathcal C_{\mathrm b}:=\mathcal C_e^++[-s,0]^3$, a box of side $n_e+2r_++s\in s\mathbb N$ partitioned into blocks of side $s$. For a translate $t\in[0,s)^3$ let $\mathcal P_t$ be its blocks translated by $t$; then $\mathcal C_{\mathrm b}+t\supset\mathcal C_e^+\supset\mathcal C_e$. Apply \cite[\SerAnum{prop:configBO}]{SerA} with the block box $\mathcal C_{\mathrm b}+t$, the blocks $\mathcal P_t$ and the electron region $\mathcal C_e$. Since $n_e\in2s\mathbb N$ and $r_+\in s\mathbb N$, the box $\mathcal C_{\mathrm b}$ centred near the electron position $0$ has corners in $\mathbb Z^3$ (as the electron box in the companion paper, where $n_e\in2s\mathbb N$ is required for the same reason), and $\mathcal C_{\mathrm b}+t$ has corners in $t+\mathbb Z^3$, which is allowed by part (iv) of that proposition. With $H'_{\mathrm{BO}}:=\Hf+W'(q)$, $W'(q)$ the electronic ground state energy at fixed configuration on $\mathcal C_e$ (Dirichlet), which does not depend on $t$, this gives (a) with
\[
J^{(t)}_q(\psi):=\Bigl(1-\frac1{\nu^2}\Bigr)T(|\psi|)+\sum_{Q\in\mathcal P_t}\Bigl[\frac{h_Q^2}{8u'_Q}+\gamma\bar\rho_Q\min(u'_Q,u_b)\Bigr]-\mathcal V_0-\mathcal D_\theta,\qquad\psi\in H^1_0(\mathcal C_e),\ \|\psi\|=1 .
\]
Blocks without electron density only add the non-negative terms $h_Q^2/(8u'_Q)$.

\emph{(b).} We need the average over $t$ with the infimum over $\psi$ taken last. This is \cite[\SerAnum{prop:configBO}(v)]{SerA} with $\mu$ the uniform probability measure on $[0,s)^3$ (only the blocks and the sinc points are translated, the field and its coordinates are not), which applies since $\mathcal C_e\subset\bigcap_{t\in[0,s]^3}(\mathcal C_{\mathrm b}+t)$: it gives $H'_{\mathrm{BO}}\ge\bar{\mathcal U}(q):=\inf_\psi\bar J_q(\psi)$ with $\bar J_q:=\mathrm{avg}_tJ^{(t)}_q$. Let $\Xi_0\in[0,1]$ be a functional of the field relative to the electron, depending only on the field in $B(0,\rho_\Xi)$ and Lipschitz, $|\Xi_0(A+B)-\Xi_0(A)|\le \LipXi\sup_{|y|\le\rho_\Xi}|B(y)|$, and assume that $\bar{\mathcal U}(q)\ge(1-\varepsilon_1)F-c_{\mathrm{BO}}$ for all $q$ and $\bar{\mathcal U}(q)\ge(1+t')F-c_{\mathrm{BO}}$ whenever $\Xi_0(\transl_{-x}A_q)>0$ (for some $t'>0$) for some $x$ in the electron box. We claim
\begin{equation}\label{eq:630P}
H(0)\ge(1-\varepsilon_1)F+t'F\,\Xi_0(A)-c_{\mathrm{BO}}-C_{\mathrm{IMS}}\frac c{n_e^2}.
\end{equation}
Indeed, let the box model be the operator $h^L$ of \cite[\SerAnum{sec:periodization}]{SerA} with $h_{\mathrm{el}}[a]=\frac c2(p+ea)^2$ on $L^2(\mathcal C_e)$ with Dirichlet conditions (the working Hamiltonian of \cite{SerA} in the dictionary of Section~\ref{ssec:inputA}, with the kinetic coefficient $c$), written in the coordinates $q$ of \cite[\SerAnum{sec:finite-model}]{SerA}; then $h^L=\Hf+\int^\oplus h_{\mathrm{el}}[B_q]\ge H'_{\mathrm{BO}}$ is the form to which (a) applies, and its version $h^\infty$ without the torus is related to the full-space model $H_{\mathrm{full}}$ of Setting~\ref{set:full} on states in $C_c^\infty(\mathcal C_e)\otimes_{\mathrm{alg}}\mathcal F_{\mathrm{fin}}(C_c^\infty)$ (to which IMS localization with a smooth localizing function reduces) by \cite[proof of \SerAnum{lem:Qform}, step~(iv)]{SerA}: in the Schr\"odinger representation of the field, $\frac c2\|(p+eA)\Psi\|^2=\int d\mu(q)\,\frac c2\|(p+eA_q)\Psi_q\|^2$, where $A_q$ is the Gaussian field configuration and $\Psi_q$ the corresponding fiber of $\Psi$ (an identity for each configuration, so it applies with the bounded potential $t'F(1-\Xi)$ added); the passage to $H(0)$ is the fiber decomposition below. In this box model the Born--Oppenheimer inequality and the case distinction give $h^L-t'F\,\Xi\ge(1-\varepsilon_1)F-c_{\mathrm{BO}}$ with $\Xi(x,A):=\Xi_0(\transl_{-x}A)$. The removal of the torus is Lemma~\ref{lem:Wpm} with the electronic operator $h'[a]=\frac c2(p+ea(x))^2+t'F(1-\Xi(x,a))\ge0$ and $\sigma=1$: the Lipschitz bound is converted by the arithmetic-geometric mean inequality into the form (W$\pm$) (with the cube enlarged to $\mathcal C_e+B(\rho_\Xi)$), and the additive error $\delta(1+2/(1+\delta))$ vanishes as $\delta\to0$ and $L\to\infty$ along odd $L$ (the bound of \cite[\SerAnum{prop:configBO}]{SerA} is stated on $\mathbb T_L$ with $L$ odd; Lemma~\ref{lem:Wpm} holds for every $L\ge L_0$, so the subsequence suffices). IMS localization of the electron in boxes $\mathcal C_e$ of side $n_e$ costs $C_{\mathrm{IMS}}c/n_e^2$, and since $\Xi$ is translation covariant the fiber decomposition gives \eqref{eq:630P}. In this transfer the blocks of every translate cover $\mathcal C_e^+$, with
\[
r_+:=s\Bigl\lceil\frac{r_W+R_c+2(b_\theta+s)}s\Bigr\rceil ,
\]
so that $\mathcal C_e^+\supset B(x,r_W+R_c)$ for every $x\in\mathcal C_e$ (the ball $B(x,r_W+R_cR_*)$ of Section~\ref{sec:constructive}), with the margin $2(b_\theta+s)$ for the support of the core kernel, and its side $n_e+2r_+$ is in $s\mathbb N$. The errors $\mathcal V_0+\mathcal D_\theta$ grow like the volume $(n_e+2r_++s)^3$ of $\mathcal C_{\mathrm b}$ (\cite[\SerAnum{prop:configBO}(ii)]{SerA}) and remain $o(F)$ since $n_e+2r_++s\lesssim\ell\log\lambda$.

\emph{(c).} Let $x:=\gamma\bar\rho_Q$. \emph{Step c1 (low branch).} For blocks with $u'_Q\le u_b$, Lemma~\ref{lem:AMGM} gives $h_Q^2/(8u'_Q)+xu'_Q=h_Q\sqrt{x/2}+\Delta_Q$.

\emph{Step c2 (high branch).} For blocks above the field ceiling, $u'_Q>u_b$, the bracket contains $xu_b$ instead of $xu'_Q$, and $h_Q^2/(8u'_Q)+xu_b$ can be smaller than $h_Q\sqrt{x/2}$. Completing the square as in \cite[\SerAnum{prop:configBO}(iii)]{SerA} we write $h_Q^2/(8u'_Q)+xu_b=h_Q\sqrt{x/2}+\Delta_Q-h_Q^2/(2u_b)$ with
\[
\Delta_Q=\Delta^{\mathrm{hi}}_Q:=u_b\Bigl(\sqrt x-\frac{h_Q}{2\sqrt2u_b}\Bigr)^2+\frac{3h_Q^2}{8u_b}+\frac{h_Q^2}{8u'_Q}\ge0,
\]
and put the ceiling loss into $\Eloss$. It is needed only for blocks with electron density: if $\bar\rho_Q=0$, then $x=0$, the bracket is $h_Q^2/(8u'_Q)$, and we take $\Delta_Q:=h_Q^2/(8u'_Q)\ge0$ (the identity of Lemma~\ref{lem:AMGM} with $x=0$), which satisfies bound (a) of Lemma~\ref{lem:AMGM} since $h_Q^2/(8u'_Q)\ge h_Q^2/(8u_t)$ for $u'_Q\le u_t$, while bound (b) concerns only $x>0$. The deficits $\Delta^{\mathrm{hi}}_Q$ satisfy the two bounds of Lemma~\ref{lem:AMGM}: for a threshold $u_t$ and $x_t:=h_Q^2/(8u_t^2)$, (a) if $u'_Q\le u_t$, then $\Delta^{\mathrm{hi}}_Q\ge3h_Q^2/(8u_b)>h_Q^2/(32u_t)$; (b) if $u'_Q\ge u_t$, $x\ge(1+\eta)x_t$ and $\eta\le3+2\sqrt3$, then $\Delta^{\mathrm{hi}}_Q\ge\frac{h_Q^2}{8u_t}(\sqrt{1+\eta}-1)^2$ --- for $u_b\ge u_t$ from the first term, since $\sqrt x-h_Q/(2\sqrt2u_b)\ge(\sqrt{1+\eta}-1)h_Q/(2\sqrt2u_t)$, and for $u_b<u_t$ from the second, since $(\sqrt{1+\eta}-1)^2\le3$.

\emph{Step c3 (ceiling loss).} As $\psi\in H^1_0(\mathcal C_e)$, at most $((n_e+2s)/s)^3$ blocks carry a ceiling loss, and by $h_Q\le\pi|Q|$ and $u_b=N_b/(\nu^2e^2)$ \cite[\SerAnum{prop:configBO}(ii)]{SerA} (with $u_b$ as defined in the hypotheses of that proposition) the ceiling loss is at most $\frac{\pi^2}2\nu^2e^2(n_e+2s)^3s^3/N_b$; it does not grow with $r_+$, and it is $o(F)$.

\emph{Step c4 (summation and rescaling).} Summing over the blocks, the lower bound on $h_Q$ in \cite[\SerAnum{prop:configBO}(ii)]{SerA} and the Cauchy--Schwarz inequality $|Q|^{1/2}m_Q(\psi)^{1/2}\ge\int_Q|\psi|$ (as in the proof of part (iii) there) give $\sum_Qh_Q\sqrt{\gamma\bar\rho_Q/2}\ge(1-\eta_{\mathrm A})^{1/2}(1-C\log(2+s)/s)_+\beta_{\mathrm{MF}}\|\psi\|_1$, with the mean-field constant $\beta_{\mathrm{MF}}$ of \cite{SerA}. $\Eloss:=\mathcal V_0+\mathcal D_\theta+\frac{\pi^2}2\nu^2e^2(n_e+2s)^3s^3/N_b$, where $\mathcal V_0,\mathcal D_\theta$ enter $J^{(t)}_q$ with a minus sign and the last term bounds the ceiling loss; $\Eloss$ does not depend on $q$, $t$ or $\psi$. Finally the scaling $\varphi(y)=R_*^{3/2}|\psi|(R_*y)$ with $R_*$ the minimizing scale of $T+\beta_{\mathrm{MF}}\|\cdot\|_1$ gives $T(|\psi|)+\beta_{\mathrm{MF}}\|\psi\|_1=F\,\mathcal E_1[\varphi]/F[1]$, $F=\inf(T+\beta_{\mathrm{MF}}\|\cdot\|_1)$; as $T=\frac c2\|\nabla\cdot\|^2$, comparing the kinetic terms gives $F=cF[1]/R_*^2$. With $1-\varepsilon_1:=\min\{1-\nu^{-2},(1-\eta_{\mathrm A})^{1/2}(1-C\log(2+s)/s)_+\}$ this proves (c). Letting $\nu\to\infty$ and $\eta_{\mathrm A}\to0$ slowly with $\ell$ (with $s=\lfloor\log\ell\rfloor$, $N_b\ge c'\eta_{\mathrm A}s/K_V$ and $K_V$ polylogarithmic as in \cite[\SerAnum{prop:configBO}]{SerA}) gives $\varepsilon_1+\Eloss/F\to0$. \qed

\subsection{Properties (T1) and (T2) of the core indicator}
Let $\tilde\chi$ pass from $1$ to $0$ on $[(1-\vartheta)g_s,g_s]$ with $\|\tilde\chi'\|\le2/(\vartheta g_s)$, and let $0<f_+<1$, $2g_s<\tau$, $g_s/(1-f_+)<\tau$. Choose the thresholds
\[
u'_s:=|Q|\,\sat_\tau^{-1}\bigl((1-\vartheta')(1-\vartheta)g_s\bigr),\qquad u_s:=|Q|\,\sat_\tau^{-1}\bigl(g_s/(1-f_+)\bigr),
\]
with $0<\vartheta'<1$. For the cube $C(y)$ of the core kernel, fractions being averages over block centres:

\begin{lemma}\label{lem:T1T2}
(T1) If $d(y)<1$, at least a fraction $p_d:=\vartheta'(1-\vartheta)g_s/\tau$ of $C(y)$ lies outside $D^-$. (T2) If $d(y)>0$, at least a fraction $q_d:=f_+$ of $C(y)$ lies in $D^+$. The ratio $u_s/u'_s$ tends to $1$ as $\vartheta,\vartheta',f_+\to0$.
\end{lemma}
\begin{proof}
(T1) $d(y)<1$ means $\bar g:=g_b(y)>(1-\vartheta)g_s$. Since $0\le \sat_\tau\le\tau$, the fraction of the cube where $\sat_\tau\ge x':=(1-\vartheta')(1-\vartheta)g_s$ is at least $(\bar g-x')/\tau\ge p_d$, and there $u'_Q>u'_s$. (T2) $d(y)>0$ means $\bar g<g_s$; by Markov the fraction where $\sat_\tau\ge g_s/(1-f_+)$ is at most $1-f_+$, and on the rest $u'_Q\le u_s$. The last claim follows from continuity of $\sat_\tau^{-1}$.
\end{proof}

\subsection{Covariance, fluctuation and global bounds (Propositions~\ref{prop:Mprime} and~\ref{prop:global})}
Let $v_Y$ be the unprojected gradient of the core-level $g_b(Y)$ as in Section~\ref{sec:proof}, so that $\nabla g_b(Y)=P_Tv_Y$, and $\bar C_g:=\sup_Y\|v_Y\|^2$; by the saturation and Schur's test for the unprojected block gradients, as in Lemma~\ref{lem:Mw}, $\bar C_g\le2\tau/(\mathrm e\,b_\theta^3)$. The $v_Y$ are supported in the cubes of side $2(b_\theta+s)$ around $Y$, so $\langle v_Y,v_{Y'}\rangle=0$ for $|Y-Y'|_\infty\ge2(b_\theta+s)$, and the contraction property of $P_T$ and Schur's test give $\|\int a\,\tilde\chi'(g_b)\nabla g_b\|^2\le64(b_\theta+s)^3\|\tilde\chi'\|_\infty^2\bar C_g\int|a|^2$. Write $r:=r_{a_c}$, $W:=W_{\mathrm{ms}}$, and abbreviate $\phi(\mathcal M(Y))$ to $\phi$; recall $a_c=W_*$, so $r(W)\ge\max(W,W_*)$, and $\phi\le\mathrm e$ since $\mathcal M\le1$.

\emph{Covariance defect.} By covariance, $X_c(\transl_aA)=\int(Y+a)\chi_W(Y+a)\phi\dd Y/r(\int\chi_W(Y+a)\phi\dd Y)$; differentiating at $a=0$ and using $\Rm=I-\partial_a(\cdot)(\transl_aA)|_{a=0}$,
\[
\Rm(X_c)=\Bigl(1-\frac W{r(W)}\Bigr)I-\frac1{r(W)}\int\bigl(Y-r'(W)X_c\bigr)\nabla\chi_W(Y)^{\mathsf T}\phi\dd Y .
\]
Here $\nabla\chi_W\ne0$ only in the edge shell $r_W-w\le|Y|\le r_W$, and $\phi>0$ requires a point with $d>0$ in $B(Y,R_c)$; so the integral vanishes outside the event $\Gamma'_R:=\{\exists y:r_W-w-R_c\le|y|\le r_W+R_c,\ d(y)>0\}$. The first term vanishes when $W\ge2a_c=2W_*$, i.e.\ outside $S_M$. Hence $\Rm(X_c)=0$ on $(\Gamma'_R\cup S_M)^c$, which is the last statement of Proposition~\ref{prop:events}(a) and, since $\Gamma'_R\subset\mathcal B_\theta$, Proposition~\ref{prop:Mprime}(b). Globally $|X_c|\le r_W$, $|Y-r'X_c|\le2r_W$ and $r\ge W_*$, so
\[
\|\Rm(X_c)\|\le1+\frac{2r_W\|\nabla\chi_W\|_\infty\mathrm e\,|B(0,r_W)|}{W_*},
\]
a polynomial in $\ell$ and $\log\lambda$.

\emph{Carr\'e du champ.} $\nabla X_c=\int \mathsf c(Y)\nabla\mathcal M(Y)\dd Y$ with $\mathsf c:=(Y-r'(W)X_c)\chi_W\phi'(\mathcal M)/r(W)$. Writing $\nabla\mathcal M$ as a ball average of $\tilde\chi'(g_b)\nabla g_b$, we get $\nabla X_c=P_T\int\tilde{\mathsf c}(y)\tilde\chi'(g_b(y))v_y\dd y$ with $\tilde{\mathsf c}$ the ball average of $\mathsf c$, and $\|\tilde{\mathsf c}\|_2\le\|\mathsf c\|_2$; the bound above gives $\Gamma_{X_c}\le64(b_\theta+s)^3\|\tilde\chi'\|^2_\infty\bar C_g\|\mathsf c\|_2^2$. From $\phi'^2\le2\phi^2+2\mathrm e^{4\Mzero}\|\zeta'\|^2_\infty1\{\Mzero\le\mathcal M\le2\Mzero\}$, $\int\chi_W\phi^2\le \mathrm e\,W$ and $W/r(W)^2\le1/r(W)\le1/W_*$,
\[
\Gamma_{X_c}\le64(b_\theta+s)^3\|\tilde\chi'\|_\infty^2\bar C_g(2r_W)^2\Bigl[\frac{2\mathrm e}{W_*}+\frac{2\mathrm e^{4\Mzero}\|\zeta'\|_\infty^2|B(0,r_W)|}{W_*^2}\Bigr]
\]
for all configurations. In the design, $b_\theta\asymp\ell$, $\bar C_g\lesssim\ell^{-5}$, $\|\tilde\chi'\|_\infty\lesssim g_s^{-1}\asymp\ell^2$, $r_W\asymp\ell\log\lambda$, $W_*\asymp\mathrm e^{4\Mzero}\ell^3$, so the prefactor is $\lesssim\ell^{3-5+4}(\ell\log\lambda)^2=\ell^4(\log\lambda)^2$, the first term in the bracket is $\lesssim\ell^{-3}$ and the second $\lesssim\ell^{-3}(\log\lambda)^3$: $\Gamma_{X_c}\lesssim\ell(\log\lambda)^5$. The second term comes from the transition band of $\zeta$ and does not vanish on the good set. This proves Proposition~\ref{prop:Mprime}(a), and with the global bound on $\Rm(X_c)$ Proposition~\ref{prop:global} for $X_c$. For $\Xw$, $\|\Rm(\Xw)\|\le\|\mathsf M\|\|\Rm(X_c)\|$ with $\|\mathsf M\|\le1+2c_\rho(\int_{\mathrm{shell}}d_w\dd Y/b^3)/M^w_*$, and the splitting of the proof of Lemma~\ref{lem:Mw} into the parts through $d_w$ and through $X_c$, with $M_*$ replaced by $a_w=M^w_*\le r_{a_w}(M)$ (and $M/r_{a_w}(M)^2\le1/a_w$) and $\|a+b\|^2\le2\|a\|^2+2\|b\|^2$, gives $\Gamma_{\Xw}\le\mathrm{poly}(\ell)$ globally. \qed

\subsection{Rare events (Proposition~\ref{prop:events})}\label{ssec:events}
The event of Proposition~\ref{prop:events} is a core-level depression in a given shell of distances from the electron; it is rare, with an exponent proportional to the width of the shell divided by $\ell$. Its proof is an Agmon estimate whose margin comes from a surgery at core level. Physically, a core-level depression is the core of the bubble, and the bubble sits on the electron: a core-level depression far from the electron costs an energy of the order $\mu(g_s)\asymp\ell^5$ of the mean-field energy, while the translation coefficient is $c\asymp\ell^7$, so the Agmon length is $\sqrt{c/\mu}\asymp\ell$ and the probability decays exponentially in the distance divided by $\ell$.

\emph{Core-level surgery margin.} Let $Q_Y:=Y+[-(b_\theta+\frac s2),b_\theta+\frac s2]^3$ be the support of the core kernel at $Y$, at distance $\ge D_s$ from an electron box $\mathcal C$ of side $2n$, and $\mathcal E:=Q_Y+B(D_\eta+\sqrt3s)$ with $t_0=\frac12$. On states with electron in $\mathcal C$, Theorem~\ref{thm:gauge} (with a constant $\eta_S$), Lemma~\ref{lem:seam} and Lemma~\ref{lem:hardy}(iii) applied with uniform weights give
\begin{equation}\label{eq:HS2}
H-E\ge(1-\delta)\mathcal V_Y(A)-S_0-\kappa_1|r_{\mathrm e}(x)|^2-\delta^{-1}T_\Sigma,\qquad S_0:=\eta_S\Ebox+c^{(r)},\ \kappa_1:=\kappa_{\eta_S},
\end{equation}
with $\mathcal V_Y\ge-B_-$, and the following lower bound on $\mathcal V_Y$ in terms of the level $\bar g:=g^{(\tau)}_{b_\theta}(Y)$.

\begin{lemma}\label{lem:thetamargin}
Let $\rho_*(1-\mathrm e^{-1})>a+2\theta_\delta$ (design), $q_{\min}:=1-(a+2\theta_\delta)/((1-\mathrm e^{-1})\rho_*)>0$ and $V_b:=|Q_Y|=(2b_\theta+s)^3$. For $0<\bar g\le(a+2\theta_\delta)g_s$,
\[
\mathcal V_Y\ge\mu(\bar g):=V_b\Bigl[\frac{c_{\mathrm{lev}}}{\bar g}-B'\Bigr],\qquad c_{\mathrm{lev}}:=\frac{\varepsilon_{\mathrm H}(1-\mathrm e^{-1})\kappa_{\mathrm e}^2\rho_{\mathrm m}^2q_{\min}^2}8,\quad B':=\frac{\varepsilon_{\mathrm H}(1+\varepsilon_{\mathrm H})\pi\rho_{\mathrm m}}4;
\]
$c_{\mathrm{lev}},B'$ depend only on the design.
\end{lemma}
\begin{proof}
For each translate of the block lattice apply Corollary~\ref{cor:margin} with weights $a_j=1$; the left side does not depend on the translate, so we may average, which turns the sum into $s^{-3}\int_{Q_Y}u'^{-1}_{Q(z)}\dd z=|Q|^{-1}\int\dd\nu(z)/u_{Q(z)}$ with $\nu:=s^{-3}1_{Q_Y}(z)\dd z$, $\nu_{\mathrm{tot}}=V_b/|Q|$. The level of $\nu$ is $\bar g$, and $\bar g\le(a+2\theta_\delta)g_s<(1-\mathrm e^{-1})\tau$ for $\tau=\rho_*g_s$, with $q_{\bar g}\ge q_{\min}$; Lemma~\ref{lem:level}(a) gives $\int\dd\nu/u_Q\ge(1-\mathrm e^{-1})q_{\min}^2V_b/(|Q|\bar g)$. The zero-point term is at most $\frac\pi4\rho_{\mathrm m}V_b$ since $\|\Kosc^{1/2}\|\le\pi$.
\end{proof}
The other conditions on $\rho_*$ ((T1), (T2) and $p_d>0$) are lower bounds, so the condition of Lemma~\ref{lem:thetamargin} only fixes $\rho_*$ large.

\emph{The window-edge estimate.} Fix design constants $a>1$, $\theta_\delta>0$, $\eta_p\in(0,1)$, $0<\eta<\frac12$, a box side $n$, a surgery distance $D_s$, and put $r^{\mathrm{eff}}_{\mathrm{in}}:=\sqrt3(2n+b_\theta+s+\rho_\theta)+D_s$ (with $\rho_\theta$ as below). For a shell $r-w\le|Y|\le r$ (relative to the electron) with $L_{\mathrm{sh}}:=r-w-r^{\mathrm{eff}}_{\mathrm{in}}>0$, a slope length $0<L_*\le L_{\mathrm{sh}}$, $R_{\mathrm{out}}:=r+L_*+b_\theta$, $\rho_\theta:=\theta_\delta g_sb_\theta/3$ and $N':=\frac{4\pi}3\bigl((R_{\mathrm{out}}+2\sqrt3n)/\rho_\theta+\frac{\sqrt3}2\bigr)^3$ (a bound for the number of candidates of Step~1 of the proof below; since $R_{\mathrm{out}}\ge r^{\mathrm{eff}}_{\mathrm{in}}\ge2\sqrt3n$, $N'\le24\cdot\frac{4\pi}3(R_{\mathrm{out}}/\rho_\theta)^3$ when $\rho_\theta\le R_{\mathrm{out}}$), set
\begin{gather*}
S'_{\mathrm{part}}:=\frac{\bar C_g}{8\theta_\delta^2g_s^2}\log^2\frac{N'(\mu(2\theta_\delta g_s)+B_-)}{\eta_p\mu(g_s)},\\
\mu'_{\mathrm{net}}(g):=(1-\delta)\mu(g+2\theta_\delta g_s)-S_0-S_{\mathrm{quad}}-\frac{cC_\chi}{2n^2}-S'_{\mathrm{part}}-\eta_p\mu(g_s),
\end{gather*}
where $S_{\mathrm{quad}}:=\kappa_1b_1^2+\delta^{-1}b_2^2$ with thresholds $b_r^2$ for the two quadratic forms of \eqref{eq:HS2}.

\begin{lemma}[Window-edge estimate]\label{lem:edge}
Assume
\[
\text{(G)}\quad0<\rho_\theta\le2n,\qquad\text{(H)}\quad\mu'_1:=\mu'_{\mathrm{net}}(ag_s)>0,\qquad\text{(L)}\quad L_*\le ag_s\sqrt{c/\bar C_g},
\]
and let $\mAg:=(1-2\eta)\mu'_1$. Then
\begin{multline*}
P_\psi\bigl(\exists Y:r-w\le|Y|\le r,\ g_b(Y)\le g_s\bigr)\le\Pi''\exp\Bigl(-2\,\frac{a-1}a\sqrt{\frac{2\eta\mu'_1}c}\,L_*\Bigr)\\
+\frac{N_{\mathrm{ov}}N'}\mAg\sum_{r'=1,2}\kappa_{r'}4N_{r'}\mathrm e^{-(b_{r'}^2-2T_{r'})/(4N_{r'})},
\end{multline*}
where $\Pi'':=(B_-+S_0+S_{\mathrm{quad}}+N_{\mathrm{ov}}N'\sum_{r'}\kappa_{r'}T_{r'}+cC_\chi/(2n^2)+S'_{\mathrm{part}}+\eta_p\mu(g_s))/\mAg$, $T_{r'},N_{r'}$ are uniform bounds on the trace and norm of the vacuum covariances of the two quadratic forms, and $N_{\mathrm{ov}}$ is the overlap number of the electron boxes.
\end{lemma}

\begin{proof}
\emph{Step 1 (multiplicative lower bound).} Localize the electron by an IMS partition subordinate to the lattice of cubes of side $n$ (loss $cC_\chi/(2n^2)$), whose supports, the electron boxes $\mathcal C$ of \eqref{eq:HS2}, have side at most $2n$ and overlap number $N_{\mathrm{ov}}$. In each box use Lemma~\ref{lem:softmax} with the candidate set $\mathcal K_{\mathcal C}$ of the points $Y$ of the lattice $c_{\mathcal C}+\rho_\theta\Z^3$ whose cube $Q_Y$ is at distance $\ge D_s$ from $\mathcal C$ and with $|Y-c_{\mathcal C}|\le R_{\mathrm{out}}+2\sqrt3n$, where $c_{\mathcal C}$ is the centre of $\mathcal C$; the cells $Y+[-\frac{\rho_\theta}2,\frac{\rho_\theta}2)^3$, $Y\in\mathcal K_{\mathcal C}$, are disjoint and lie in $B(c_{\mathcal C},R_{\mathrm{out}}+2\sqrt3n+\frac{\sqrt3}2\rho_\theta)$, so $|\mathcal K_{\mathcal C}|\le N'$, with the gap $\theta_\delta g_s$ in place of $\delta$ in Lemma~\ref{lem:softmax} (the letter $\delta$ in $\mu'_{\mathrm{net}}$ and $S_{\mathrm{quad}}$ remains the seam constant of \eqref{eq:HS2}) and $\beta_{\mathrm{sel}}=\log(N'(\mu(2\theta_\delta g_s)+B_-)/(\eta_p\mu(g_s)))/(\theta_\delta g_s)$; then apply \eqref{eq:HS2} for each box and candidate. Averaging over the translations of the box lattice by $t\in[0,n)^3$ (the candidate lattices move with the boxes, so the construction has period $n$) and passing to the fiber (Setting~\ref{set:full}) gives $H(0)-E\ge V-\kappa_1\mathcal W_1-\delta^{-1}\mathcal W_2$, where $\mathcal W_{r'}$ are averages, with weights summing to at most one and at most $N_{\mathrm{ov}}N'$ non-zero terms per point, of $J$-real quadratic forms. If some $Y$ with relative distance in $[r^{\mathrm{eff}}_{\mathrm{in}},R_{\mathrm{out}}]$ has $g_b(Y)\le g$, the nearest lattice point is a candidate (its cube is at distance $\ge D_s$ from $\mathcal C$ by the margin $\sqrt3(s+\rho_\theta)$ in $r^{\mathrm{eff}}_{\mathrm{in}}$, and it lies within $R_{\mathrm{out}}+\sqrt3n+\frac{\sqrt3}2\rho_\theta\le R_{\mathrm{out}}+2\sqrt3n$ of $c_{\mathcal C}$ by (G)) and has level at most $g+\theta_\delta g_s/2$ and the selection loses at most $\theta_\delta g_s$; the bad weight costs at most $\eta_p\mu(g_s)$ by the choice of $\beta_{\mathrm{sel}}$, and $S'_{\mathrm{part}}$ is the selection loss. By convexity, $\mathcal W_{r'}\le b_{r'}^2+\mathcal L_{r'}$ with $\mathcal L_{r'}$ the average of $(W-b_{r'}^2)_+$. Hence $V-\sum\kappa\mathcal W\ge\mu'_{\mathrm{net}}(g)-\sum\kappa_{r'}\mathcal L_{r'}$, and $V\ge-\mAg\Pi''$ always.

\emph{Step 2 (the Agmon function).} Let $\Upsilon$ be the radial profile vanishing on the shell $[r-w,r]$, linear with slope $ag_s/L_*$ from $0$ to $ag_s$ on the inner slope $[r-w-L_*,r-w]$ and on the outer moat $[r,r+L_*]$, rising with the same slope to $ag_s+\delta_0$ beyond them and constant afterwards ($\delta_0>0$ small, with $\delta_0L_*/(ag_s)\le b_\theta$); thus $|\Upsilon'|\le ag_s/L_*$ everywhere, and $\Upsilon<ag_s$ only on the slope, the shell and the moat. With $B:=B(0,R_{\mathrm{out}})$ let $\tilde G:=-\frac1{\beta'}\log\bigl(|B|^{-1}\int_Be^{-\beta'(g_b(Z)+\Upsilon(|Z|))}\dd Z\bigr)$, an integral soft-minimum: $\min_B(g_b+\Upsilon)\le\tilde G\le\min_B(g_b+\Upsilon)+\Delta_{\beta'}$ with $\Delta_{\beta'}\to0$ as $\beta'\to\infty$ uniformly in the configuration, since $g_b+\Upsilon$ is Lipschitz with a constant independent of the configuration ($|g_b(Y)-g_b(Y')|\le\sqrt3\,\tau|Y-Y'|/(2b_\theta+s)$ because $0\le\sat_\tau\le\tau$, and $|\Upsilon'|\le ag_s/L_*$). Its translation derivative is $\langle\nabla\Upsilon\rangle_p$, the average of $\nabla\Upsilon$ with respect to the Gibbs weight $p\propto\mathrm e^{-\beta'(g_b+\Upsilon)}$, plus a boundary flux at $|Z|=R_{\mathrm{out}}$, where $\Upsilon=ag_s+\delta_0$ (as $\delta_0L_*/(ag_s)\le b_\theta$); since $\int_B\mathrm e^{-\beta'(g_b+\Upsilon)}=|B|\mathrm e^{-\beta'\tilde G}$, on $\{\tilde G<ag_s\}$ the flux is at most $3\mathrm e^{-\beta'\delta_0}/(\beta'R_{\mathrm{out}})$, which vanishes as $\beta'\to\infty$ uniformly in the configuration. The soft minimum of Proposition~\ref{prop:shell} has the same form. Its carr\'e du champ is at most $\bar C_g$ (Jensen). Let $k':=\sqrt{2\eta\mu'_1/c}\;L_*/(ag_s)$, let $F$ be $C^1$, equal to a constant $F_{\mathrm{net}}\ge k'((a-1)g_s-3\Delta)$ on $[0,g_s+\Delta]$, with $F'=-k'$ on $[g_s+2\Delta,ag_s-\Delta]$, $|F'|\le k'$ everywhere and $F=0$ on $[ag_s,\infty)$ (rounding the corners in intervals of length $\Delta$), and $f:=F(\tilde G)$.

\emph{Step 3 (the Agmon condition).} On $\mathcal Z\subset\{\tilde G<ag_s\}$ a minimizer $Z_0$ of $g_b+\Upsilon$ over $B$ has $\Upsilon(|Z_0|)\le\tilde G<ag_s$, hence lies on the slope, the shell or the moat, at relative distance in $[r^{\mathrm{eff}}_{\mathrm{in}},R_{\mathrm{out}}]$, and $g_b(Z_0)<ag_s$; Step~1 with $g=ag_s$ gives $V-\sum\kappa\mathcal W\ge\mu'_1-\sum\kappa\mathcal L$. Since $|\langle\nabla\Upsilon\rangle_p|\le ag_s/L_*$ and $\Gamma_{\tilde G}\le\bar C_g$,
\[
\tfrac12\GAg_f\le\tfrac12k'^2\Bigl(c\,\frac{a^2g_s^2}{L_*^2}+\bar C_g\Bigr)+o(1)\le c\,k'^2\frac{a^2g_s^2}{L_*^2}+o(1)=2\eta\mu'_1+o(1)
\]
by (L). Hence on $\mathcal Z$, $V-\sum\kappa\mathcal W-\frac12\GAg_f\ge\mAg-\sum\kappa\mathcal L-o(1)$, the $o(1)$ vanishing as $\beta'\to\infty$. A linear profile suffices because the window constant $K$ is chosen at the end (Remark~\ref{rem:order}), so the value of the rate constant $c_{\mathrm{ev}}$ plays no role.

\emph{Step 4 (conclusion).} The event lies in $\{\min_B(g_b+\Upsilon)\le g_s\}\subset\{f=F_{\mathrm{net}}\}$, and $f\le F_{\max}=F_{\mathrm{net}}$. Corollary~\ref{cor:agmonquad} with $F=F_{\mathrm{net}}$ (so that no factor $\mathrm e^{2(F_{\max}-F)}$ appears), $\langle W\rangle\le T_{r'}$ for each of the at most $N_{\mathrm{ov}}N'$ forms (Corollary~\ref{cor:agmonquad} with constant majorants $\bar a=1$, since the weights depend on the configuration) and the tail bound of Lemma~\ref{lem:tail} give the claim after the limits.
\end{proof}

\emph{Asymptotics.} Take $\mu$ to be the lower bound of Lemma~\ref{lem:thetamargin}. In the design, $V_b=(2b_\theta+s)^3\asymp\ell^3$, $g_s\asymp\ell^{-2}$, $\mu(g_s)\asymp V_bc_{\mathrm{lev}}/g_s\asymp\ell^5$. Condition (G) holds for $\ell\ge\ell_1$, since $\rho_\theta=\theta_\delta g_sb_\theta/3\asymp\ell^{-1}$ and $n=K_n\ell$ give $\rho_\theta/n=O(\ell^{-2})$. Moreover $S_0\le\eta_S\Ebox+c^{(r)}$ with $\Ebox\asymp\ell^5$ and $\eta_S$ a small constant, $cC_\chi/n^2\lesssim\ell^5/K_n^2$ for $n=K_n\ell$, $S'_{\mathrm{part}}/\mu(g_s)\lesssim\ell^{-2}\log^2\ell$, $\kappa_1b_1^2$ is super-polynomially small for $D_s:=\ell^{1/2}$ (Corollary~\ref{cor:far}; $D_s\le\ell$ keeps $r^{\mathrm{eff}}_{\mathrm{in}}\asymp\ell$) and $\delta^{-1}b_2^2\lesssim\delta^{-1}b_\theta^4\log\ell$ (Lemma~\ref{lem:trace}) is $O(\ell^{-1}\log\ell)$ relative to $\mu(g_s)$. Since $(1-\delta)\mu((a+2\theta_\delta)g_s)\ge(1-\delta)V_bc_{\mathrm{lev}}/((a+2\theta_\delta)g_s)-O(V_b)$, (H) holds for $\ell\ge\ell_1$ once $\eta_S,\delta,K_n^{-1}$ are small constants and $\eta_p<1/(4(a+2\theta_\delta))$ (with $\delta\le\frac12$), with $\mu'_1\asymp\ell^5$; the constants $\eta_S,K_n$ are chosen after the auxiliary constants of (D$'$) (Lemma~\ref{lem:design}). Then $\sqrt{\eta\mu'_1/c}\asymp\ell^{-1}$, while the right side of (L) is $\asymp\ell^{-2}(\ell^7/\ell^{-5})^{1/2}=\ell^4$. For $L_*=L_{\mathrm{sh}}=K\ell\log\lambda$ the bound of Lemma~\ref{lem:edge} is at most $\mathrm{poly}(\ell)\lambda^{-c_{\mathrm{ev}}K}$, and the tail terms are at most any power of $\ell$ by the choice of $b_{r'}^2$.

\emph{Proof of Proposition~\ref{prop:events}.} Let $r_{\mathrm{max}}:=r_W+R_c$ and $r_{\mathrm{near}}:=r^{\mathrm{eff}}_{\mathrm{in}}+L_{\mathrm{near}}$ with the window length $L_{\mathrm{near}}=K\ell\log\lambda$. (b) is Lemma~\ref{lem:edge} applied once to the shell $[r_{\mathrm{near}},r_{\mathrm{max}}]$ (that is, $r=r_{\mathrm{max}}$, the shell width $r_{\mathrm{max}}-r_{\mathrm{near}}$ in place of $w$, and $L_{\mathrm{sh}}=L_*=L_{\mathrm{near}}$), with $R_{\mathrm{out}}=r_{\mathrm{max}}+L_{\mathrm{near}}+b_\theta$: the width of the shell enters only through $R_{\mathrm{out}}$, that is, polynomially through $N'$ and logarithmically through $S'_{\mathrm{part}}$, and neither (G), (H), (L) nor the exponent. (a): outside $\mathcal B_\theta$ every point $y$ with $d(y)>0$ and $|y|\le r_{\mathrm{max}}$ has $|y|<r_{\mathrm{near}}$. Since $\phi(\mathcal M(Y))>0$ requires such a point in $B(Y,R_c)$ and $|Y|\le r_W$, the weights of $X_c$ live in $B(0,r_{\mathrm{near}}+R_c)$, and as $W_{\mathrm{ms}}/r_{a_c}(W_{\mathrm{ms}})\le1$, $|X_c|\le r_{\mathrm{near}}+R_c<\rrel$. Moreover $d(y)>0$ means $g_b(y)<g_s$, and $[r_W-w-R_c,r_W+R_c]\subset[r_{\mathrm{near}},r_{\mathrm{max}}]$ by the design inequality $r_W-w-R_c\ge r_{\mathrm{near}}$, so $\Gamma'_R\subset\mathcal B_\theta$. The vanishing of $\Rm(X_c)$ outside $\Gamma'_R\cup S_M$ was shown above. For each $\kappa_p$ the constant $K$ is chosen after $\kappa_p$. \qed

\subsection{The mass event}
Proposition~\ref{prop:SM} is Proposition~\ref{prop:SMexpl}, proved in Section~\ref{sec:constructive}; Remark~\ref{rem:Mw} holds by definition of $S_M$ (Definition~\ref{def:good}).

\section*{Statements and Declarations}

\subsection*{Tool and computational resource disclosure}
This work was carried out with substantial assistance from Claude models
(Anthropic), accessed through the Claude Code agent in 2026. The AI system was
used to explore strategies of proof, to write out detailed arguments and
computations, to check them, and to prepare the code in the ancillary files.
The author has read and checked the entire paper and takes full
responsibility for its content. No AI system is an author.

\subsection*{Funding}
This research received no external funding.

\subsection*{Competing interests}
The author has no competing interests to declare that are relevant to the content of this article.

\subsection*{Data and code availability}
No data were generated or analysed in this study.

The script used to evaluate the constants of Part~I in interval arithmetic (Appendix~\ref{app:constants}; Python with \texttt{mpmath}) is available as an ancillary file and at \url{https://github.com/tabayashi0117/pauli-fierz-uv-limit-code}. It evaluates the closed-form constants of Corollary~\ref{cor:partI-explicit} in interval arithmetic and checks two closed-form integrals by quadrature; no other statement of the paper depends on a computation.

\end{document}